%% file: main.tex
\def\ExaNeStPaperMode{TR}

\documentclass[acmsmall,nonacm]{acmart}

\usepackage{environ,etoolbox}

\usepackage{multirow}
\usepackage{hyperref}
\usepackage{listings}
\usepackage{amsmath}
\usepackage{algorithmic}
\usepackage{array}
\usepackage{comment}
\usepackage{todonotes}
\usepackage[font=small]{subcaption}

\NewEnviron{TRonly}{\ifdefstring{\ExaNeStPaperMode}{TR}{\BODY}{}}
\NewEnviron{Jonly}{\ifdefstring{\ExaNeStPaperMode}{J}{\BODY}{}}

\newcommand{\AI}[1]{}
\newcommand{\FC}[1]{}

\AtBeginDocument{%
  \providecommand\BibTeX{{%
    \normalfont B\kern-0.5em{\scshape i\kern-0.25em b}\kern-0.8em\TeX}}}

\begin{document}

\title{The ExaNeSt Prototype:\\
       Evaluation of Efficient HPC Communication Hardware\\
       in an ARM-based Multi-FPGA Rack}


\subtitle{\textbf{Technical Report FORTH-ICS / TR-488  -  July 2023}}

\author{Manolis Ploumidis}
\email{ploumid@ics.forth.gr}
\author{Fabien Chaix}
\email{chaix@ics.forth.gr}
\author{Nikolaos Chrysos}
\email{nchrysos@ics.forth.gr}
\author{Marios Assiminakis}
\email{marios4@ics.forth.gr}
\author{Vassilis Flouris}
\email{vflouris@ics.forth.gr}
\author{Nikolaos Kallimanis}
\email{nkallima@isi.gr}
\authornote{Currently with: Industrial Systems Institute (ISI) / Athena RC \& University of Ioannina}
\author{Nikolaos Kossifidis}
\email{mick@ics.forth.gr}
\author{Michael Nik-\\oloudakis}
\email{michnik@ics.forth.gr}
\author{Polydoros Petrakis}
\email{ppetrak@ics.forth.gr}
\author{Nikolaos Dimou}
\email{ndimou@ics.forth.gr}
\author{Michael Gianioudis}
\email{yanoudis@ics.forth.gr}
\author{George Ieronymakis}
\email{ieronym@ics.forth.gr}
\author{Aggelos Ioannou}
\email{aggelos.ioannou@gmail.com}
\authornote{Currently with: Dept. of Computer Science, Lawrence Berkeley National Laboratory, CA, USA}
\author{George Kalokerinos}
\email{kalokerinos@ics.forth.gr}
\author{Pantelis Xirouchakis}
\email{pxirouch@ics.forth.gr}
\author{George Ailamakis}
\author{Astrinos Damianakis}
\email{a.damianakis@deddie.gr}
\authornote{Currently with the Hellenic Electricity Distribution Network Operator S.A.}
\author{Michael Ligerakis}
\email{ligeraki@ics.forth.gr}
\authornote{Currently also with EXAPSYS plc.}
\author{Ioannis Makris}
\email{makrisj@ics.forth.gr}
\authornotemark[4]
\author{Theocharis Vavouris}
\email{vavouris@ics.forth.gr}
\author{Manolis Katevenis}
\email{kateveni@ics.forth.gr}
\authornote{Also with the Dept. of Computer Science, University of Crete}
\author{Vassilis Papaefstathiou}
\email{papaef@ics.forth.gr}
\authornote{Currently also with the Dept. of Computer Science, University of Crete}
\author{Manolis Marazakis}
\email{maraz@ics.forth.gr}
\author{Iakovos Mavroidis}
\email{jacob@ics.forth.gr}
\authornotemark[4]
\affiliation{%
  \institution{Computer Architecture and VLSI Systems (CARV) Laboratory, Institute of Computer Science (ICS), Foundation for Research and Technology -- Hellas (FORTH)}
  \streetaddress{N. Plastira 100}
  \city{Vassilika Vouton, Heraklion, Crete}
  \country{Greece}
  \postcode{GR-700 13}
}
\renewcommand{\shortauthors}{Ploumidis, Chaix, et al.}
\renewcommand{\shorttitle}{The ExaNeSt Prototype: Efficient HPC Communication in a Multi-FPGA Rack}

\begin{abstract}

\newpage

{\bf ABSTRACT:}

We present and evaluate the ExaNeSt Prototype,
a liquid-cooled rack prototype consisting of 
256 Xilinx ZU9EG MPSoCs, 4 TBytes of DRAM, 16 TBytes of SSD,
and configurable interconnection 10-Gbps hardware.
We developed this testbed in 2016-2019
to validate the flexibility of FPGAs for experimenting with
efficient hardware support for HPC communication
among tens of thousands of processors and accelerators
in the quest towards Exascale systems and beyond.
We present our key design choices reagrding overall system architecture, 
PCBs and runtime software, and summarize insights resulting from measurement 
and analysis.

Of particular note, our custom interconnect includes a low-cost 
low-latency network interface, offering user-level zero-copy RDMA,
which we have tightly coupled with the ARMv8 processors in the MPSoCs.
We have developed a system software runtime on top of these features,
and have been able to run MPI. We have evaluated our testbed through 
MPI microbenchmarks, mini, and full MPI applications.
Single hop, one way latency is $1.3$~$\mu$s;
approximately $0.47$~$\mu$s out of these are attributed to
network interface and the user-space library that exposes its 
functionality to the runtime. Latency over longer paths increases as expected,
reaching $2.55$~$\mu$s for a five-hop path.
Bandwidth tests show that, for a single hop,
link utilization reaches $82\%$ of the theoretical capacity.
Microbenchmarks based on MPI collectives reveal that
broadcast latency scales as expected
when the number of participating ranks increases.
We also implemented a custom Allreduce accelerator
in the network interface,
which reduces the latency of such collectives by up to $88\%$.
We assess performance scaling through weak and strong scaling tests for HPCG, LAMMPS, and the miniFE mini application;
for all these tests, parallelization efficiency
is at least $69\%$, or better.

\end{abstract}




\keywords{FPGA, High-speed Network, Low-latency, High-density, co-design, PGAS, MPI, exascale.}

\maketitle


\include{1_introduction}

\include{2_related_work}

\include{3_platform}

\include{4_network}

\include{5_software}
\include{6_evaluation}
\include{7_hardware_accelerators}

\include{8_conclusion}


\begin{Jonly}
\bibliographystyle{ACM-Reference-Format}
\end{Jonly}
\begin{TRonly}
\bibliographystyle{abbrv}    
\end{TRonly}

\bibliography{paper}


\end{document}

%% file: 1_introduction.tex
\section{Introduction}
High Performance Computation (HPC) is a key necessity
in numerous domains of modern society, such as
weather forecast and the environment,
molecular dynamics, chemistry, biology and medicine,
fluid dynamics, engineering and industrial design,
astronomy and physics,
data analytics, economy and sociology, and so on.
All of these heavily depend on the processing capabilities
offered by parallel machines.
The increasing demand for both processing and storage
leads to the growth of HPC machines in both size and complexity, resulting in problems of scalability.
To overcome these challenges,
the development of novel interconnect technologies
for building efficient and scalable HPC systems is essential.
The design and tuning of interconnects for many thousands of nodes,
the network interfaces, software drivers,
operating system (OS) libraries and runtime systems
are all crucial for the success of HPC machines.

In modern HPC systems,
the complexity of the machine increases sharply,
requiring Graphics Processing Units (GPUs)
or Field-Programmable Gate Arrays (FPGAs)
to accelerate critical computation and communication tasks,
thus offloading the Central Processing Units (CPUs).
The memory hierarchy also deepens, driving the need for high network bandwidth and efficient interfaces. Managing the complexity of such systems while delivering high performance requires tuned and efficient runtimes.
Additionally, energy consumption is a major challenge
in large and extreme scale systems\cite{6012897}. Hence, a parallel approach, based on low-power ARM processors and reconfigurable logic,
has demonstrated promising results\cite{9041710}. Finally, low-power FPGAs are emerging as potential competitors
to accelerators such as GPUs and Vector processors
for a wide range of applications
\cite{Blott16, Cilardo18, 10.1145/3149457.3149479, Escobar16}.

In this paper, we present and evaluate
the multi-FPGA \emph{ExaNeSt} prototype,
which we designed and built as an experimentation vehicle
for the development of complex large-scale systems.
The platform is a result of 4 years of collaboration \cite{7723536} between academic institutes and industrial partners in Europe under the 
\emph{European Exascale System Interconnect and Storage} (ExaNeSt) 
EU project~\cite{exanesturl}.
This effort was motivated by the trend towards HPC systems
built using micro-servers \cite{Oleksiak2019,montblanc2020}
or processors commonly used in embedded SoCs \cite{1592896},
so as to reduce power consumption.
The design of the ExaNeSt compute node,
the \emph{Quad FPGA Daughter Board (QFDB)}\cite{qfdb_paper},
was based on tightly-coupled low-power ARM-based
Multi-Processor Systems-on-Chip (MPSoCs)
and liquid cooling \cite{trets20},
to improve compute density and reduce energy consumption.

The ExaNeSt prototype rack consists of
48 QFDB modules (each containing 4 FPGAs),
hosted in 12 blades, thus aggregating 192 FPGA MPSoCs.
It is split into two sub-systems,
with the larger of them consisting of 8 blades.
The topology implemented is a three-dimensional (3D) torus.
While the ExaNeSt prototype is not an actual HPC machine,
we use it to emulate and evaluate HPC systems
and their custom interconnect technologies.
The reconfigurable nature of FPGAs
allows for efficient customization of the communication infrastructure,
which is a crucial factor for overall system performance.

We leveraged the ExaNeSt platform to develop
a low-cost, high-performance, low-latency interconnect, \emph{ExaNet},
suitable for both application (HPC) and storage traffic.
ExaNet consists of Network Interfaces (NIs), developed within this Lab,
and Routers, derived from APENET \cite{ammendola2004apenet}
and developed in INFN, Italy. 

The ExaNet NIs are simple, efficient,
and integrated inside the same FPGA chips as the ARMv8 processors,
in every  node of the system (in each of the 4 FPGAs in every QFDB). 
In our implementation,
the NI occupies less than 10\% of the Zynq Ultrascape+ Xilinx FPGA \cite{ZynqTRM}.
It was co-designed together with the software MPI runtime,
with simplicity and high performance in mind.
Integration of the NI in the same chip with the processors
offers advantages:
\emph{(i)} it obviates the need for bulky and costly
Network Interface Cards (NICs) connected over PCI; and
\emph{(ii)} tight integration reduces message latency.
Despite its low-cost, the ExaNet NI supports
user-level, protected, zero-copy Remote Direct Memory Access (RDMA)
that bypasses the OS kernel,
in order to achieve low latency and low CPU overhead.
It leverages the Input/Output Memory Management Unit (IOMMU)
that accompanies the ARM processors
to translate user-level Virtual Addresses into Physical Addresses,
thus avoiding the need for virtual-physical-pair
page registrations in the NIC.
Effectively, payload destination addresses (80 bits)
route packets across the protected Global Virtual Address Space
offered by the testbed.

\begin{figure}[b]
    \includegraphics[width=0.5\textwidth]{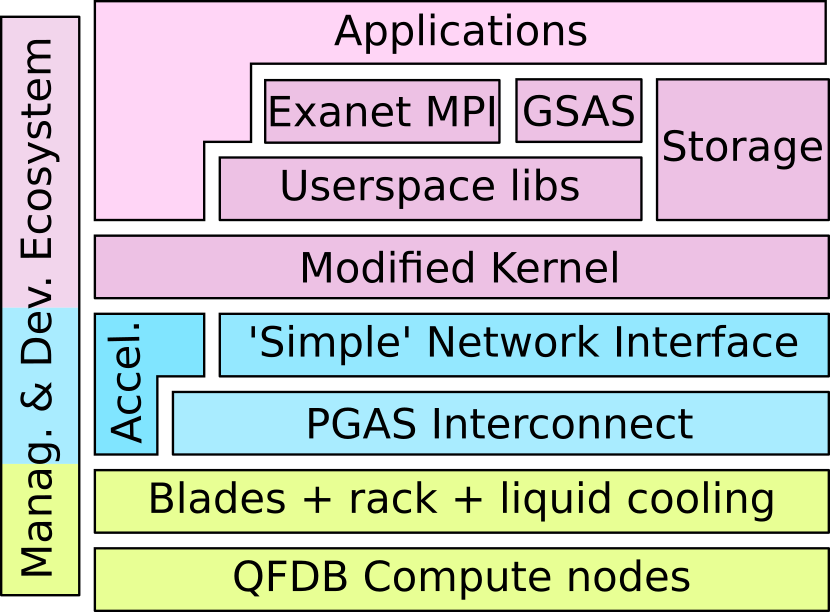}
\caption{Hardware and Software components of the ExaNeSt platform}
\label{fig:platform}
\end{figure}

The ExaNet NI offers 1024 RDMA hardware channels to user processes
for high-throughput reliable transfers
between local and remote virtual addresses.
These RDMA channels implement an one-to-one  transport
for large transfers.
The NI further provides channels suitable for
many-to-one communication of short messages at low latency
(\emph{packetizer-mailbox}).
In order to utilize these channels,
key tweaks were made to the Linux kernel and drivers,
and user space libraries have been implemented.
On top of that, we have implemented
an efficient partial implementation of the MPI standard
in a Global Shared Address Space (GSAS) environment. This implementation offers
the most commonly used point-to-point and collective primitives.
Figure~\ref{fig:platform} presents a high level, bottom up view of the full system derived, including
main hardware components, FPGA functionality and the corresponding software ecosystem.

We evaluated the proposed ExaNeSt communication infrastructure
using MPI micro-benchmarks, a miniAPP, and two full MPI applications.
The point-to-point based microbenchmark
showed a one-way single hop latency of $1.3$~$\mu$s
with a hardware-related contribution of approximately $0.47$~$\mu$s.
The achieved throughput over a single link reached $82\%$ of the theoretical capacity for large message sizes. A simple model for broadcast latency showed that the observed latency scaled as expected. Most notable deviations correspond to small messages and can be
at most $15\%$. Note though that, for small messages, actual transfer times are quite
small and are thus, more susceptible to system noise. 

The performance of the proposed interconnect and software stack
was assessed through a mini application (miniFE) 
and two full MPI applications (HPCG and LAMMPS), 
with weak and strong scaling tests for all applications, 
and with a varying number of processes 
--from one process up to full prototype capacity (i.e. $512$ processes).
Results showed that efficiency was at least $70\%$ 
for all HPCG and LAMMPS tests, 
and $69\%$ for the miniFE's weak scaling test with 512 processes, 
where the highest percentage of communication time was observed. 
Overall, the latency and bandwidth results are quite promising 
for the efficiency of the network interface, 
and there is potential for even better performance 
if the NI gets integrated in ASIC form 
together with the processor cores. 
The tests performed on the prototype revealed 
quite promising scaling properties 
across the available range of process counts.

We consider the following to be the main contributions 
of the ExaNest prototype:
\begin{itemize}
    \item {\bf Multi-processor testbed 
    with configurable hardware support for communication} - 
    The first platform to integrate 256 FPGAs and offering 
    a multi-processor environment with configurable communication
    \item {\bf Dense node-level and system-level integration} - 
    At the node level, the ExaNest prototype integrates 
    4 FPGAs (ZU9EG MPSoCs), 4 DRAM modules (DDR4 SODIMMs), 
    and an M.2 Solid-State Disk (SSD) 
    on a 120mm x 130mm x 21mm compute node. 
    At the system level, the ExaNest prototype integrates 
    four such compute nodes 
    into a liquid-cooled thin (47mm thickness) blade 
    for optimal thermal management and high-density computing
    \item {\bf Low-cost, low-latency Network Interface} 
    offering virtualization and fast completion notifications, 
    tightly coupled with the ARMv8 processing cores
    \item{\bf A custom MPI runtime} leveraging Network Interface's capabilities for user-level
        zero-copy transfers
\end{itemize}

The remaining paper is structured as follows. 
In Section~\ref{sec:related_work}, the state of the art is presented 
and our contributions are placed in that context. 
In Section~\ref{sec:platform}, 
the components of the ExaNeSt platform are described: 
they provide a unrivaled level of flexibility 
for experimenting with software and firmware configurations. 
In Section~\ref{sec:net_interface}, 
the hardware elements of the ExaNeSt-based interconnect are discussed; 
and Section~\ref{sec:kernel_mods} 
details the various software layers that leverage the interconnect. 
Section~\ref{sec:evaluation} presents experimental results 
that highlight the merits of the ExaNeSt fabric, and 
Section~\ref{sec:hardware_accelerators} illustrates 
the use of hardware accelerators in the ExaNeSt platform. 
Finally, Section~\ref{sec:conclusion} concludes the paper.

%% file: 2_related_work.tex
\section{Related Work}
\label{sec:related_work}

Compared to currently available platforms, ExaNeSt mainly focuses on providing a flexible inter-processor network connectivity in a large-scale system. To the best of our knowledge, ExaNeSt is the first prototype employing up to 128 tightly coupled ZU9EG MPSoCs, each one integrating an ARM Cortex-A53 quad-core processor and an FPGA in the same chip package. The ExaNest architecture gives us the opportunity to emulate and test the efficiency and scalability of different network interfaces and topologies. The ZU9EG FPGA can emulate any network interface directly attached to the ARM processor, since the communication path between the processor and the FPGA includes only a high-performance AXI interconnect instead of a more complex PCIe interface. Moreover, tens of 10~Gb/s inter-MPSoC links are utilized to set up different network topologies.

Although there is significant research on FPGA-based multi-node systems \cite{Escobar16}, so far there are not many large FPGA-based infrastructures. Some example multi-FPGA platforms are the following: the multi-FPGA MareNostrum Experimental Exascale Platform \cite{meep} designed to explore future RISC-V supercomputers, the Novo-G\# system \cite{7761639}, which integrates 448 Stratix FPGAs, the Rivyera cluster from SciEngines employing up to 128 Xilinx Spartan FPGAs per machine \cite{rivyera}, the 64-FPGA Maxwell supercomputer at the University of Edinburgh \cite{maxwell}, the 512-FPGA Cube cluster at the Imperial College \cite{cube}, and lately the AMD HACC clusters~\cite{amd-hacc}, among which the Paderborn University cluster hosts 48 Xilinx Alveo U280 and 32 BittWare 520N FPGA boards. Moreover, there are commercial multi-FPGA platforms such the Maxeler MPC nodes \cite{maxelerMPC-web}, which accelerate applications on reconfigurable Data Flow Engines (DFEs) \cite{maxelerDFE, maxelerDFE-SIGARCH, maxelerDFE-web}, the Amazon EC2 F1 instances containing up to eight FPGAs \cite{amazon-ec2-f1}, and the Convey HC-2 system \cite{Convey}, which incorporates a hybrid-core computer that tightly integrate two industry standard Intel Xeon processors with four Xilinx Virtex FPGAs, along with a very powerful memory subsystem.

Several of these systems employ commercial FPGA boards, such as the Xilinx Alveo accelerator cards \cite{alveo-web}, the Bittware FPGA accelerator boards \cite{bittware-web}, the Hitech Global development boards \cite{hitechglobal-web} or the Digilent boards \cite{digilent-web}. These systems use mainly PCIe in order to transfer data between the host processors and the FPGA boards, while the inter-node communication is usually based on standard Ethernet or InfiniBand, limiting flexibility for testing new network interfaces and infrastructures.

The FPGAs in these multi-FPGA systems are mainly used as acceleration engines. Indeed, there exist several commercial parallel systems demonstrating that FPGAs constitute an energy efficient and flexible choice for several parallel applications. Microsoft uses FPGAs in its Bing search engine \cite{catapultACM, catapultISCA} under the Catapult project to achieve 95\% higher performance at the cost of 10\% higher power consumption. Cray delivers a CS500 cluster system with Stratix FPGAs for the Noctua project \cite{Plessl18}. Baidu is using low-cost FPGAs to accelerate Deep Neural Networks \cite{Ouyang14}, while IBM deploys FPGAs for large NoSQL data stores \cite{noSQL}.

Using the ExaNeSt configurable network infrastructure, we developed the ExaNet interconnect, which was evaluated on 128 ZU9EG to better understand the scalability and the effectiveness of the proposed solution. ExaNet is a low-latency HPC interconnect based on a Partitionned Global Address Space (PGAS) and lean Network Interfaces. For medium and large-scale systems, the user traffic between processors, such as MPI communication between nodes and I/O, usually transits over a low-latency, high throughput network (e.g. Mellanox InfiniBand, Cray Aries, Intel Omni-Path Architecture (OPA)). InfiniBand is the most widely used network in large supercomputers, such as the Summit supercomputer \cite{summitNetwork}. 
The latest commercially available InfiniBand products are based on the NDR (Next Data Rate) standard, supporting up to 400~Gb/s per port. In 2015, Intel introduced Omnipath, an HPC interconnect running at 100~Gb/s per port. It was available both as standalone PCIe adapters but also integrated into certain Skylake and Knights Landing CPUs, as an extra chip. Lately, ATOS introduced the Bull eXascale Interconnect (BXI) \cite{bxi}, which is integrated in the Atos BullSequana XH \cite{sequana-web}.

Compared to the aforementioned Network Interfaces, ExaNet provides a customizable solution which offers low latency, scalability and coherent communication.
In this work, we propose simple but generic network interface primitives that can be integrated in the same chip with the main processor. In order to achieve low latency and low CPU overhead, the network interface supports {\it virtualized, user-level initiated,
  protected and reliable} bulk transfers
that completely bypass the kernel. 
Integrating the Network Interface in the same chip (or package) with the processors and the memory interconnect offers the possibility to use the same block both for
on-chip and system level communication, thus saving cost and reducing
the silicon area footprint. In our scheme, the network interface
exploits the IOMMU, which is close to the processor, in order to translate process-level
virtual addresses to physical memory pages, thus avoiding the need for
a separate, synchronized translation mechanism inside the network interface card. Moreover, we do not need to pin the pages involved in the communication \cite{psistakis2020part}, since we handle
the occasional page faults that may occur in RDMA transfers by
retransmitting  the failing packets in hardware. 

We have further evolved this Network Interface architecture with congestion management capabilities \cite{congestion18} and we have further improved its functionality by offloading more tasks in hardware, without significantly impacting the small footprint of the design.

%% file: 3_platform.tex
\section{The ExaNeSt platform}
\label{sec:platform}

The design of an entire High Performance Computing system is extremely demanding, due to the variety of the required skills and the scores of pitfalls that take shape during the development process. In this Section, we present important components of the ExaNeSt platform, ranging from compute hardware to system and management software, with a focus on the caveats that we faced and the techniques that we employed to address them. Overall, the design process was driven by the following goals:

\begin{itemize}
    \item Low power and high node density targeting HPC demands
    \item High flexibility for both software and hardware infrastructures
    \item Lean high-speed interconnection network
\end{itemize}

\subsection{The ExaNeSt compute node: Quad-FPGA Daughter Board}
\label{sec:qfdb}

\begin{figure}[hbt]
    \includegraphics[width=\textwidth]{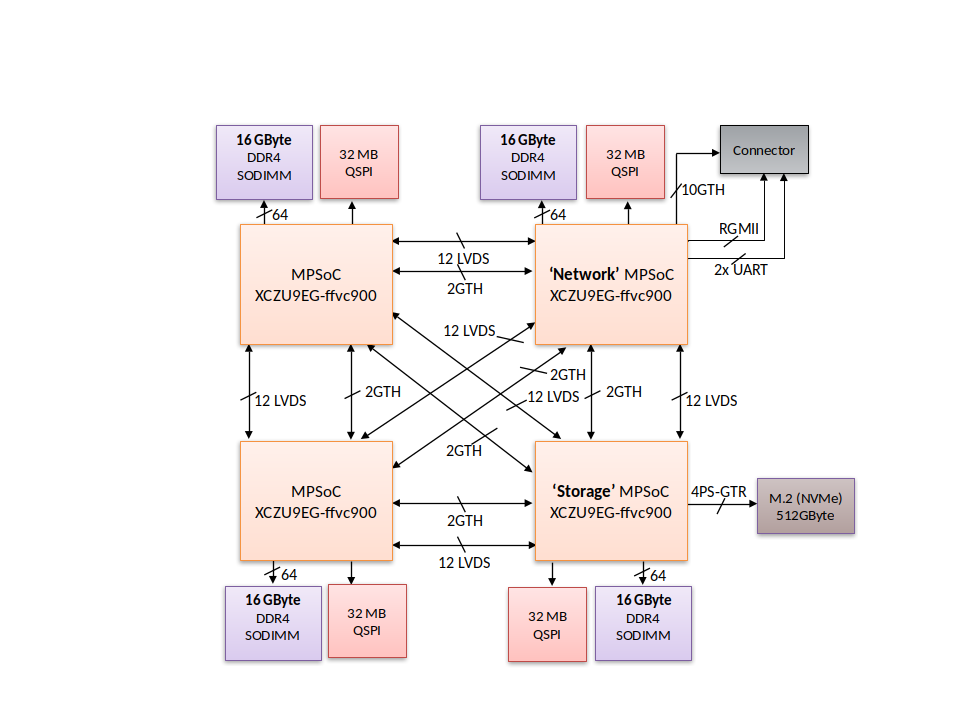}
    \caption{Top-level diagram of the Quad-FPGA Daughter Board compute node}
    \label{fig:qfdb_diagramm}
\end{figure}

In order to enable the development of custom hardware architectures, the ExaNeSt compute node includes Field Programmable Gate Arrays (FPGAs). We selected the Xilinx Zynq Ultrascale$^+$ device, which, at that time, was the state-of-the-art device, providing both a large Programmable Logic Array and a tightly coupled multi-core ARM processor in the same package. Furthermore, many aspects of the board design were driven by our density goals, and, to date, the ExaNeSt compute node is denser than any other FPGA-based daughter board publicly available. Each compute node, named Quad-FPGA Daughter Board (QFDB), hosts 4 Xilinx ZU9EG MPSoCs. A top-level diagram of the QFDB is presented in Figure~\ref{fig:qfdb_diagramm}. In addition to the four FPGAs connected to each other directly, each QFDB supports a large amount of memory and an SSD storage, all within a small footprint (120~mm $\times$ 130~mm). The top right MPSoC in Figure~\ref{fig:qfdb_diagramm}, referred to as the “Network MPSoC”, is responsible for cross-QFDB connectivity.

\begin{figure}[hbt]
    \includegraphics[width=0.8\textwidth]{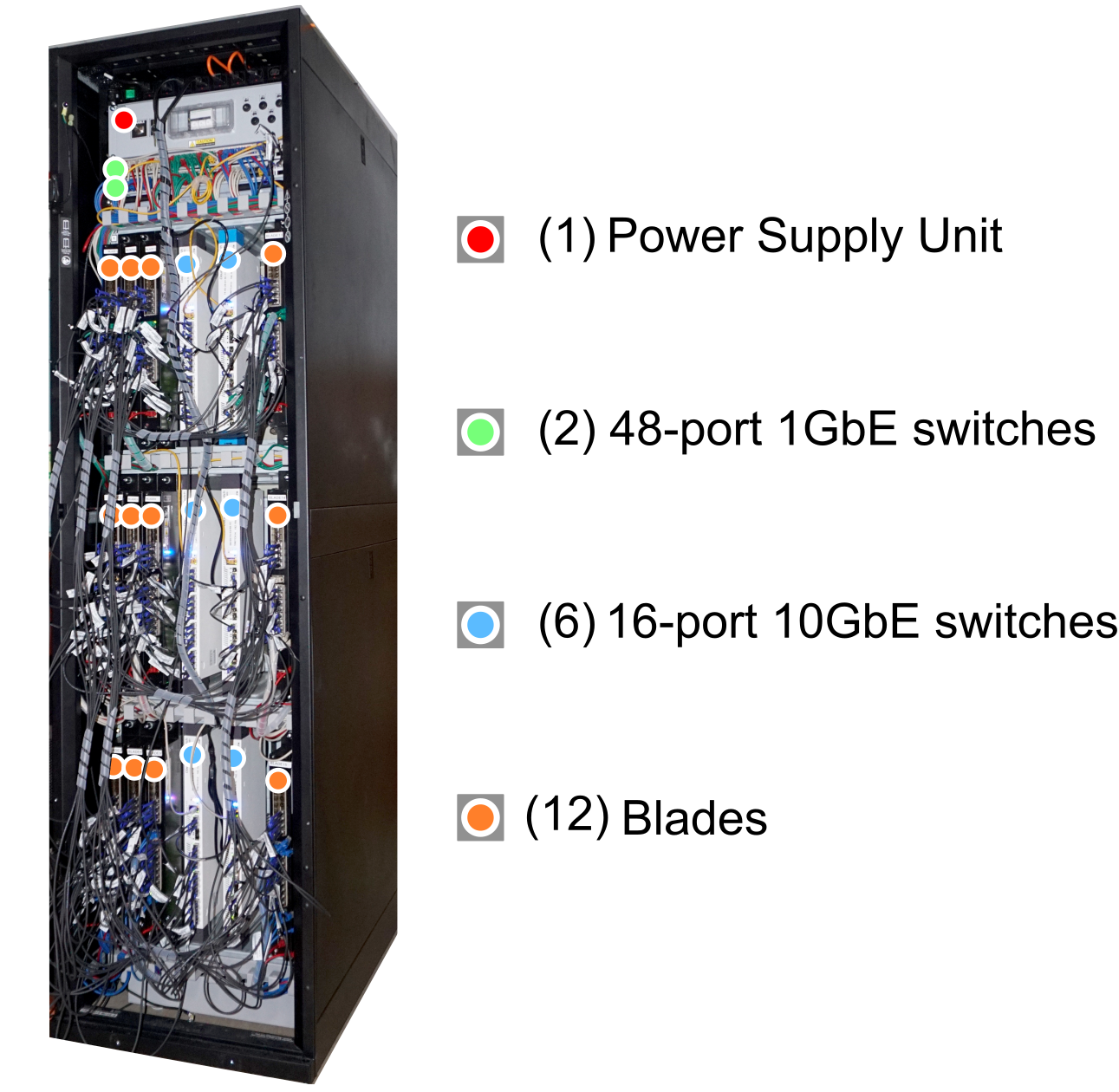}
    \caption{The liquid-cooled testbed and its components}
    \label{fig:proto_photo}
\end{figure}

Each MPSoC integrates a quad-core ARM Cortex-A53 processor, with CPU frequency at up to 1.3 GHz, numerous peripherals, and reconfigurable logic. A 16-GByte DDR4 memory module is connected to the ARM processor of each MPSoC. 16 LVDS pairs and 2 high-speed Xilinx GTH transceivers between each pair of MPSoCs provide a low-latency communication channel and a high-bandwidth communication channel, respectively.
The LVDS pairs offer a total bandwidth of up to 12.8 Gb/s in each direction between each
MPSoC pair, while two GTH links offer a total bandwidth of up to 32 Gb/s, as each GTH
transceiver can reach up to 16 Gb/s. The "Network MPSoC" provides connectivity to the external world through 10 GTH high-speed links. Section~\ref{sec:topology} further describes the topology implemented using those high-speed links. The bottom right FPGA, referred to as the “Storage MPSoC”,
provides connectivity to the NVMe storage through the Xilinx PS-GTR transceivers implementing a
4xPCIe Gen 2.0 channel.

The PCB supports hundreds of high-speed signals (96 LVDS pairs, four 72-bit
SODIMMS, 22 16~Gb/s high-speed links, etc.) and thus the routing of the high-speed signals, including the DDR and the high-speed serial signals, was very challenging.
Short traces for the high-speed signals were possible with the appropriate component placement.
The PCB stack-up includes 16 layers: 3 layers for routing the high-speed signals, 3 layers for routing the low-speed signals and 10 layers for ground and voltages. The PCB layout is hierarchical, with the bottom level including a module which contains one MPSoC, its DDR memory, regulators, sensors and clock generators. The top level layout design includes 4 interconnected
bottom-level modules and the rest of the components, such as the connectors, additional regulators and sensors.

The board uses a 48V power input, while a series of voltage regulators provide the appropriate voltages to various components of the board. Cyntec and Texas Instruments regulators were used since they offered high efficiency (about 95\%) and compact solutions. Twelve power sensors were placed at the output of the high-power regulators. Overall, each QFDB consumes between 20 Watts when idle and 200 Watts, for the most demanding hardware accelerators.

Due to the board density, bring-up activities required substantial time and effort, in large part because of the board density, which impeded many measurements and workarounds at the PCB-level. In order to mitigate those limitations, we leveraged the flexibility of the FPGAs to experiment and monitor the board. Finally, since our HPC testbed consists of dozens of QFDBs, we faced manufacturing concerns, and in turn we built a semi-automated test procedure to validate each QFDB upon procurement or during maintenance phases.

\subsection{High-density Liquid-cooled ExaNeSt cabin}\label{sec:liquid_cooled_cabin}
\begin{figure}[b]
    \includegraphics[width=0.6\textwidth]{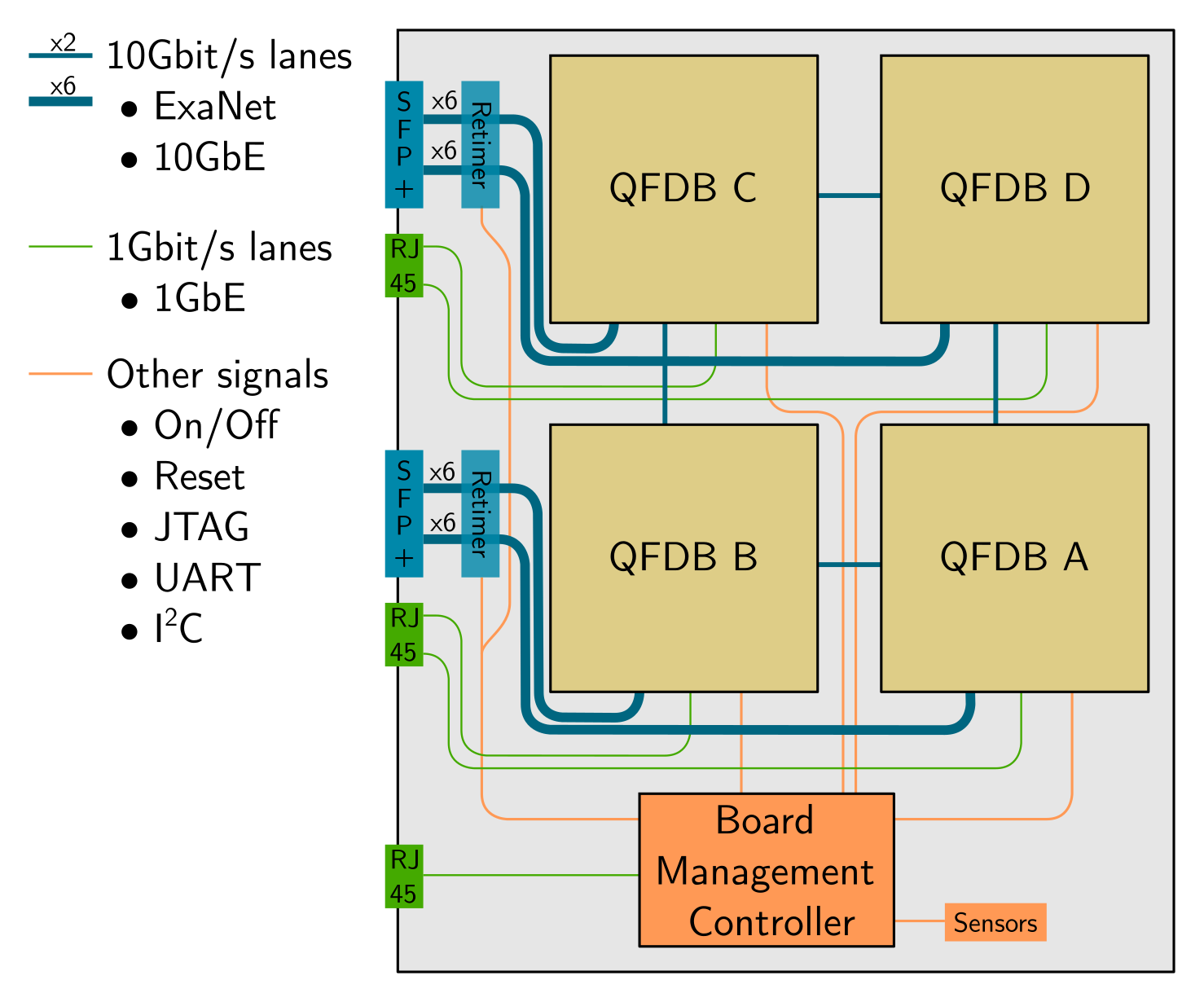}
\caption{Diagram of the custom blade}
\label{fig:ice_blade}
\end{figure}

In order to improve the compute and power density, a liquid-cooled cabin was designed to host the system, as shown in Figure~\ref{fig:proto_photo}. The mechanical chassis, power supply and liquid cooling system were provided by ICEOTOPE~\cite{iceotope}. In order to densely host compute nodes, ICEOTOPE also developed a custom liquid-cooled blade to interconnect and manage multiple QFDBs. Figure~\ref{fig:ice_blade} gives an overview of this board. Each blade hosts 4 QFDBs and a Board Management Controller (BMC), and offers power delivery and highly efficient cooling. The BMC is responsible for managing and controlling the QFDBs (e.g. power cycling, accessing serial ports and JTAG programming) as well as for configuring the blade. In addition, the blade provides high-speed connectivity among the QFDBs as well as among the multiple blades. The connectivity towards other blades is offered through a score of SFP+ connectors that provide six 10~Gb/s links per QFDB. In order to improve the signal integrity, active retimers intervene between the QFDBs and the SFP+ connectors. The board also hosts the Ethernet infrastructure (1GbE and 10GbE), used for management and legacy software. Ethernet connectivity (1GbE) is provided between the BMC and the QFDBs for management purposes.  

The cabin can contain up to 27 blades, i.e. 108 QFDBs that aggregate a total of 432 MPSoCs. Twelve blades are currently hosted in the cabin, each one containing four QFDBs, for a total of 192 interconnected MPSoCs. An air-cooled subsystem also exists, that hosts four additional blades than can be interconnected to the liquid cooled ones. The latter is used to better aid specific hands-on development purposes.

\begin{TRonly}
\subsection{System Boot and Management}\label{sec:dev_ecosystem}

There were several challenges in the development of the system boot software and management. In particular, as the system is quite complex and heterogeneous, significant effort was put for bring up and management, primarily in the following areas: (a) the boot process, including workarounds to cope with functional limitations of the prototype (e.g. DDR channel issues, boot sequence dependencies), (b) the infrastructure and tools developed to build and deploy new hardware/software designs, (c) the custom Linux distribution we developed, tailored to the characteristics of the platform, and (d) the management setup to support multiple users. 

The booting process has been modified several times to deal with various platform limitations during the development of the platform. As an example, the DDR memory in the first version of the QFDB was not functional. Since the time to fix the PCB design and manufacture new functional boards was long, we had to develop a boot image as well as specific hardware interfaces in order to connect the QFDB to another FPGA board and transparently "borrow" its memory~\cite{qfdb_paper}. Another important limitation, imposed by the dense QFDB architecture, was that only one out of four MPSoCs on the QFDB, designated as the Network MPSoC, has Ethernet connectivity and 10~Gb/s high-speed links to the external "world", i.e. connected to the external connector of the board, while the remaining three MPSoCs should route the traffic through the Network MPSoC. Hence, the Ethernet connectivity of the four MPSoCs is configured and managed by the Network MPSoC. Initially, we implemented a software bridge on the Network MPSoC to route the traffic to/from the other three MPSoCs of the QFDB. Next, to achieve better performance, we implemented 10GbE interfaces on the FPGA fabric of each MPSoC, using DMAs and custom Ethernet drivers to transfer Ethernet packets. An Ethernet router was developed in the Network MPSoC, which routes traffic from all four MPSoCs to a custom 10G MAC. An outgoing link of the QFDB is then used, that employs custom flow control to avoid losses, and allows interconnectivity between different QFDBs through an external commercial rack-level 10 GbE switch.

As the platform grew up, two additional challenges appeared. First, we had to designate a unique network address automatically for each MPSoC. We designed a secure scheme that uses persistently programmable parts of the FPGA, based on the Zynq Ultrascale+ e-FUSEs and the ARM Trusted Firmware (ATF), thus giving a distinct "naming" for each QFDB in the system. Second, we had to deal with boards reliability. In effect, some MPSoCs were subject to voltage variations, which created instability. In addition, a thermal protection mechanism was desirable to avoid disaster scenarios. Hence, we developed a software program that runs in the Platform Management Unit (PMU) of the MPSoC, in order to monitor internal voltages and temperatures and power down the MPSoC when needed.

\begin{figure}[b]
    \includegraphics[width=0.4\textwidth]{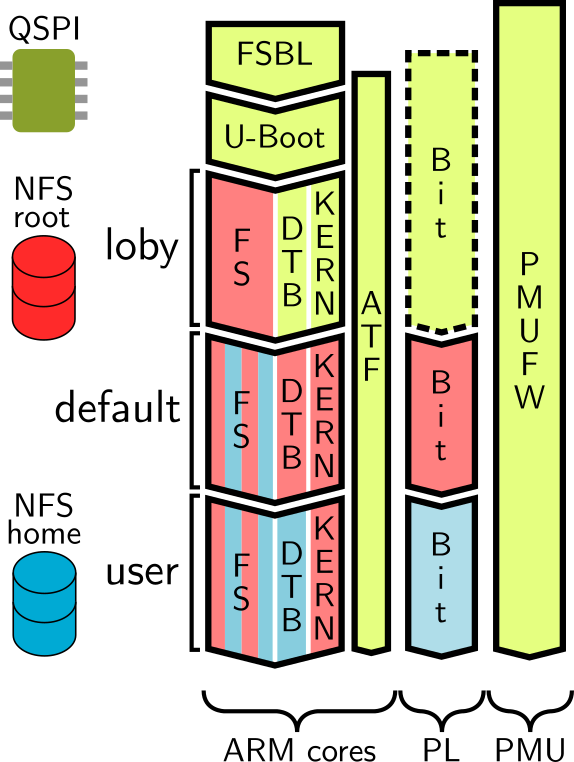}
\caption{The flexible QFDB boot process}
\label{fig:boot_process}
\end{figure}

The development of a robust platform requires the customization of every component in the boot sequence. As existing tools were not efficient, we had to develop a custom tool, named YAT (Yet Another Tool), which automatically generates a boot image based on user profiles including the artifacts required for the initial booting stages (First Stage Boot Loader, U-boot, ARM Trusted Firmware, PMU firmware and Linux kernel). These binaries are aggregated into a single boot image, which is flashed into the QSPI memory attached to each MPSoC. An outline of the final boot process is presented in Figure~\ref{fig:boot_process}. Optionally, this boot image can also incorporate the FPGA programming file, i.e. the bitstream, designated as "Bit" in the figure.

Each platform user may have different FPGA programming firmware and software configurations. In order to support a multi-tenant architecture, these configurations have to be deployed swiftly onto the multi-FPGA platform, and in a manner that allows the sharing of the platform resources efficiently. Thus we developed a two-stage boot process. A first "minimal" Linux kernel is loaded from local storage (QSPI flash memory), and mounts a root Network File System shared among all MPSoCs ("NFS root" in Figure~\ref{fig:boot_process}). This share is read-only in order to improve the system security. Memory-based read/write overlays are created automatically for specific linux folders, to provide a read/write file system. Once the root File System (rootFS) is loaded, a script fetches a new set of binaries, initiates the execution of a second "fully-featured" kernel through kexec~\cite{kexec}, and optionally (re-)programs the FPGA. The second kernel loads the same rootFS, as well as a second share (NFS home in Figure~\ref{fig:boot_process}), which contains the home folders of the users. Optionally, users may initiate additional stages to change the MPSoC configuration, without rebooting the whole board. In order to automate the development of new boot images with user configurations, YAT was extended to generate the so-called "boot packages", which packs together the kernel, Device Tree Blob (DTB) and the FPGA bitstream along with user boot packages.

In order to have a Linux distribution tailored to the needs of the ExaNeSt platform, such as the network connectivity and the booting environment, we created a Linux distribution, called Carvoonix and, which is based on Gentoo Linux. Carvoonix combines the flexibility of Gentoo and the convenience of distributions based on binary packages. In effect, a “builder” package distribution library is used for preparing, compiling and uploading optimized packages to a binary package repository, while a “deploy” package distribution library uses the binary repository for the final distribution. In case new packages are requested, we build and test them using the “builder” library and then make them available to the users through the “deploy” library. This increases the overall flexibility of the Linux environment.

Finally, the management of the QFDBs required significant effort. In order to support the ExaNeSt diskless environment, a management Virtual Machine (VM) was initially created. This VM provided the necessary network services (NFS, DHCP, DNS, NTP) to QFDBs as well as a user gateway functionality. By  using a single management VM, the deployment of other QFDB-based systems was significantly simplified. However, as the platform grew larger, those functionalities were migrated to a physical machine in order to improve the performance. 
The NFS shares are backed up by a redundant "ZFS" file manager pool. We implemented rolling snapshots at different time frequencies (from an hour up to a month), which proved to be very effective. Finally, "Ansible" automation software was leveraged in order to manage large sets of QFDBs more efficient.

Overall, we have learned a lot during the development of the ExaNeSt platform. Foremost, we found out that deep knowledge across domains (electronics, FPGA firmware and system software) was merely a prerequisite. On many occasions, the tight collaboration between those fields of expertise has been crucial to the progress of the developments. 

Eventually, our approach to the system boot software and management environment can be adapted to other prototypes, where flexibility and customizability at a medium scale are priorities. In effect, our approach automates the generation of all necessary FPGA firmware and software artifacts, as well as their selective deployment onto the system; while offering users a relatively stable work environment.

\end{TRonly}

%% file: 4_network.tex
\section{The ExaNeSt network communication} 
\label{sec:net_interface}

In this section, we present the {\it ExaNet} interconnect, developed in ExaNeSt and EuroEXA projects.
We focus on the ExaNet {\it Network Interface (NI)},  which is implemented in the Programmable Logic (PL) and connected to four (4) ARMv8 cores in the Programming System (PS) segment of the Xilinx MPSoC.
The ExaNet also features (Torus) network routers that evolved from the APENET architecture ~\cite{ammendola2004apenet}. 

The design principles behind the ExaNet interconnect are summarized below: 
\begin{itemize}
    \item Low-latency user-level initiation with complete OS bypass in order to minimize latency and software overheads \cite{5289226, 569696},
    \item Custom "cell-based" switch interconnect in order to reduce latency,
    \item Flow control for (nearly) lossless operation and congestion management to protect victim flows under congestive episodes, with shallow buffers inside the network core to reduce latency and cost \cite{kornaros1998implementation, chrysos2015discharging},  
    \item Co-existence of compute and storage traffic over a converged interconnect (see Section \ref{sec:converged_network}).
    \end{itemize}

\subsection{Network Topology} \label{sec:topology}
\begin{figure}[t]
\centerline{\includegraphics[width=0.9\textwidth]{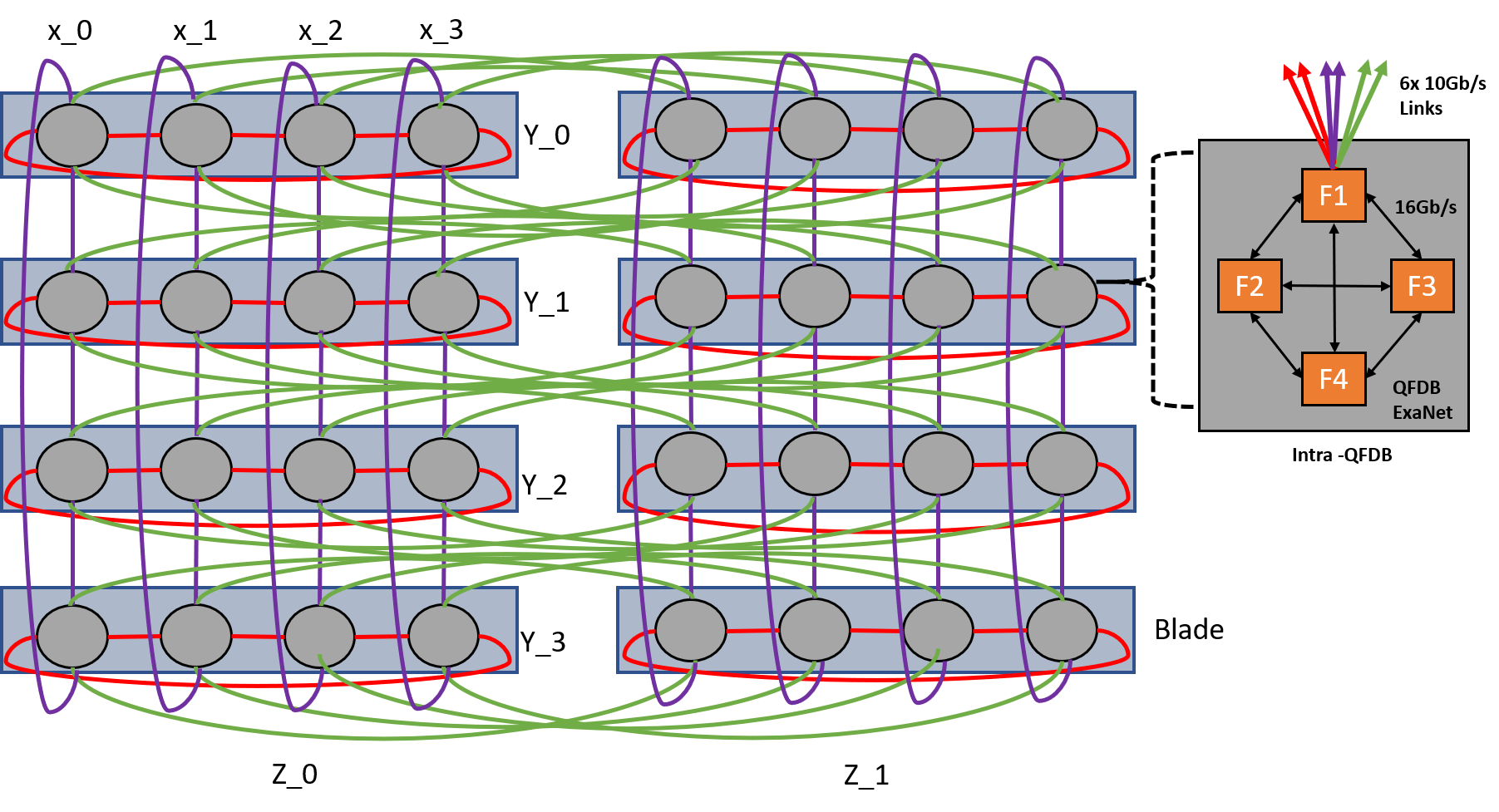}}
\caption{HPC prototype topology}
\label{fig:proto_topo}
\end{figure}

The current topology of the ExaNeSt prototype is a 3D-Torus as depicted in Figure \ref{fig:proto_topo}.
The four Xilinx MPSoCs on a QFDB are fully connected through 16~Gb/s links, as shown on the right side of the figure.  The "Network MPSoC" (orange F1 block in the figure) is responsible for the external connectivity, as explained in Section \ref{sec:qfdb}. Only the firmware of the "Network MPSoC" can route ExaNet packets to different destinations and thus the network traffic from the remaining three MPSoCs of the QFDB is first forwarded to the "Network MPSoC".

The four QFDBs inside a blade are connected in a ring topology through 10~Gb/s links on the X axis (red links). The QFDBs between four blades are also connected in a ring topology on the Y axis, as shown in the figure (purple links), creating a quad-blade group. Finally, the symmetrical QFDBs between two quad-blade groups are connected together as shown in the figure (green links). 

A first implementation of the ExaNet network  architecture is currently installed in the full-scale ExaNeSt prototype that comprises 8 blades, or 8x4x4x4 ARMv8 cores, and runs real HPC applications, using the custom MPI library, outlined in Section~\ref{sec:mpi_runtime}.
 
\subsection{Overview of interconnect}
Internally, the ExaNet interconnect features shallow buffers at network links (4~KBytes) and link-level flow control in order to avoid packet drops. The network packets (cells) are also small. Every cell carries up to 256 bytes of payload, split in 128-bit words. Additionally, each  cell carries 32 bytes of control information used for transport, network routing, and link-level protocols, which is split in a 16 bytes header and a 16 bytes footer.  The small packet size minimizes the latency of high priority traffic in front of busy links. We try to optimize the end-to-end latency, but we nevertheless perform store-and-forward  when reading the payload of a cell from its source memory, in order to avoid memory-induced network hiccup, e.g.  lines hitting in the cache subsystem may overtake older memory requests.

All switches and routers in ExaNet perform cut-through. In particular, there is a small input-queued  switch in every FPGA  to route packets between the local NI endpoints and  the local links, with a latency of 2 clock cycles. 
Additionally, the network FPGA of each QFDB also contains a Network Router  \cite{ammendola2004apenet} with connections to remote QFDBs.  
The QFDBs are connected in a 3D-Torus topology through HSS links, as discussed in Section~\ref{sec:topology}, and  the Network Router  guarantees  deadlock freedom using single-path (dimension ordered) routing.

The ExaNet Network Interface (NI) offers two main primitives for  user-level communication: 1) an RDMA engine, suitable for bulk remote-memory transfers; and 2) packetizer-mailbox pairs suitable for sending and receiving small messages with ultra-low latency. Regarding the connection of the NI with the ARMv8 cores and the host memory, we use a high-speed AXI-4 master interface that offers separate read and write channels (128-bit data width each). The NI and the switches implemented in the and Programmable Logic (PL part of our FPGA) run  at 150 MHz, thus every read or write channels offers a raw memory bandwidth of  19.2~Gb/s. 
The base latency of a request round-trip between the PS (ARMv8 cores and memory subsystem) and the PL part of the FPGA ranges between 100-150ns, depending on the request.  We have managed to saturate the paths to and from memory using eight (8) outstanding AXI burst transactions (16 x 128-bit words). All ExaNet links inside the FPGAs run at the same speed (128 bits at 150MHz), and incur an overhead of two words (header+footer) for every 16 words of payload (efficiency 16/18).

We have further explored the generic ExaNet architecture in order to implement and evaluate a number of novel technologies for interconnects such as: 
 
\begin{enumerate}
   \item End-to-end transfer resiliency in hardware, allowing to fully offload the CPU. Our reliable RDMA transport features dynamic connections, with fast (hardware-level) recycling of hardware contexts;
   \item Fast completion notifications at the receiver, suitable for optimized MPI runtimes \cite{ploumidisExacom}; 
    \item Accurate congestion management in hardware, replacing sub-optimal (software-based) solutions, such as TCP \cite{giannopoulos2018accurate}, \cite{chrysos2015discharging};
    \item  Page pinning avoidance by dynamically handling page-faults during RDMA transfers. For this purpose, we reuse the reliable hardware RDMA transport to replay faulting pages  \cite{psistakis2020part}, \cite{psis_tpds}; 
    \item Tightly-coupling new versions of the network interface with RISC-V cores \cite{redsea2022}. 
\end{enumerate}

\begin{figure}[ht!]
\centerline{\includegraphics[width=0.9\textwidth]{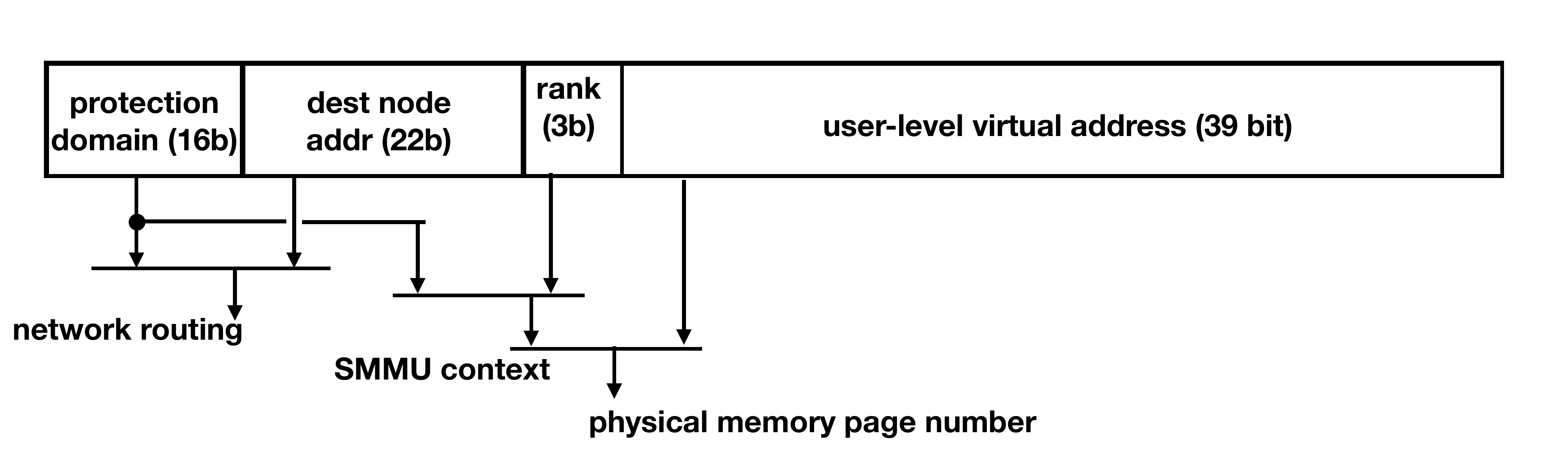}}
\caption{Overview of ExaNeSt 80-bit Global Virtual Address.}
\label{fig:global-address}
\end{figure}

\subsection{Global Virtual Address Space}
\label{sec:global_virtual_address}
All NI endpoints and memory locations inside the ExaNeSt prototype belong to a Global Virtual Address Space (GVAS) and can be addressed directly by any node through the  interconnect.
Figure~\ref{fig:global-address} depicts the GVAS address, here used as a destination address of a network packet, i.e. the destination address of the first byte of the packet payload. An address in  GVAS  is 80 bits long.  Part of the GVAS  is used to route packets inside the physical interconnect, and part to identify a memory location within a node  (belonging to some process).  In order to route a network packet in the ExaNeSt GVAS, the packet address includes the following fields:

\begin{itemize}
    \item  {\it The protection domain id (PDID)}, 16 bits in the ExaNeSt prototype, is used by the hardware to check the initiator's access rights
on particular GVAS locations. The PDID  carried by network packets mirrors the value set in special hardware registers by the system software. PDIDs allow the system administrator to create virtual groups of processes that can safely read and write their memory. With 16-bit PDIDs, we can have up to 64K process groups.
    \item  {\it The destination node  address}, 22 bits, allows up to 4M nodes in the system. In principle, the node ID could also be virtual in order to allow process migration \cite{katevenis2007interprocessor}.However, in our current prototype we static map  GVAS node IDs to physical nodes -- i.e. endpoints of the interconnect.
    \item  {\it The rank}, 3 bits,  is used to identify a specific local port, such as a peripheral/accelerator or a user-level process (e.g. MPI rank) with private memory space within a node.
    \item {\it The user-level virtual address}, 39 bits,  identifies a virtual address within a process, in the case of RDMA. Together with the rank, it can be used to compose a 42-bit {\it node-level virtual address (VA)}. 
\end{itemize}

\subsection{ExaNet Lean Network Interface}\label{sec:network_interface_hw}
The ExaNet network interface offers virtualized hardware blocks that support user-level initiated, protected transfers that completely bypass the kernel. Virtualization is provided by offering multiple channels that can be allocated to different processes. In this way, processes  can safely initiate multiple concurrent transfers, without any kernel involvement. Eventually, the network interface serializes the requests from different applications, on a per-packet/cell  basis.

\begin{figure}[t]
\centerline{\includegraphics[width=0.7\textwidth]{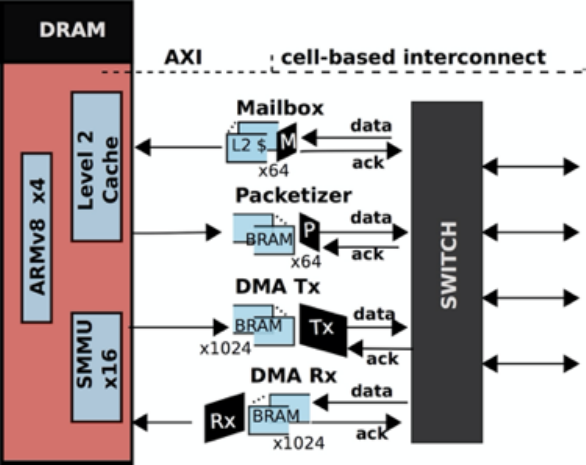}}
\caption{The hardware endpoints of the network interface. All blocks are virtualized and can serve requests from multiple concurrent  processes.}
\label{fig_NI_blocks}
\end{figure}

The main building blocks of the ExaNet network interface are depicted in Figure~\ref{fig_NI_blocks}. The packetizer and mailbox primitives can be used for operations equivalent to remote store and load (from an incoming queue) operations, respectively. Simple remote store instructions cannot directly be routed in ExaNet. The  packetizer-mailbox primitives that we use instead offer a reliable transport for small latency-sensitive messages. 

\noindent {\bf Packetizer:} The packetizer is a hardware  block connected to the ARM core via an AXI-4 memory-mapped interface, and to the network using an ExaNet interface. It is virtualized,  offering 64 virtual interfaces, each one accessible through a private memory page. Each interface further provides four (4) memory-mapped channels.  Each channel can be used by the thread or process that owns the interface to transfer a packet and to monitor its execution by polling specific status bits in the page of the interface. Each channel can be in one of the following states: ongoing, acknowledged, negatively acknowledged, timed out.
A user program stores the payload of the message into the channel address using (posted) store commands. In the final store command, the process writes the payload size and the destination address on a designated address of the channel, which prompts the packetizer engine to create a network packet, carrying the user-defined payload and the destination address, as well as the appropriate PDID. 

\noindent {\bf Mailbox:} The virtualized mailbox is a hardware block that includes 64 memory-mapped virtual interfaces. Multiple sources can concurrently write in the same mailbox interface. For ultra-low read latency, mailbox data that arrive from the network are written in the Level-2 cache (and main memory) of the ARM processor, utilizing a coherent (ACE AXI) interface between the PL and PS in the Xilinx MPSoC. The tail pointers of the incoming queues are maintained and updated in the FPGA (PL), while the head pointers are maintained and updated by the runtime. A process can request a new mailbox from a special  driver, which allocates one mailbox interface and associates it with the process PDID.  The process can then start polling for new arrivals as  discussed in Section~\ref{sec:mpi_runtime}.  

The mailbox hardware compares the PDID of incoming packets with the PDID of the targeted interface, generating a network negative acknowledgement (NACK) if these do not match. A NACK is generated also in case of packet error or full mailbox. Otherwise, the packet is enqueued into the  virtual mailbox and an ACK is generated and routed  to its source. The packetizer of the transfer maintains hardware timers, allowing the source to re-transmit the packet if the ACK is not received on time.

\begin{figure}[b]
    \centerline{\includegraphics[width=0.98\textwidth]{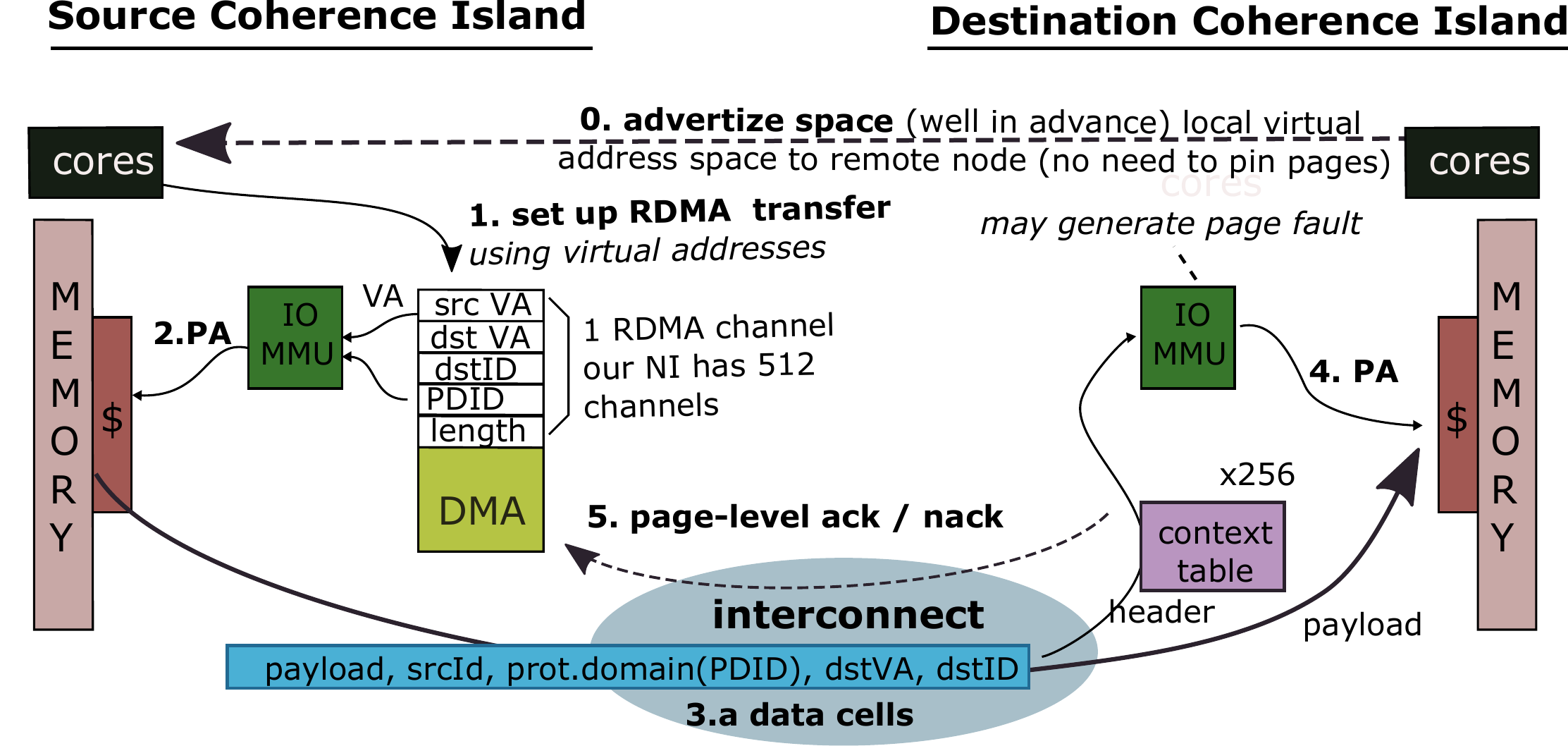}}
\caption{Overview of user-level RDMA operation}
\label{fig:rdma_overview}
\end{figure}

\subsection{RDMA engine}
\label{sec:virt_rdma}
For bulk transfers, we have developed an RDMA engine, which is optimized for zero-copy, user-level communication,  and which can offload the host end-nodes from  transfer-induced overheads. As shown in Figure~\ref{fig:rdma_overview}, the Exanet RDMA engine consists of two main blocks, one at source node for reading data from memory and sending them to the network (Send unit),  and one at destination node, responsible for receiving  network data, writing them to memory, and generating end-to-end acknowledgments and completion notifications (Receive unit).

The Exanet RDMA Send unit offers 16 pages that can be allocated to different processes. Each page offers 32 write and 32 read channels (see  Section~\ref{sec:rdma_read}).  An RDMA write operation commences when a local process writes a 64-byte descriptor inside an available channel at the source node. 
The initiator must know the RDMA destination address, which must belong to a (remote) process with the same PDID. 
Special NI firmware running in the R5 co-processor of the MPSoC discovers new transfers, splits them in consecutive (16~KBy) blocks or transactions, 
 waits for end-to-end acknowledgements. The R5 co-processor may re-transmit blocks that do not receive acknowledgement. 

The transmission of individual blocks is  handled by a hardware Send Engine, which splits blocks in cells. The hardware Send unit uses the SMMU  (the ARM-based IOMMU available in our Xilinx MPSoC) for virtual address translation,  and fetches the cell payload from memory, issuing an (IO-coherent) AXI-read burst operation inside ARM's memory subsystem. In this way, data can be read from caches, thus obviating the need to flash caches before transfers.  The Send Engine waits the entire payload from memory in order to inject the cell.  

The Receive unit consists of a hardware engine, which forwards the payload of each incoming cell as it arrives from the network directly to its final memory destination. 
These AXI-write requests pass through the SMMU and the IO-coherence interconnect (see steps 3 and 4 in Figure~\ref{fig:rdma_overview}). The Receive engine also maintains dynamic contexts in order to issue block-level acknowledgements and notifications. 

\subsubsection{RDMA Read} 
\label{sec:rdma_read} 
Our RDMA  supports {\it Read operations}. Essentially, the issuer of the operation generates an RDMA Read Request and uses its packetizer to forward it to a special mailbox allocated to the RDMA Send unit in the node with the source data. The Send unit then  allocates an available Read channel in the targeted page (protected by the  PDID in the request), and completes the RDMA Read operation with an  RDMA write. We use completion notification to inform the issuer that the Read data have arrived.

\subsubsection{RDMA R5 co-processor}
In the Exanet RDMA, a significant portion of the Sender unit is implemented through special firmware in the R5-Cortex co-processor IP of the Xilinx MPSoC. The premise of this approach is that the co-processor handles only transaction-level operations (initiation, ACK/NACKs re-transmissions), thus does not need to run at (cell) network speed. On the other hand, this software solution offered flexibility, development speed, and programmability benefits \cite{paraskevas2018virtualized}.   Nevertheless, even our optimized R5 implementation consumes in 2-4 ~$\mu$s every time it is invoked. This  dominates the interconnect (and MPI) base latency. Inside the RED-SEA project,  we redesign and implement these functions in hardware, which  can now handle small transfers in back-to-back cycles while  reducing the base latency of RDMA operations well below 1$\mu$s. 
\subsubsection{SMMU \& No page pinning}
\label{sec:smmu}
In ExaNet, we handle page faults that may occur in RDMA transfers {\it dynamically} using special software and hardware at the endpoints, in order to re-transmit blocks that experienced page-faults. In this way, there is no need to pin pages prior to transfer; in turn, this allows to expose the full address space of applications through RDMA operations  \cite{psistakis2020part, psis_tpds}. Instead of designing a custom hardware for memory address translations 
at source and destination nodes, we reuse the ARM System Memory Management Unit (SMMU)~\cite{ZynqTRM}. The SMMU is provided by ARM, but, operation-wise, is similar to other IOMMUs. Traditionally,  IOMMUs are responsible for memory requests to and from I/O devices that are memory-mapped to special (pinned to memory) virtual memory regions. In ExaNet, we reuse ARM's SMMU for accesses coming from the network.

As shown in Figure~\ref{fig:rdma_overview}, a SMMU in  source and destination nodes is in charge of the address translation of global virtual addresses to local physical memory addresses (PA), as also described in Section~\ref{sec:global_virtual_address}. For this to work, we had to "convince" the SMMU to use  pointers to process page tables inside the SMMU channels (context banks). Effectively, the SMMU hardware walks and fetches translations from the OS page tables of user processes. The SMMU further includes a Translation Lookaside Buffers (TLBs) for recent page translations. Similar to a CPU TLB miss, a SMMU  (TLB) miss can be handled by the hardware {\it Page Table Walker} without any software (or OS) intervention if there is no page fault. In case of page-fault, the SMMU generates an interrupt which is handled by OS modules. 

\subsection{Hardware complexity of ExaNet Network Interface}

The hardware resources required by the Network Interface in the Xilinx FPGA are minimal. The Packetizer and the Mailboxes consume 20K FPGA LUTs (5.5\%) and 8 FPGA BRAMs (1\%) ~\cite{ZynqTRM}\footnote{Mailbox payload buffers are implemented in host memory.}. The RDMA engines of the Send and Receive units consume 33K LUTs (12\%) and 19 BRAM blocks (2\%)in total.

\subsection{Allreduce accelerator}\label{sec:all_reduce}
Another important component of the Network Interface is an accelerator for the \textit{MPI\_Allreduce} collective primitive.
The accelerator was implemented with Xilinx Vivado High-Level Synthesis and is currently able to handle the
following cases:
\begin{itemize}
\item Sum, min, max arithmetic operations
\item Int, float, double datatypes
\item Up to 1024 MPI ranks
\item Vector size up to 256 bytes (Bigger vector sizes require multiple operations on 256-bytes vectors)
\item At most 1 MPI rank per FPGA/SoC
\item Whole QFDBs must be part of the Allreduce operation, i.e. the total number of MPI ranks must be a multiple of 4
\end{itemize}
For other use cases, the Allreduce primitive is handled in software through the algorithms supported
by ExaNet-MPI (Section \ref{sec:runtimes})

\begin{figure}[t]
\centerline{\includegraphics[width=1.0\textwidth]{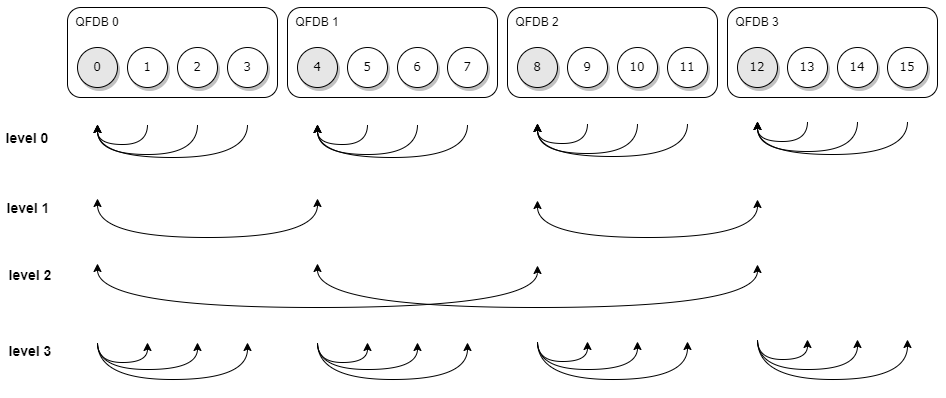}}
\caption{An example of 16 ranks running the Allreduce accelerated operation. 'Network' FPGAs are highlighted with gray color.}
\label{fig:allredue_algorithm}
\end{figure}

For the purposes of our implementation and by taking into account the unique properties of our system,  we came up with an Allreduce algorithm tuned specifically to QFDB characteristics. The algorithm tries to minimize \textbf{a)} the network traffic between the nodes involved and \textbf{b)} the processor~$\leftrightarrow$~Network~Interface interaction. In effect, those interactions take place only at the beginning and at the end of the Allreduce operation, while all the arithmetic calculations and communication patterns are performed and determined inside the hardware. Moreover,
we designed 2 distinct modules in the accelerator, in order to minimize the intra-QFDB traffic and optimize the inter-QFDB communication. The Allreduce {\em client} module resides inside all FPGAs but 'Network' FPGA. It is responsible for fetching and sending data to the 'Network' FPGA at the beginning of the Allreduce algorithm and updating the software at the end of the operation. The Allreduce {\em server} resides only in the 'Network' FPGA and is responsible for inter-QFDB communication and for performing arithmetic calculations. This algorithm is illustrated in Figure~\ref{fig:allredue_algorithm} for an example of 16 ranks. Overall, the implemented Allreduce algorithm is the following:

\begin{itemize}
\item Initialization: At the beginning of an Allreduce operation, the software programs every module with the arithmetic operation type, datatype, vector size. Two pointers are also provided, one with the memory address of the data and one with the memory address of the table that contains the \{MPI rank, network address\} pairs. The rest of the algorithm can be visualised as a graph of $log_{2}(N)$ levels, where $N$ is the  number of ranks involved in the Allreduce collective.
\item Level $0$: Each module performs a DMA operation to fetch its own vector from the local memory. The {\em client} modules send their data to the {\em server} module in the 'Network' FPGA; and the {\em server} module calculates the reduced vector for all the MPI ranks that reside in that QFDB.
\item Levels $1$ to $log_{2}(N)-1$: Each {\em server} module exchanges its local vector with another {\em server} module, and performs reduction with its own local vector. With each level, the rank difference between the {\em servers} increases as a power of two (i.e., 4,8,16,32,etc). At the end of level $log_{2}(N)-1$, each {\em server} module has calculated the final reduced vector of all the ranks that take part in the Allreduce operation.
\item Level $log_{2}(N)$: The {\em server} module broadcasts the vector to its intra-QFDB {\em client} modules and each module updates the memory with the reduced vector and notifies the software that the Allreduce operation has completed.
\end{itemize}

%% file: 5_software.tex
\section{The ExaNet system software}
\label{sec:kernel_mods}
Apart from the development of the ExaNeSt hardware platform, the custom interconnect and the network interface, significant effort was also
put on developing and tuning the system software stack. The main goal was to derive
runtimes that efficiently take advantage of all the hardware capabilities while addressing any
potential platform limitation. Along this direction, HW components of the ExaNeSt network interface
and SW components of the system software were codesigned. 
For example, in order to support efficient remote data transfers the system software should take advantage of the HW primitives of the ExaNet network interface, which was designed based on the requirements of the applications.

\subsection{ExaNet user-space communication API} \label{sec:userspacelibs}
At the top of the ExaNeSt system software stack, the ExaNeSt user-space communication API
allows user-level access to the hardware blocks of the network interface (see Section \ref{sec:network_interface_hw}), such as, the RDMA engine, the virtualized packetizer and the virtualized mailbox. The simple hardware blocks of the network interface allow for a minimal but efficient user-space API, as shown in Section \ref{sec:evaluation}.

The virtualized mailbox/packetizer hardware blocks, that are accessed through the user-space API, offer the capability of user-level low-latency atomic message delivery. In particular, the API provides to runtimes and user applications the functionality for allocating virtual interfaces of the virtualized mailbox and packetizer that reside on the local compute node. The linux kernel is only involved in the allocation/deallocation of virtual interfaces. The kernel driver that is responsible for the mailbox gives direct access to user applications to a specific set of hardware registers. By using this set of registers, the applications can directly send and/or receive small messages to/from remote processes in user space. In this way, the actual data transfers are initiated at user-level without requiring any kernel involvement. The API allows applications to atomically send messages of up to 64-bytes to any virtual interface (or process address) of any remote node. Furthermore, if an application has acquired an interface of the virtualized mailbox, it can receive messages sent by remote nodes of the same protection domain. The messages generated by the packetizer wait for an end-to-end acknowledgment, and can be re-transmitted after the expiration of a timeout period.

This API also allows applications and runtimes (e.g. MPI) to perform RDMA protected transfers between local and remote virtual addresses. The API for RDMA is quite minimal, providing calls for accessing a remote virtual address, inserting a descriptor, and polling for completion. Both RDMA read and RDMA write operations are supported.

\subsection{Low-overhead runtimes} \label{sec:runtimes}

\subsubsection{The ExaNet-MPI runtime} \label{sec:mpi_runtime}

 MPI is the de facto standard
 for a large number of parallel HPC applications, allowing to efficiently exploit the resources available on modern HPC systems. Thus, a major component of the ExaNeSt system software is
 a custom, platform-specific MPI library that takes advantage of the features of the ExaNet interconnect. We have developed the \textit{ExaNet-MPI} runtime, which partially implements the MPI standard, including almost all point-to-point and collective primitives. However, 
 its current version lacks support for RMA operations and non-blocking collectives. The ExaNet-MPI runtime implements all
 the primitives required by the MPI microbenchmarks and applications that were used for the platform evaluation, while it does not support rarely used primitives (e.g. MPI\_Scan, persistent requests).
 Collectives where implemented using the same algorithms that were used by MPICH, version 3.2.1.

The ExaNet-MPI was designed to take advantage of the features provided by the ExaNet network interface in order to improve performance.
In particular, it takes advantage of the capabilities of the virtualized mailbox and packetizer to offer efficient
atomic, low latency message delivery required by latency sensitive traffic, such as, control messages for MPI protocol handshaking, as well as, acknowledgments. Moreover, the packetizer can be used for eager data transfers in order to send directly payload or control data in a single message instead of using the RDMA engine. In the current ExaNet version, the maximum
payload size that can be send from a packetizer to a mailbox is $56$ bytes. ExaNet-MPI also takes advantage of
the high bandwidth RDMA engine, described in Section, and specifically its capability to deliver a completion notification in parallel
with the actual data. This notification can be sent to any virtual address. 
 
The key feature of the network interface that ExaNet-MPI exploits, is that the mailbox, the packetizer and the RDMA engine are virtualized and thus they allow for
user-level data transfers. This is implemented using the ExaNet communication API,
described in Section \ref{sec:runtimes}, which provides to the upper layers (i.e. the MPI runtime) of the system software user-level access to the NI hardware blocks.
In this way, the kernel is involved only in the initialization of a transfer and not in the actual data transfer, avoiding the overhead of system calls, which may become an important performance bottleneck.

In order to implement the ExaNet-MPI runtime we considered two different approaches. The first one was to modify a well known
MPI implementations, such as OpenMPI \cite{4100410} or MPICH \cite{GROPP1996789}, in order to support the ExaNet communication API. For the OpenMPI runtime, this approach would
require the development of a new byte transfer layer (BTL) that uses the API to transfer traffic through the ExaNet network interface. Moreover, this approach would require
extra effort to experiment with different messaging protocols for the point-to-point messages.
The second approach, which was eventually followed, was not based on any existing MPI implementation. Instead, ExaNet-MPI
consists of two different libraries, one that implements the MPI point-to-point (\textit{pt2pt})
primitives and one that implements
the MPI collective functions using the point-to-point primitives. Any primitive that is not implemented is delegated to an MPICH library (version $3.2.1$). The only modification required to this
library was to export the context ID of a communicator in a 16bit value. This was needed since packetizer
messages have limited space for control data.
This approach gave us more flexibility to experiment with different implementations and protocols, such as different versions of the rendez-vous protocol for pt2pt messages.
 \begin{figure}
    \includegraphics[width=\textwidth,scale=0.8]{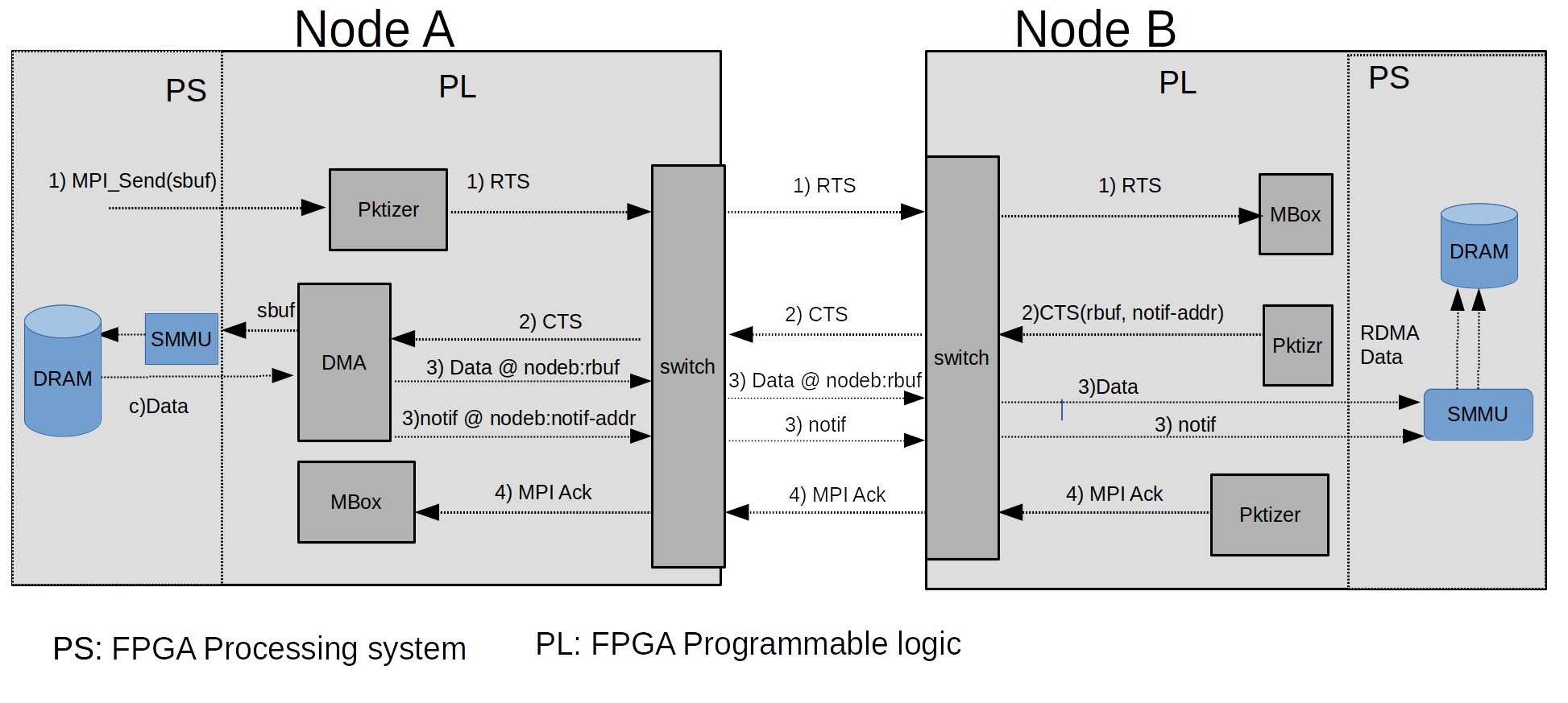}
    \caption{ExaNet-MPI: Rendez vous protocol for matching send/recv}
    \label{fig2:exanet-mpi-send-recv}
\end{figure}

Figure \ref{fig2:exanet-mpi-send-recv} depicts the hardware blocks and the message exchanges involved in the ExaNet-MPI rendez-vous
protocol. The scenario in the figure includes the communication between an MPI process A running on node A and an MPI process B running on node B.
In this scenario, process A issues an MPI\_Send before process B issues the corresponding MPI\_Receive.
The buffer that contains the data of the sender is denoted as \textit{sbuf}. First, on the sender's side, the pt2pt library
sends a \textit{request-to-send (RTS)} message that includes control information, such as, the data size of the MPI transfer, the communicator,
the source and the destination rank.
The RTS message is sent through the sender's packetizer to the receiver's mailbox. In step 2 of the figure, when the matching receive is
posted,
ExaNet-MPI on the receiver's side, sends a \textit{clear-to-send (CTS)} message that specifies
the virtual address of the receiver buffer (\textit{rbuf}) for the data transfer and the virtual address (\textit{notif-addr}) of the transfer completion notification that process A should send to process B. The \textit{notif-addr} address can be any virtual address of the receiver process. The CTS is also a small message,
which is sent through the packetizer and targets the mailbox of senders's RDMA.
In step 3 of Figure \ref{fig2:exanet-mpi-send-recv}, process A setups the RDMA transfer in order to write
the data in the receiver's buffer (\textit{rbuf}). Note that \textit{rbuf} is preprocessed in order to incorporate information
that indicates the target node transforming it into a \textit{remote virtual address}. The same applies to the virtual
address provided by the receiver where the completion notification should be sent. Let us also note that, data issuing and
notification delivery take place concurrently. On node A, the System MMU (SMMU) translates the virtual address of the sender's \textit{sbuf} buffer address to physical address (also explained in Section~\ref{sec:virt_rdma}). On node B, when the request to access the remote virtual address enters the receiver's network interface logic,
any information related to the node id is removed while the virtual address of the receiver's \textit{rbuf} buffer is translated to
physical address through the SMMU.
Once process B has sent the CTS message, it can periodically poll the
notification address for the completion notification. When the notification arrives, the receiver issues an acknowledgment back to the sender, in step 4 of the figure, and the message exchange completes.

The point-to-point protocol described in the aforementioned scenario is sender-initiated and is actually read-based
with \textit{read} being emulated through a DMA write from the receiver. We are currently evaluating a write-based
variant of the rendez-vous protocol that is both sender- and receiver-initiated with matching also taking place
at both sides.

\subsubsection{The GSAS environment}

GSAS \cite{GSAS,GSAS2} (\textit{Global Shared Address Space}) is an efficient and flexible shared memory abstraction on top of the ExaNeSt platform. The GSAS environment defines an application programming interface (API) that provides a shared memory abstraction model to distributed applications. More specifically, GSAS gives the ability to processes running across remote servers/nodes to communicate in a way resembling a system that provides coherent shared memory communication. It allows the applications to allocate/de-allocate memory (local and remote), and to perform several atomic operations (i.e. read, write, Fetch\&Add, etc.) on the allocated space by using the appropriate API calls that the environment provides. The API of GSAS provides most of the functionality that PGAS languages and runtimes provide. The current implementation of the GSAS environment uses a minimal set of hardware primitives for efficiently supporting the use-case of a shared memory abstraction on top of message passing system.

\begin{TRonly}
\subsection{ IP-over-ExaNet service} \label{sec:converged_network}

While the focus of this testbed is on HPC runtimes that directly instantiate RDMA transfers, some applications and services typical to HPC systems demand legacy networking interfaces. Primarily, Distributed File Systems (DFS) are implemented in the kernel, or, rely on socket semantics to perform communications. There is also a variety of management services such as resource management or monitoring that may suffer from the management network limitations in large systems. As a consequence, modern HPC systems strive to adapt their High-Performance fabric to also support legacy socket interfaces, which leads to the convergence of a variety of applications and services onto the same network.

In order to support those socket-based services, we built a minimal IP overlay on top of the ExaNet interconnect presented in Section~\ref{sec:net_interface}.
The architecture of our solution is presented in Figure~\ref{fig:converged_network}. For sake of simplicity, this service is implemented by a user-space program that tunnels packets between the Linux kernel and the ExaNet fabric. In order to deal with the lifecycle of nodes robustly, Out-of-Band (OOB) communication is established using the Ethernet Management Network upon start. Each node regularly emits an heartbeat message and reacts to messages from other nodes.

Once a remote node is detected, receiving and transmitting buffer rings are allocated by the user-space program, and a handshake takes place using the ExaNet interconnect. Afterwards, IP packets are transferred over the ExaNet fabric, from the transmitting buffer of an endpoint to the receiving buffer of the other endpoint. In order to improve performance, multiple packets may be sent in a single RDMA transfer, and we leverage the RDMA notification mechanism to efficiently synchronize the transmitter and the receiver. Finally, it is worth noting that we take advantage of the properties of the ExaNet fabric (data protection, in-order reliable delivery) to reduce the burden on the CPU cores.

In order to interact with Linux kernel networking, the program instantiates a TUN socket, which allows simple user-space programs to act as a complete IP interface. Hence, the Linux kernel routes packets into the user-space program and packets received by the program are forwarded into the kernel. In this implementation, we relied on the $read()$ and $write()$ system calls. With newer Linux kernels, \textit{liburing} may be leveraged to improve performance.
 \begin{figure}[h]
    \includegraphics[width=\textwidth,scale=0.8]{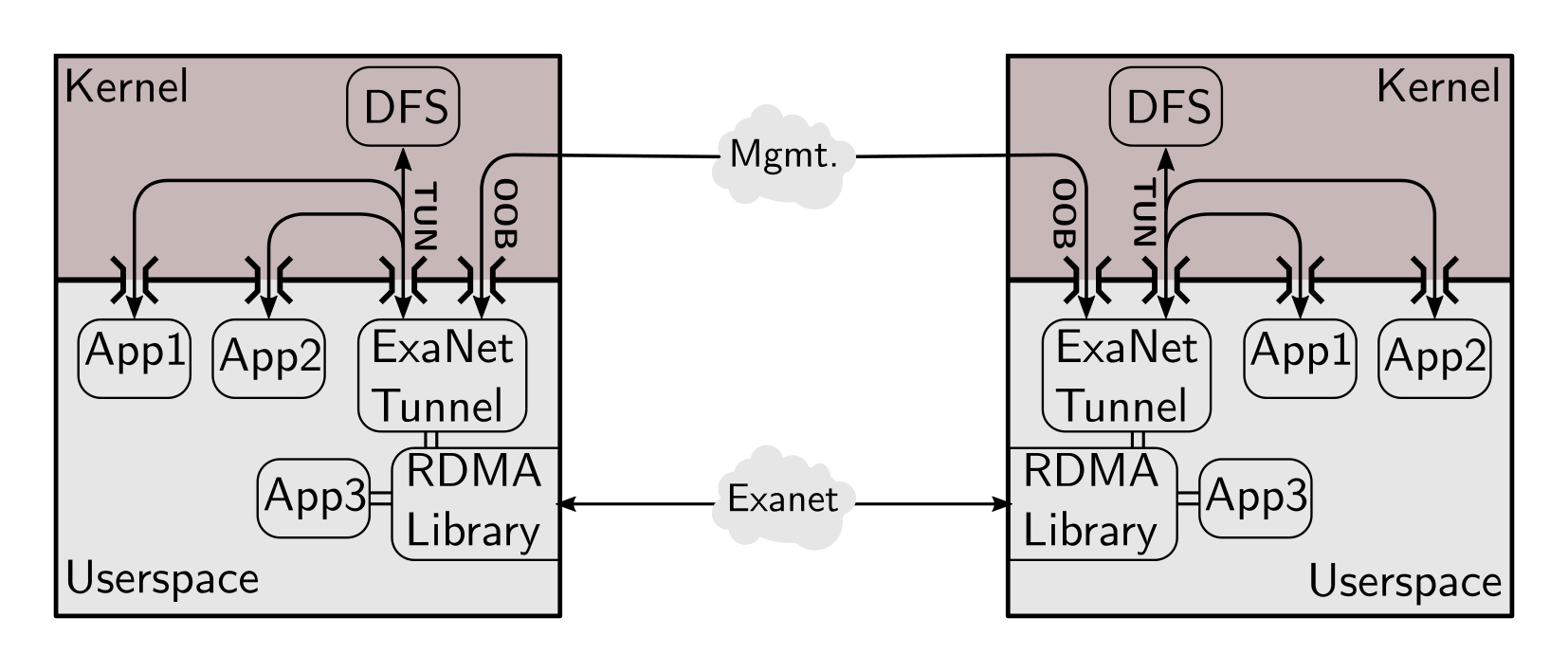}
    \caption{Principle of the IP-over-ExaNet service}
    \label{fig:converged_network}
\end{figure}

\pagebreak 

In order to assess the performance of this solution, we used two nodes with a network distance of 5 hops, and we compared with the 10 Gigabit Ethernet management interface. It is important to note that both management interface links and cross-QFDB ExaNet links have the same lane rate (10Gbit/s).  For throughput measurements, the \textit{iperf3} benchmark was used. Results are presented in Figure~\ref{fig:converged_network_throughput}. The converged network service consistently offers better throughput. In particular, for large UDP packets, our solution achieves a throughput of 4.7~Gb/s while the baseline reaches only 1.3~Gb/s. Latency-wise, the structure of our solution hinders its reactivity to sporadic messages. When the user-space program actively polls for incoming messages, average Round-Trip Time (RTT) is 90 $\mu$s; while baseline average RTT
is 72 $\mu$s. Another approach relies on adaptive sleep periods to provide very similar throughput with much smaller processing core usage. However, in that case, average RTT reaches 2.2 ms; which might be prohibitive for delay-sensitive applications.

 \begin{figure}[h]
    \includegraphics[width=\textwidth,scale=0.8]{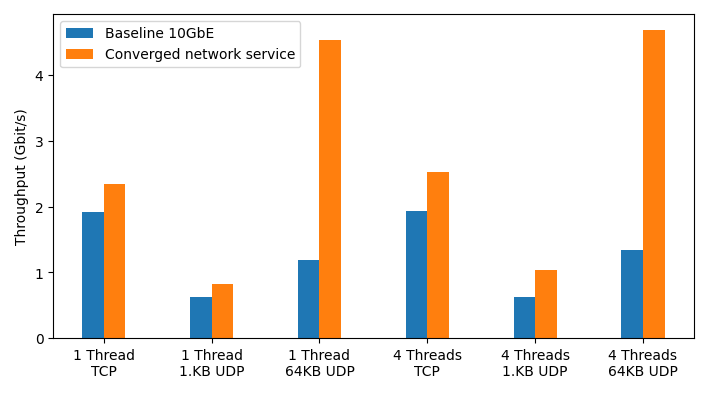}
    \caption{Throughput for different IP traffic scenarios using the converged network service over 10Gbit/s high-speed links}
    \label{fig:converged_network_throughput}
\end{figure}

\end{TRonly}

%% file: 6_evaluation.tex
\section{Evaluation} \label{sec:evaluation}
In this section we evaluate the performance of the ExaNet communication infrastructure.
The evaluation includes results from microbenchmarks to real-world MPI scientific applications, using the MPI runtime described in Section \ref{sec:mpi_runtime}.

We first employ {\bf point-to-point based microbenchmarks}, from the  OSU suite \cite{osumicrobenchs}, to
explore latency and bandwidth for the different types of links and paths available in the platform.
From the OSU suite, we also employ the broadcast and allreduce collective microbenchmarks.
Both of them are used as a first step towards assessing how performance scales with an increasing number of
processes. The allreduce benchmark is also used to measure the performance benefit of the allreduce accelerator
that was presented in Section~\ref{sec:all_reduce}.
We further explore how the performance of the MPI runtime and the platform scales through two MPI applications and a
\textit{miniapp}.
Miniapps are relative short codes that are aimed to capture key application performance characteristics \cite{jcc2015}.
For all of them, we perform weak and strong scaling tests.
\textit{MiniFE} is a miniapp from the \textit{Mantevo} project 
\cite{minifetechreport}, that mimics the finite element generation, assembly and solution for an unstructured grid problem.
Several scientific problems require the formulation of a nonlinear system of equations, which in turn is solved by a Conjugate Gradient solver. In that sense, miniFE is representative of MPI applications targeting such problems.
The first application measured is LAMMPS \cite{LAMMPS}.  This is a widely-used tool for particle-based modeling of materials at scales ranging from atomic to mesoscale, to continuum.
The next MPI benchmark used is HPCG. HPCG \cite{doi:10.1177/1094342015593158} is intended to provide a new metric for HPC systems ranking
and constitutes a complement to the
High Performance LINPACK (HPL) benchmark. It generates a synthetic
discretized three-dimensional Partial Differential
Equation (PDE) model problem, and computes preconditioned Conjugate Gradient iterations for the resulting sparse linear system.
PDEs are also common in various scientific problems. Thus, the computation and communication patterns generated by HPCG
are also representative of various scientific applications.

\subsection{Microbenchmark-level results}

\subsubsection{Osu latency}

First, we focus on the ExaNet interconnect latency
and bandwidth, across different paths with different number of hops and different types of links in the ExaNeSt platform.
Thorough deployment of the microbenchmarks
is quite important, since deriving accurate performance measurements for complex large scale systems
is challenging \cite{1592958,7426807,4536494}. 
As mentioned above, we used the OSU microbenchmarks suite \cite{osumicrobenchs}.
Benchmarks in this suite have several desirable features. First, they perform a set of \textit{warm-up}
MPI operations in order to discard any machine setup overheads.
For the case of MPI collectives,
these benchmarks issue a barrier before calling the MPI collective primitive to synchronize the process execution \cite{4536494,7426807}. Finally, each benchmark is executed several times in order to analyze variances in performance and identify deviations over time~\cite{7426807}.
These benchmarks
are compiled and executed in the MPI runtime environment presented in Section \ref{sec:mpi_runtime}.
\begin{table}[]
\begin{tabular}{|l|c|c|c|}
\hline
\multirow{2}*{Type} & Number
& HSS Links & \multirow{2}*{Example}\\ 
& of hops & \& Capacity & \\ \hline
\textcircled{a} Intra-QFDB-sh & 1 & $1\times{}16$~Gb/s & M1QAF1 - M1QAF2 \\ \hline
\textcircled{b} Intra-mezz-sh & 1 & $1\times{}10$~Gb/s & M1QAF1 - M1QBF1 \\ \hline
\textcircled{c} Intra-mezz-mh(2) & 2 & $1\times{}10$~Gb/s $+ 1\times{}16$~Gb/s & M1QAF1 - M1QBF2 \\ \hline
\textcircled{d} Intra-mezz-mh(3) & 3 & $1\times{}10$~Gb/s $+ 2\times{}16$~Gb/s & M1QAF2 - M1QBF3 \\ \hline
\multirow{2}*{\textcircled{e} Inter-mezz($i,j,k$) } & \multirow{2}*{$i+j+k$ } & $(i+j)\times{}10$~Gb/s & MmQAF1 - MnQAF1, \\ 
& & $+ k\times{}16$~Gb/s &  $m \ne n$ \\ \hline
\end{tabular}
\caption{Different ExaNet paths between MPSoCs}
\label{tab:diff_exane_paths}
\end{table}

In Table \ref{tab:diff_exane_paths}, we list the different types of the {\bf evaluated paths}.
The ExaNeSt MPSoCs, as referenced in the "example" column, are named denoting the mezzanine index, the QFDB index, and the MPSoC index. For example,
\textit{M1QBF1} corresponds to the network MPSoC (F$1$) of the second QFDB (QB) on mezzanine one (M$1$).
The row \textcircled{a} corresponds to a path between two MPSoCs on the same QFDB (Intra-QFDB). In effect, since all network traffic inside a QFDB is routed directly (see Section \ref{sec:topology}), we have only single hop paths for each MPSoC of a QFDB. The links connecting MPSoCs within a QFDB are 16~Gb/s serial links.
Paths \textcircled{b}-\textcircled{d} listed in the table (\textit{Intra-mezz}) correspond to paths between MPSoCs that reside on the same mezzanine but on different QFDBs, connected through 10~Gb/s serial intra-mezzanine links. 
For example, the single-hop path \textit{Intra-mezz-sh}
corresponds to a link between the network MPSoCs of two different QFDBs on the same mezzanine.
A path between two MPSoCs other than F$1$ includes both intra-QFDB and inter-QFDB links. For example, the path from \textit{M1QAF2} to \textit{M1QBF3}
corresponds to entry \textcircled{d} (Intra-mezz-mh(3)) as it includes an intra-QFDB link between \textit{M1QAF2} and \textit{M1QAF1} (F2-to-F1), one intra-mezzanine 10~Gb/s link between \textit{M1QAF1} and \textit{M1QBF1} (QA-to-QB) and an intra-QFDB link between \textit{M1QBF1} and \textit{M1QBF3} (F1-to-F3).
Paths across mezzanines include at least one inter-mezzanine 10~Gb/s link.
The last table entry (\textcircled{e}) shows a generic path between two MPSoCs that includes $i$ inter-mezzanine, $j$ intra-mezzanine and $k$ intra QFDB links.
\begin{figure}[b]
 \centering
 \begin{subfigure}{0.45\textwidth}
     \includegraphics[width=\textwidth]{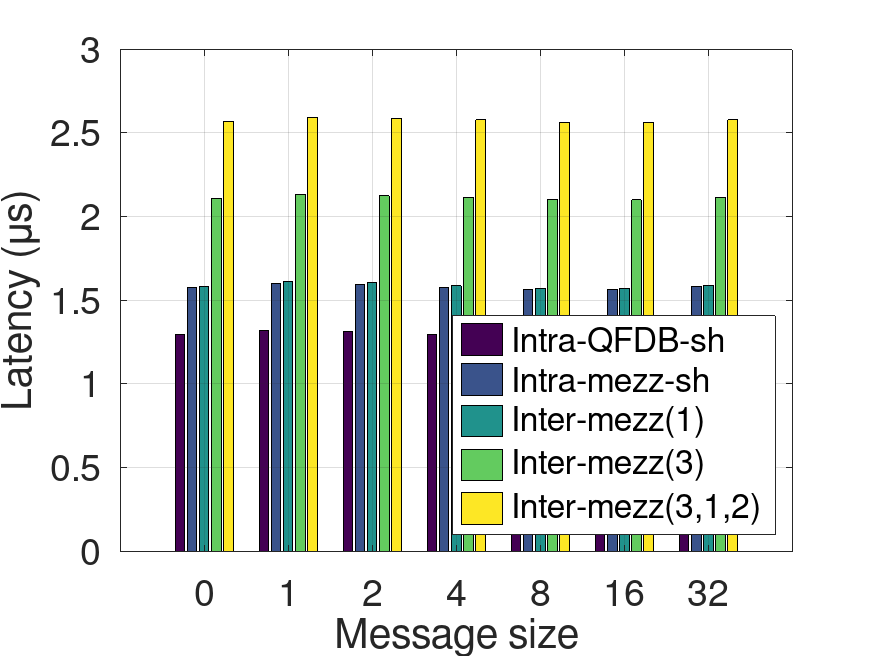}
     \caption{Eager protocol (Size $\leq$ 32B)}
     \label{fig:osu_latency_small}
 \end{subfigure}
 \hfill
 \begin{subfigure}{0.50\textwidth}
     \includegraphics[width=\textwidth]{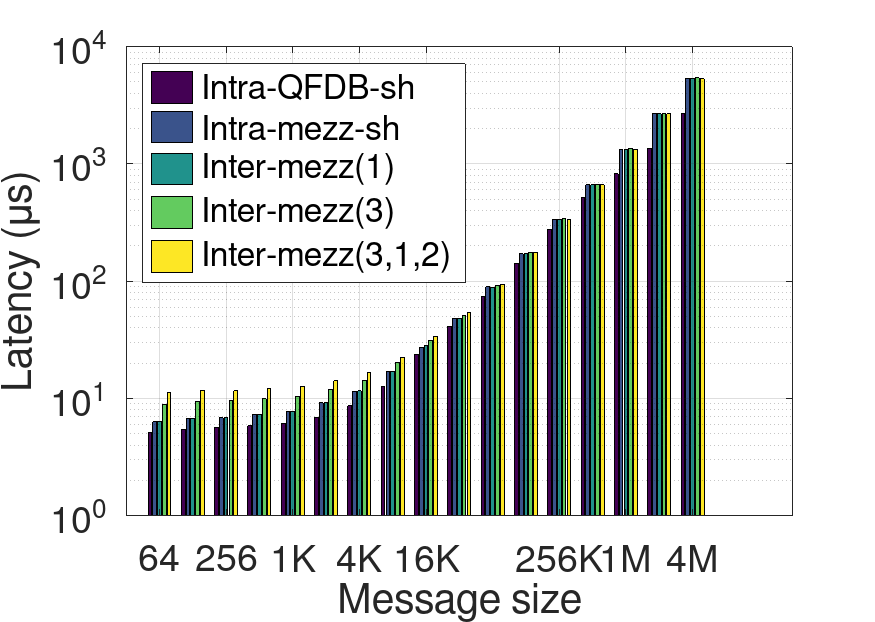}
     \caption{Rendez-vous protocol (Size $\geq$ 64B)}
     \label{fig:osu_latency_large}
 \end{subfigure}
 \caption{ExaNet-MPI: Osu\_latency results for the ExaNet paths in Table~\ref{tab:diff_exane_paths} and different message sizes}
 \label{fig:osu_latency}
\end{figure}

Figure \ref{fig:osu_latency} shows the network latency between two MPSoCs for various paths and message sizes, using the 
\textit{osu\_latency} microbenchmark.
Small messages of up to 32 bytes are transferred using the ExaNet-MPI eager data transfer protocol, which involves only the packetizer and the mailbox, as described in Section \ref{sec:mpi_runtime}.
For single-hop paths between two MPSoCs within a QFDB, the latency of an empty message (i.e. the payload size is 0) is $1.293$~$\mu$s.
Messages of up to $32$ bytes have a latency lower than $1.320$~$\mu$s. This includes MPI-related
processing, such as management and recording of MPI transactions, on both endpoints; as well as the delay
to transfer the message from main memory of the sender to main memory of the receiver. At the source node, the message has to be copied from the main memory to the packetizer; while at the destination node, the message has to be copied from the mailbox to the main memory. 
Using the Xilinx Chipscope logic analyzer core and tool \cite{xchipscope}, we have measured $100\sim{}150$~ns (approximately $15\sim{}23$~ FPGA clock cycles)
for each such copy operation on the ZU9EG MPSoC. The remaining message transfer time includes the time for the packetizer initialization, the link latency and the link transfer time.
The duration of the physical transfer of messages of a few bytes is negligible. In order to measure the link latency more accurately, we ran the \textit{osu\_latency} benchmark with two MPI ranks on the same MPSoC. The measured latency includes all the above mentioned components but the link latency, since the message is transferred inside the same MPSoC. The latency of an empty message is $1.17$~$\mu$s and thus the link latency is about $120$~ns ($1.293$-$1.17$). The time consumed by the MPI-related processing is quite noticeable because
the ZU9EG MPSoCs incorporates slow in-order ARM Cortex-A53 cores. Moreover, the latency is further affected by the long hardware path between the CPU and the memory controller inside the MPSoC. Overall, the execution time of the MPI-related processing depends on the processor speed and the implementation of the MPI library; while our main focus is on the optimization and evaluation of the ExaNet interconnect, including the Network Interface described in Section \ref{sec:network_interface_hw}.
Therefore, we created a custom micro-benchmark that directly performs a ping-pong message test directly from user space, without any kernel intervention, utilizing
the packetizer of the sender  and the mailbox of the receiver. Direct access to these hardware blocks is achieved
through the user-space libraries discussed in Section~\ref{sec:userspacelibs}. In these tests, $1000$ ping-pong messages are transmitted between two adjacent MPSoCs on the same QFDB. The average measured one-way message latency was around $470$~ns.
Another interesting observation is that the average latency increases from $1.293$~$\mu$s for $0$ bytes to $5.157$
for $64$ bytes. In effect, small messages go through the packetizer, which supports payloads up to $64$ bytes. Larger messages are sent using the RDMA engine, which
incurs various overheads. The rationale is that those overheads are amortized for larger transfers.
Furthermore, we have compared the latency between two adjacent MPSoCs on the same QFDB (\textit{Intra-QFDB-sh} path)
with the latency between two adjacent MPSoCs on different QFDBs on the same Mezzanine (\textit{Intra-mezz-sh} path). For a message of zero bytes, which carries
only control data of the MPI protocol, the corresponding latencies are $1.293$ and $1.579$~$\mu$s respectively.
The reason for this discrepancy is that the router for the communication of the MPSoCs on different QFDBs is slightly more complex, since it supports network deadlock avoidance,
and thus incurs slightly higher latency (approximately $280$ns).

We further measured the single-hop \textit{communication latency}, which includes the inter-mezzanine link latency and the message processing required by the routing logic, between two MPSoCs on different neighboring mezzanines. The measured communication latency is about $409$~ns. 
Note that, for a path of $N$ intra- or inter-mezzanine hops, the message passes through $N+1$ MPSoCs and thus through $N+1$ ExaNet hardware switches, which are responsible for routing the packets.
Based on the measured results, we have derived that the approximate link latency is $120$~ns, and thus the ExaNet routing block latency (denoted as $L_{ER}$) is approximately $(409-120)/2 \simeq 145$~ns.
Also, the link latency and the ExaNet switch latency ($L_{ER}$) are almost the same for intra- and inter-mezzanine single-hop paths.

We have next analyzed and measured the latency of an empty message transfer over the shortest
and over the longest ExaNet paths shown in Figure \ref{fig:osu_latency}. The right-most bars depicted in this
figure correspond to the latency of a path that
consists of three inter-mezzanine hops over 10~Gb/s links, one 10~Gb/s intra-mezzanine link
and two 16~Gb/s intra-QFDB links (denoted as \textit{Inter-mezz(3,1,2)}).
The latency for an empty message over this path is $2.555$~$\mu$s. 
Based on the aforementioned measured results, the expected latency over this path is $1.17 + 5*L_{ER} + 4*L_{l} + 2*L_{l}$~$\mu$s, where $L_{l}$ is the link latency.
Note that $1.17$~$\mu$s is the part of the intra-QFDB single hop latency excluding only $L_{l}$.
In the previous equation, term $2*L_{l}$ refers to the latency contributed by the two intra-QFDB hops.
Thus, the expected latency over the aforementioned path  is $2.615$~$\mu$s, which accurately matches  the observed value.

Finally, we have measured the latency of large messages using the OSU microbenchmark. The latency of a $64$ bytes message over an intra-QFDB path is $5.157$~$\mu$s, while the corresponding latency of a $4$ MBytes message
is $2689.4$~$\mu$s, which implies that the throughput achieved by the DMA engine (discussed in
Section~\ref{sec:net_interface}) is approximately $12.475$~Gb/s. This result matches the \textit{osu\_bw}
throughput reported by the benchmark, as explained below.
As a brief overview, Table~\ref{tab:sum_osu_latency} summarizes osu latency results derived through ExaNet-MPI
and the case of zero byte messages. Apart from the paths listed in Table~\ref{tab:sum_osu_latency}, we also list
the case of a path that consists of a single hop inside the same FPGA.

\begin{table}[]
\begin{tabular}{|l|c|}
\hline
Type & Num of osu\_latency ($\mu$s) \\ \hline
\textcircled{a} Intra-QFDB-sh & 1.293 \\ \hline
\textcircled{b} Intra-mezz-sh & 1.579 \\ \hline
\textcircled{c} Intra-mezz-mh(2) & 2 \\ \hline
\textcircled{d} Intra-mezz-mh(3) & 2.111 \\ \hline
\textcircled{e} Inter-mezz(3,1,2) & 2.555  \\ \hline
\textcircled{f} Intra-FPGA & 1.17  \\ \hline
\end{tabular}
\caption{Summary of ExaNet MPI osu\_latency results: zero byte message and paths in Table~\ref{tab:diff_exane_paths}}
\label{tab:sum_osu_latency}
\end{table}

\begin{figure}[b]
 \centering
 \begin{subfigure}{0.45\textwidth}
     \includegraphics[width=\textwidth]{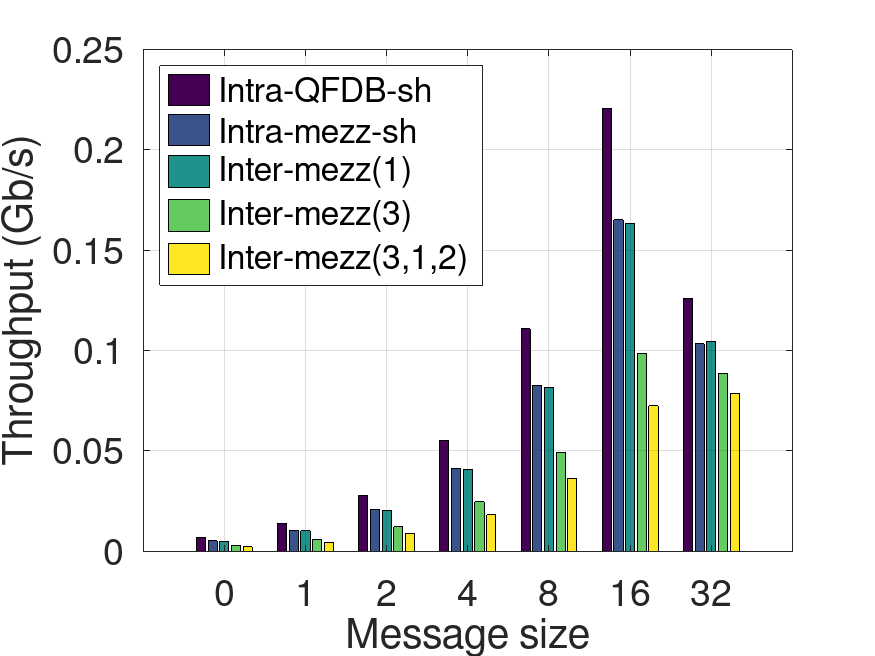}
     \caption{Eager protocol (Size $\leq$ 32B)}
     \label{fig:osu_bw_small}
 \end{subfigure}
 \hfill
 \begin{subfigure}{0.50\textwidth}
     \includegraphics[width=\textwidth]{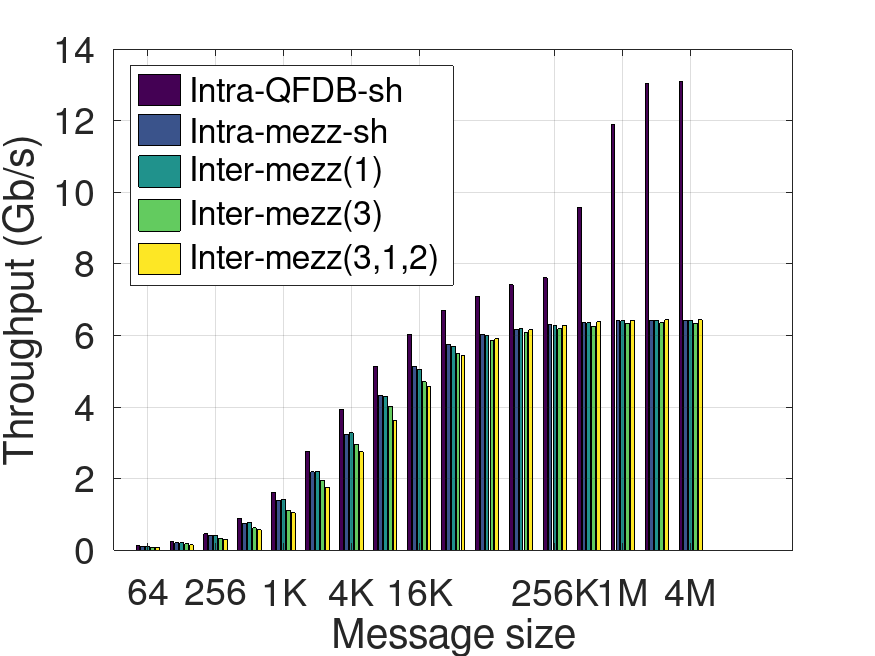}
     \caption{Rendez-vous protocol (Size $\geq$ 64B)}
     \label{fig:osu_bw_large}
 \end{subfigure}
 \caption{ExaNet-MPI: Osu\_bw results for the ExaNet paths in Table~\ref{tab:diff_exane_paths} and different message sizes}
 \label{fig:osu_bw}
\end{figure}

\subsubsection{Osu bandwidth and bidirectional bandwidth}
Figure \ref{fig:osu_bw} plots bidirectional bandwidth results from the \textit{osu\_bw} microbenchmark.
Similar to the case of latency, we consider the different paths listed in Table~\ref{tab:diff_exane_paths}. As expected, 
the capacity of the link is  under-utilized for small messages.
First, in addition to the actual payload, each message carries control information, such as the MPI tag, size, and communicator. The size of this control information is negligible in large messages that may include MBytes of payload; but not so for sizes presented in Figure~\ref{fig:osu_bw_small}. Additionally, each message exchange that follows the rendez-vous point-to-point protocol requires some
control packets, such as, Request-to-Send (RTS) and acknowledgements. For a \textit{Intra-QFDB-sh} path, a single 16~Gb/s link is traversed.
For the largest message size of $4$ MBytes, the measured throughput is approximately $13$~Gb/s. This
corresponds to a theoretical link utilization equal to $81.9\%$. For single-hop inter-mezzanine links, the link
capacity is $10$~Gb/s and the throughput achieved for messages of $4$ MBytes is $6.42$~Gb/s which
corresponds to a $64.3\%$ link utilization. This noticeable disparity in the link utilization is due to the ExaNet routing logic between different QFDBs, which transfers some control data along with each data packet, mostly for implementing flow control. The overhead of control data
would be negligible in large ExaNet packets. However, we have decided to use ExaNet packets of up to $256$ bytes in order to favor the latency over the throughput, since the latency is paramount for small control messages.

We have also run the \textit{osu\_bibw} microbenchmark for the same paths, where both
sides send messages to each other. The benchmark measures the
the maximum sustainable aggregate bandwidth between two processes \cite{osumicrobenchs}. Each side first issues multiple MPI non-blocking receive requests and then multiple MPI non-blocking send requests for a batch of messages. It then
waits for these transfers to complete before proceeding to a new batch of requests. 
For message sizes of $1$ to $4$ MBytes, the throughput reported by \textit{osu\_bibw} is almost twice the corresponding
value reported by \textit{osu\_bw}.
This is expected since the high-speed serial links connecting MPSoCs, QFDBs and mezzanines are bidirectional.
There are some small deviations though because of the sharing of the ExaNet routing logic resources and memory subsystems by both sides.
As an example, for a message of $1$ Mbyte and an \textit{Intra-QFDB-sh} link, the throughput
achieved by \textit{osu\_bibw} is $5.9\%$ lower than the expected value based on \textit{osu\_bibw}.
For small messages, where the size of the control information is comparable to the size of the payload, this deviation
can become as large as nearly $40\%$, while for an mid-size message of $4K$ over an \textit{Intra-QFDB-sh} link it is $18.3\%$.
Finally, the difference of the \textit{osu\_bibw} throughput from the expected value, based on \textit{osu\_bibw}, follows similar deviations
regardless of the type of the ExaNet path involved in the test.

\subsubsection{Osu collective microbenchmarks} 
\begin{figure}[h]
    \includegraphics[width=\textwidth,scale=0.13]{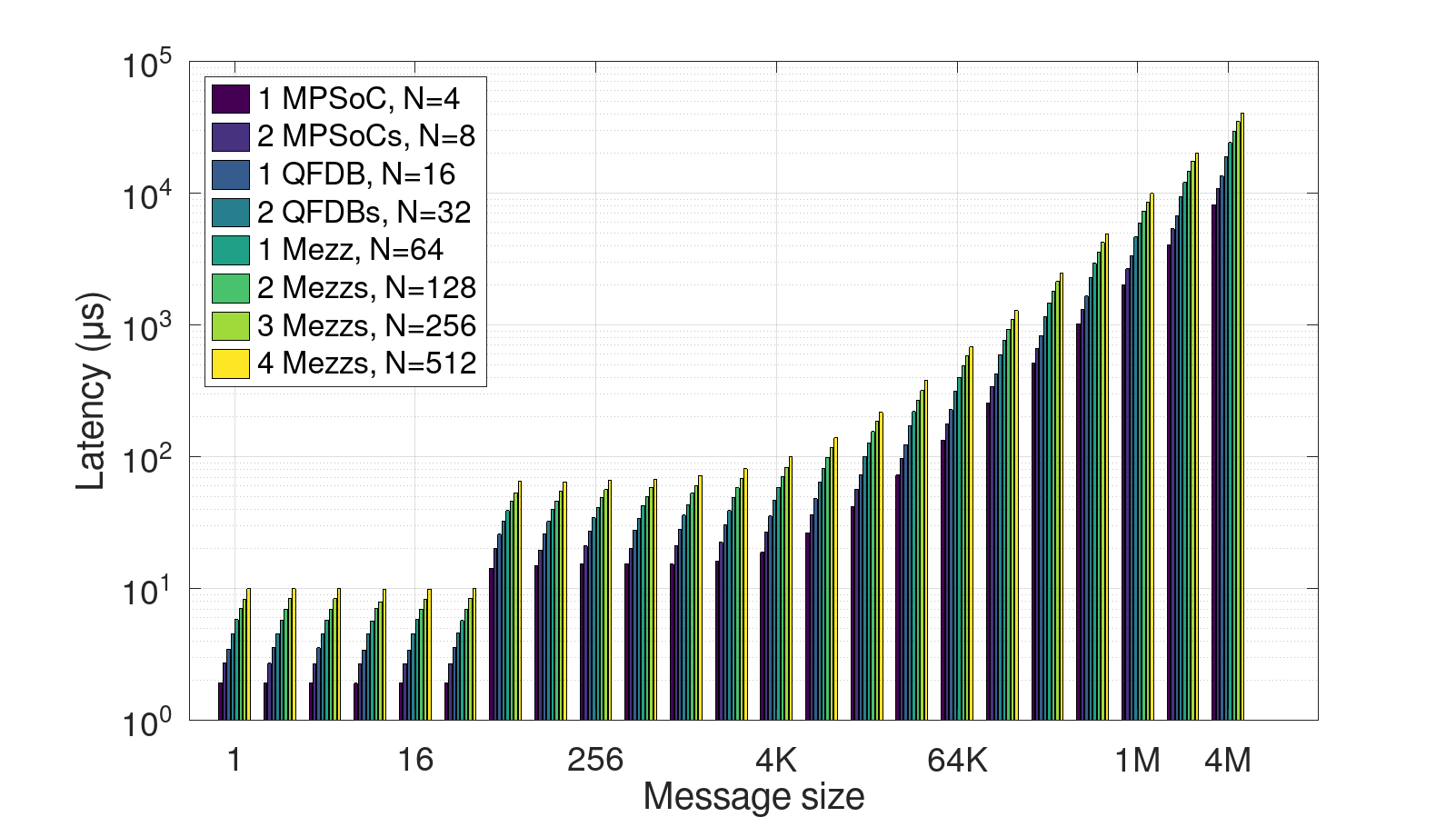}
    \caption{ExaNet-MPI: Osu\_bcast results for different number of processes}
    \label{fig:osu_bcast}
\end{figure}

The results reported so far concerned benchmarks that use point-to-point MPI primitives to exchange traffic.
Results from the \textit{osu\_bcast} benchmark that uses the
MPI\_Bcast primitive (i.e. message broadcast) will be now presented. As already discussed in Section~\ref{sec:runtimes}, part of ExaNet-MPI is a library that implements
collective primitives on top of point-to-point ones. For the case of broadcast, a binomial tree is used.
In Figure \ref{fig:osu_bcast}, we present average broadcast latency for different message sizes and different number of ranks. For
deriving the average latency, $30$ different
executions were conducted for each setup.
We spawn one process per MPSoC, so the maximum size setup contains $512$ processes, involving all eight
mezzanines of the current prototype.
Since broadcast is unrolled as a loop of point-to-point exchanges in ExaNet-MPI, there are two discrete
groups of measurements for all different setups. For messages up to $32$ bytes, the eager protocol is
used and so latency for such messages is noticeably lower than for larger messages.
For the case of $1$ byte long messages and $4$ processes, average broadcast latency is $1.93$~$\mu$s. Similar latency values are observed for all messages up to $32$ bytes, since they are
transferred through a single message in accordance to ExaNet-MPI's eager protocol.
For messages up to $1$ or $2$ kilobytes, transfer time is dominated by RDMA startup latency.
For larger messages though, doubling message sizes also resulted in doubling broadcast latency.

\begin{figure}[h]
    \includegraphics[width=\textwidth,scale=0.15]{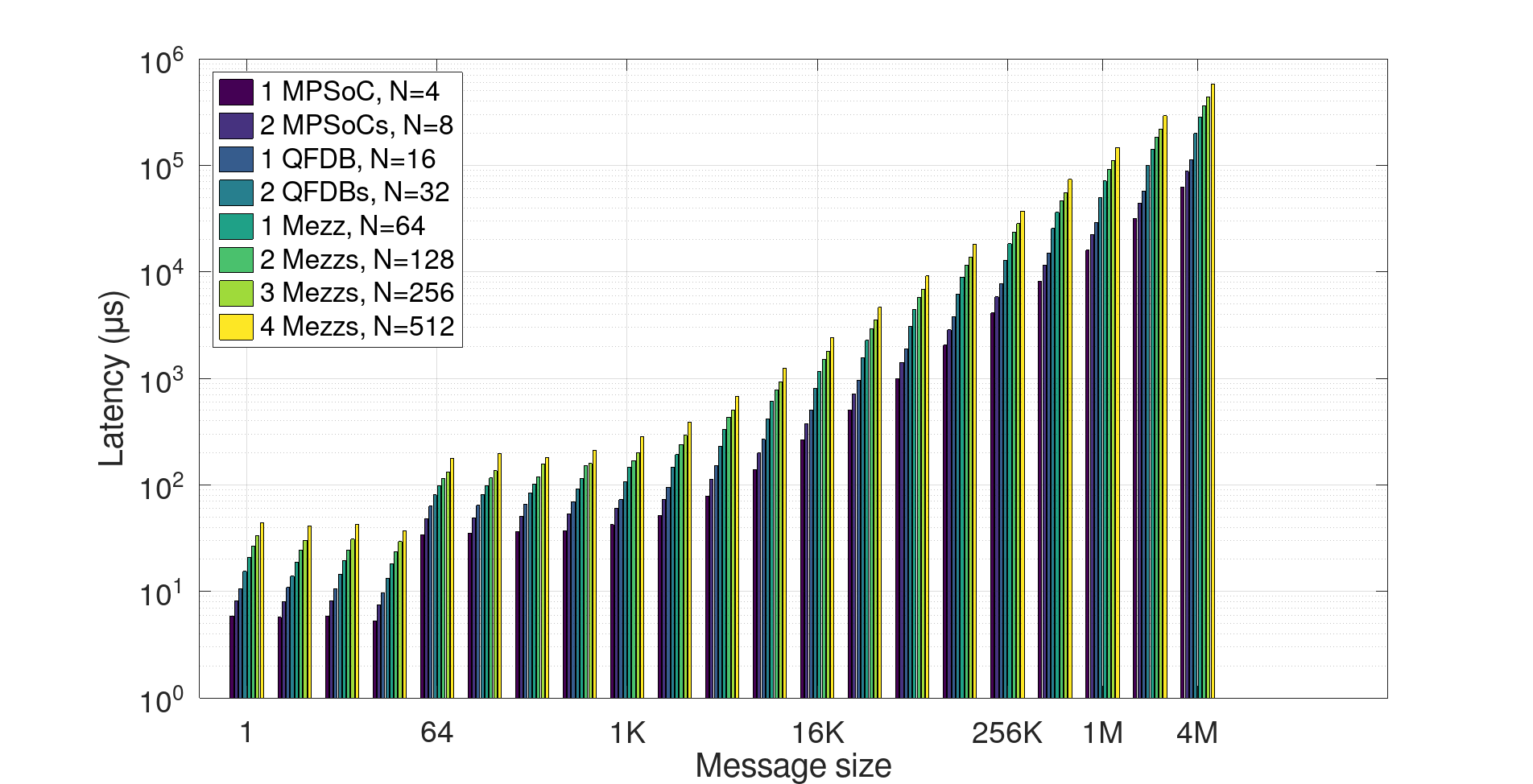}
    \caption{ExaNet-MPI: Osu\_allreduce results for different number of processes}
    \label{fig:osu_allreduce}
\end{figure}
The last microbenchmark considered is  \textit{osu\_allreduce}. The allreduce primitive in
ExaNet-MPI is implemented through the recursive doubling algorithm, which performs an
\textit{MPI\_sendrecv} call followed by an \textit{MPI\_Reduce\_local} for every scheduled step.
Apart from the communication, it also requires allocation of a temporary buffer where intermediate allreduce
results will be stored. Finally, it requires one \textit{memcopy} at the beginning to populate the
intermediate buffer with the contents in the send buffer, and a second one at the end to store
the final output in the receive buffer.
Figure \ref{fig:osu_allreduce} depicts average \textit{osu\_allreduce} latency (over 30 different executions) for a different number of 
participating processes.
Allreduce latency for the case of $4$ ranks residing on the same QFDB is $5.34$~$\mu$s. The corresponding value becomes
$33.62$~$\mu$s for the case of $64$ byte messages. The reason is again that point-to-point primitives switch to the rendez-vous
protocol for message sizes larger than $32$ bytes. As opposed to the eager protocol, the rendez-vous one relies
on the RDMA engine for message transfers, which exhibits slightly higher startup latency.
Compared to the corresponding broadcast latency for $4$ ranks, it is slightly higher due to the \textit{MPI\_Reduce\_local} and the \textit{memcopies} involved.

\subsubsection{Osu broadcast performance scaling}
Results from microbenchmarks on collectives are also used to assess the scaling properties of the network
interface and the corresponding software stack. 
Focusing on broadcast, we derive a simple model for the expect broadcast latency for a varying number of
participating processes.
The expected broadcast latency for a message of size $s$ and $N$ participating ranks is denoted as $L_{exp}(N,s)$ and is calculated through
the following equation:
\begin{equation}
L_{exp}(N,s) = Ns_{MPSoC} * L_{MPSoC}(s) + Ns_{QFDB} * L_{QFDB}(s) + Ns_{mezzanine} * L_{mezzanine}(s)
\label{eq:exp_bcast_latency}
\end{equation}
In the above equation, $L_{MPSoC}(s)$ denotes the one way latency observed for a message of $s$ bytes over a single hop intra-MPSoC path.
$ L_{QFDB}(s)$ denotes the corresponding latency for a single intra-QFDB and $L_{mezzanine}(s)$ is the corresponding latency for
a single hop intra- or inter-mezzanine path.
For the rest of the section, we will use the term \textit{inter-QFDB} to refer to intra- or inter-mezzanine paths.
Finally, $Ns_{MPSoC}$ and $Ns_{QFDB}$ capture the number of single-hop intra-MPSoC or intra-QFDB paths
traversed, while $Ns_{mezzanine}$ depicts the number of inter-QFDB paths.

For each different number of participating processes $N$, $Ns_{MPSoC}$, $Ns_{QFDB}$, and $Ns_{mezzanine}$ are extracted by identifying the pairs of communicating processes
for each step of the broadcast schedule.
For the case of $4$ ranks as an example, broadcast completes in two steps, with all four ranks residing on the same
MPSoC. For the case of $512$ ranks, the corresponding
schedule consists of five steps involving inter-QFDB transfers ($Ns_{mezzanine}=5$), two steps that involve intra-QFDB transfers ($Ns_{QFDB}=2$)
and finally, two steps with intra-MPSoC message exchanges ($Ns_{MPSoC}=2$).
Values for one-way latencies in Equation \ref{eq:exp_bcast_latency} are derived through a point-to-point microbenchmark quite similar
to \textit{osu\_latency}, named \textit{osu\_one\_way\_lat}. In that case, each rank either performs a blocking send or a blocking receive, as is the case in each 
step of the broadcast schedule.
\begin{figure}[b]
 \centering
 \begin{subfigure}{0.45\textwidth}
     \includegraphics[width=\textwidth]{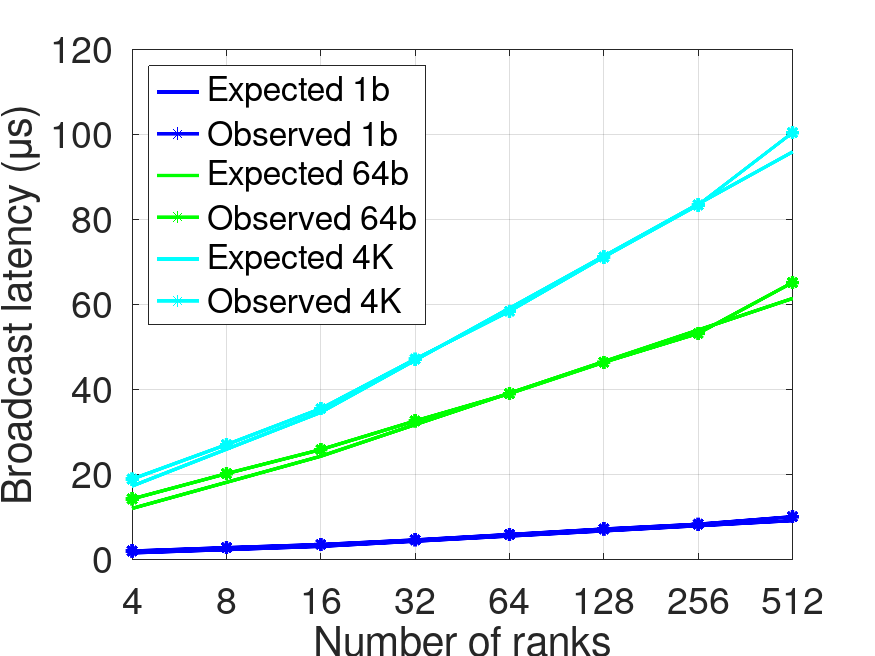}
     \caption{Small messages}
     \label{fig:bcast_scaling_small}
 \end{subfigure}
 \hfill
 \begin{subfigure}{0.45\textwidth}
     \includegraphics[width=\textwidth]{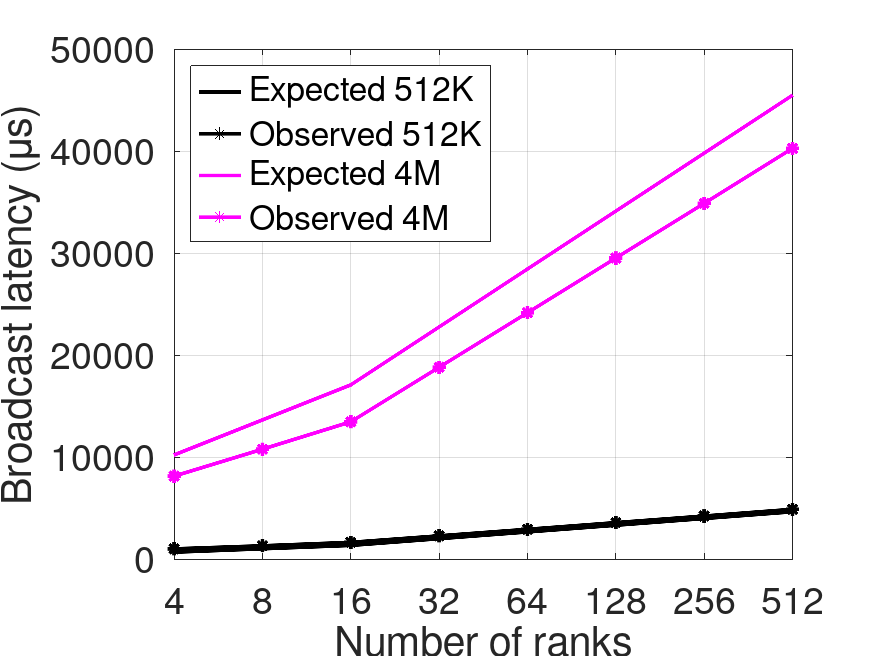}
     \caption{Large messages}
     \label{fig:bcast_scaling_large}
 \end{subfigure}
 \caption{Expected vs observed ExaNet-MPI broadcast latency}
 \label{fig:bcast_scaling}
\end{figure}

The main use of the expected broadcast latency expressed through Equation~\ref{eq:exp_bcast_latency} is to assess how
broadcast latency scales for an increasing number of participating processes. In the ideal case, expected and observed
broadcast latency values should exhibit insignificant deviations. If instead, observed latency is larger than the expected
and the gap grows with the number of participating processes, then, broadcast latency does not scale as expected.
Such an unexpected increase would be a sign of performance inefficiency involving several aspects of the
corresponding platform, including, the MPI library itself, the hardware blocks of the network interface and the
exanet interconnect.

Figures \ref{fig:bcast_scaling_small}, \ref{fig:bcast_scaling_large} present expected and actual broadcast latency for a different number of participating ranks
and indicative message sizes. Regarding the small messages in Figure \ref{fig:bcast_scaling_small}, the expected latency provided by Equation 
\ref{eq:exp_bcast_latency} is quite
close to the observed one. For the case of e.g. 1 byte, the most notable deviation corresponds to the case of $4$ ranks where the actual broadcast
latency is underestimated by approximately $24\%$. However, for small messages, one way latency values can be as low as $750$~ns, constituting broadcast
measurements more sensitive to phenomena like system noise and late arrivals.
System noise (due to OS or running daemons) can have a significant effect on performance, especially for collectives, and can introduce variance 
in the performance observed \cite{1592958,10.1007/11602569_31,Jones03impactsof}. 
Let us note that \textit{osu\_bcast} microbenchmark
performs a quite fine grain timing by acquiring one timestamp before and one after each \textit{MPI\_Bcast} execution. Moreover, a barrier is used after each
broadcast call. Disabling the barrier invocation and averaging latency over all iterations, results in an average broadcast latency
of $1.7319$~$\mu$s. The corresponding underestimation of the actual broadcast latency becomes approximately $15\%$, instead of $24\%$.
In Figure \ref{fig:bcast_scaling_small} we also see that as the number of ranks grows, the expected latency quite closely matches the observed one. The corresponding underestimation is $9.3\%$ for the case of $512$ ranks.

Another interesting observation regards the case of $4K$ messages. For scenarios with $4$ ranks only, the expected latency is $8.6\%$ lower than the actual one.
Apart from system noise and barrier effects, there is an additional reason that explains this gap. For messages that are larger than $32$ bytes, the actual data
mover used is the RDMA engine (instead of the packetizer). When multiple ranks in the same MPSoC issue concurrent RDMA transfers, they share its bandwidth.
This effect though is not captured by \textit{osu\_one\_way\_lat} which measures one way latency with a single pair of active processes per MPSoC.
Accurately capturing the drop in throughput due to bandwidth sharing is not that easy depends on the overlap in time exhibited by the
two concurrent transfers. It is also worth mentioning that, in the case of $4K$ messages and $16$ or more ranks, the absolute value of the deviation is less than 
$5\%$.
A pattern similar to that of $4K$ messages applies for the larger messages depicted in Figure \ref{fig:bcast_scaling_large}.
Equation \ref{eq:exp_bcast_latency} underestimates the observed latency by $32.4\%$ and $15.4\%$ for the case of $4$ ranks and messages of $512K$ and $4M$
respectively. For larger messages though, the effect of RDMA bandwidth sharing is expected to be more prominent.
It is also worth mentioning that the deviation between the expected and actual broadcast latency is less than $12\%$ for most of the scenarios
explored and plotted in Figure \ref{fig:bcast_scaling_large} (especially for large numbers of participating processes).
Overall, broadcast latency seems to increase in a predictable manner as the number of participating ranks increases.

\subsubsection{Accelerated allreduce collective}
In this section, we report on the evaluation based on MPI microbenchmarks. The main target of this process is to assess the benefit of the allreduce
accelerator which is part of the Network Interface (more details in Section \ref{sec:all_reduce}).
For that reason, we derive \textit{osu\_allreduce} latencies for two different ExaNet-MPI configurations: in the first one, the Allreduce collective
is handled in software and is implemented through a recursive doubling algorithm, while in the second configuration, Allreduce is handled
by our custom accelerator block. We explore four different experimental setups, involving $16$, $32$, $64$ and $128$
ranks. The main reason for having a maximum tested number of ranks $128$, is that the current version of the allreduce accelerator can be used by at most one rank per MPSoC, thus one MPI process is spawned in each MPSoC. The message sizes involved in the allreduce operation range from $4$ bytes to $4$ KBytes. For messages
larger than $4$ KBytes, the accelerator offers no benefit. The reason is that its current version divides messages in blocks of $256$ bytes, since it
is also the maximum packet size that can traverse the ExaNet interconnect. The Allreduce Accelerator is triggered once for each block
of $256$ bytes, which implies increased overhead for larger messages.

\begin{figure}[h]
    \includegraphics[width=\textwidth,scale=0.35]{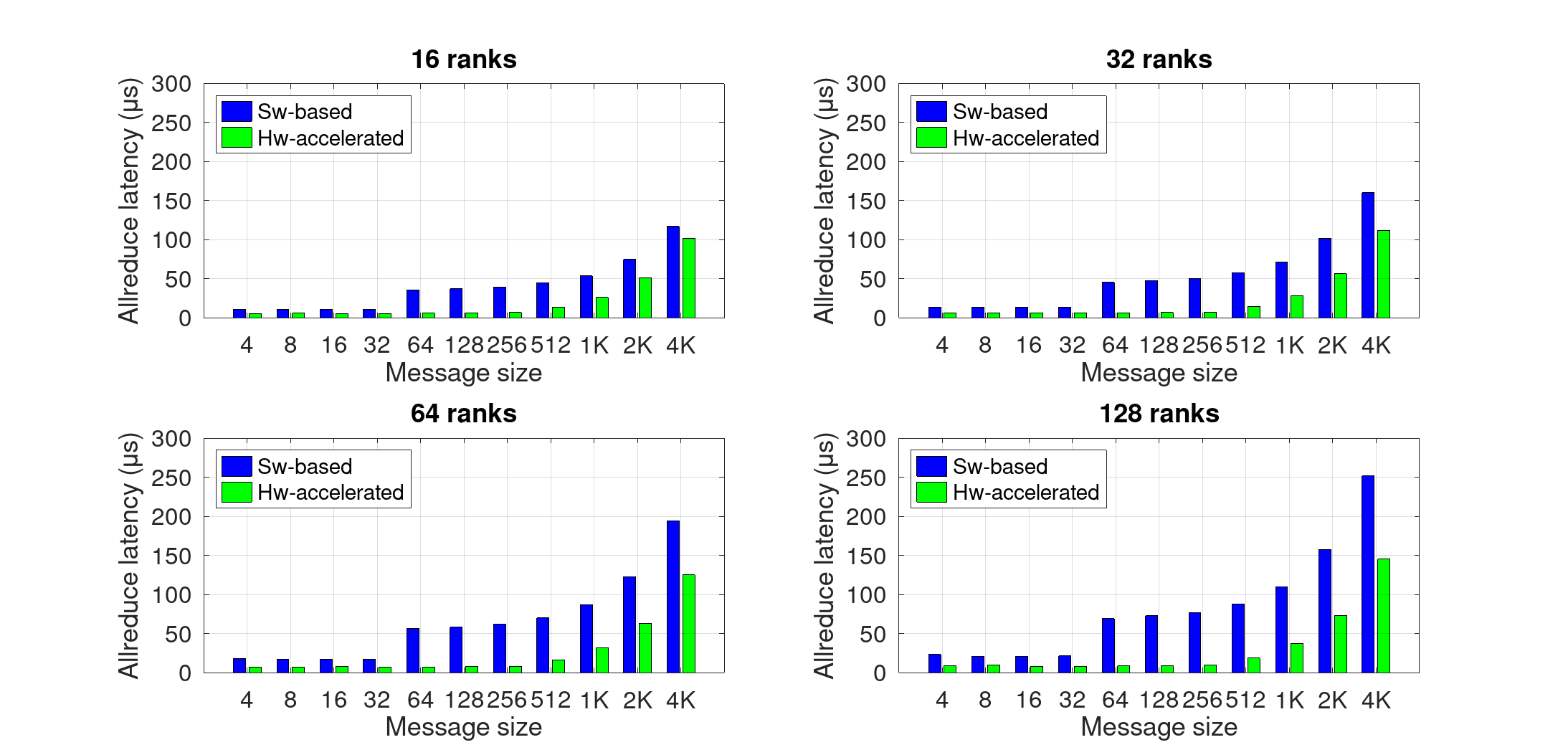}
    \caption{Benefit  of hardware acceleration for the Allreduce collective}
    \label{fig:osu_allreduce_accel}
\end{figure}
In Figure~\ref{fig:osu_allreduce_accel}, we plot the compared \textit{osu\_allreduce} latency for the two configurations of ExaNet-MPI.
In all setups presented, the accelerator achieves lower Allreduce latency than the software-based implementation. This drop is more profound for messages larger than $32$ bytes, where the software-based allreduce will use
the RDMA engine as the data mover. Recall that RDMA offers increased throughput but incurs higher latency, when compared to the packetizer.
Another interesting observation is the increased latency when the message size doubles. For the case of $16$ ranks and a message of
$256$ bytes for example, the accelerator achieves an Allreduce latency of $6.79$~$\mu$s. The corresponding values for $512$ and $1024$ byte
messages are $13.38$ and $26.11$~$\mu$s, respectively. This effect of doubled latency is expected though, since as 
mentioned the logic of the accelerator is triggered once for each block of $256$ bytes. 
Let us note that, the maximum achieved improvement in Allreduce latency reaches
$83.4\%$, $86.2\%$, $87.1\%$, and $87.9\%$ for the four rank setups depicted in Figure~\ref{fig:osu_allreduce_accel}.
Finally, we note that, the accelerated Allreduce collective seems to scale better when the number of participating ranks
increases. For the case of $256$-bytes messages, the average \textit{osu\_allreduce} latency is respectively $6.79$~$\mu$s for $16$ ranks and $9.61$~$\mu$s for $128$ ranks, when using hardware acceleration. The corresponding values for the software-based implementation are $39.7$ and $76.9$~$\mu$s, respectively. While the latency of the hardware-accelerated solution is almost stable, the latency of the software implementation  almost doubles.
\subsection{Full MPI application results}
\begin{figure}[b]
 \centering
 \begin{subfigure}{0.45\textwidth}
     \includegraphics[width=\textwidth]{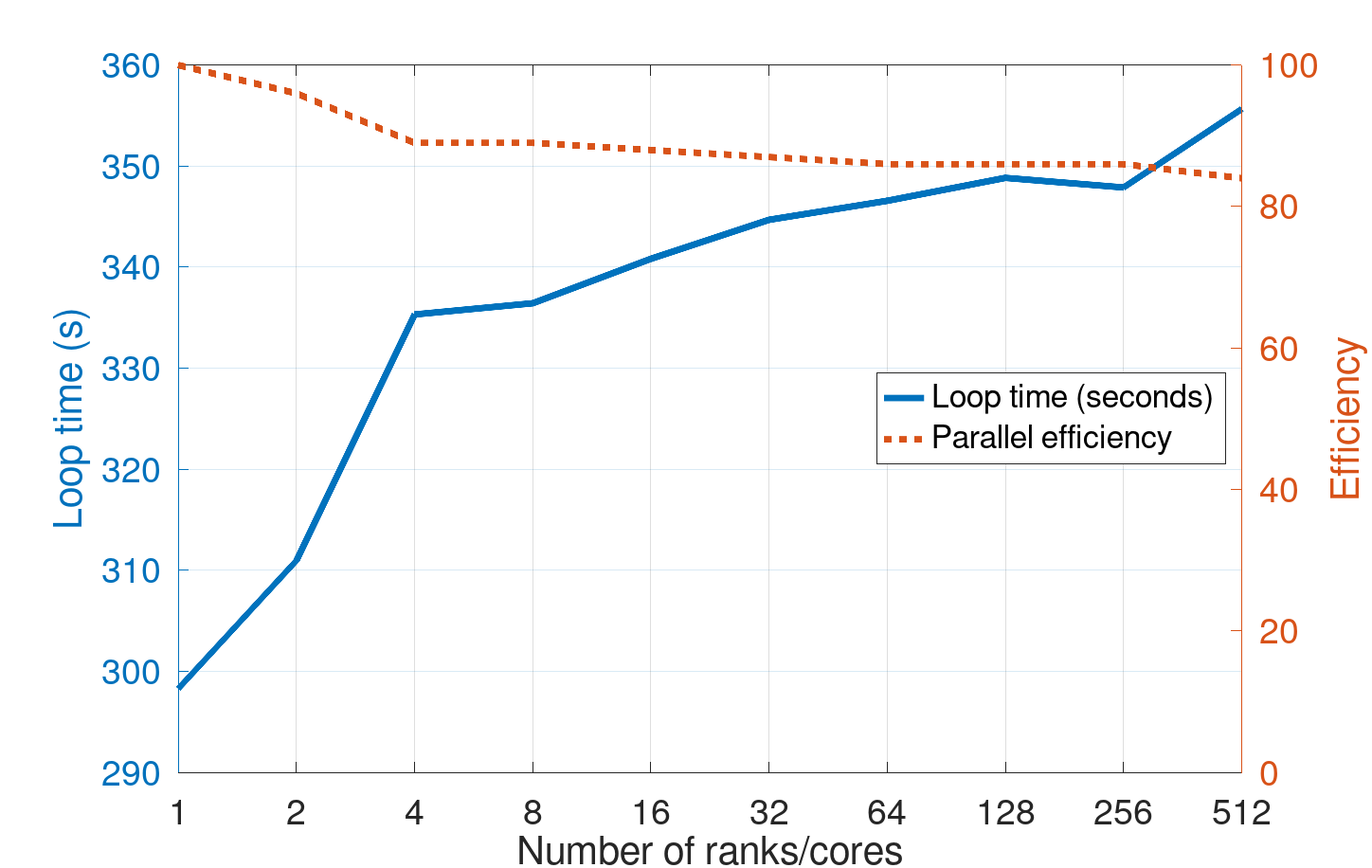}
     \caption{Weak scaling}
     \label{fig:lammps_weak_scaling}
 \end{subfigure}
 \hfill
 \begin{subfigure}{0.45\textwidth}
     \includegraphics[width=\textwidth]{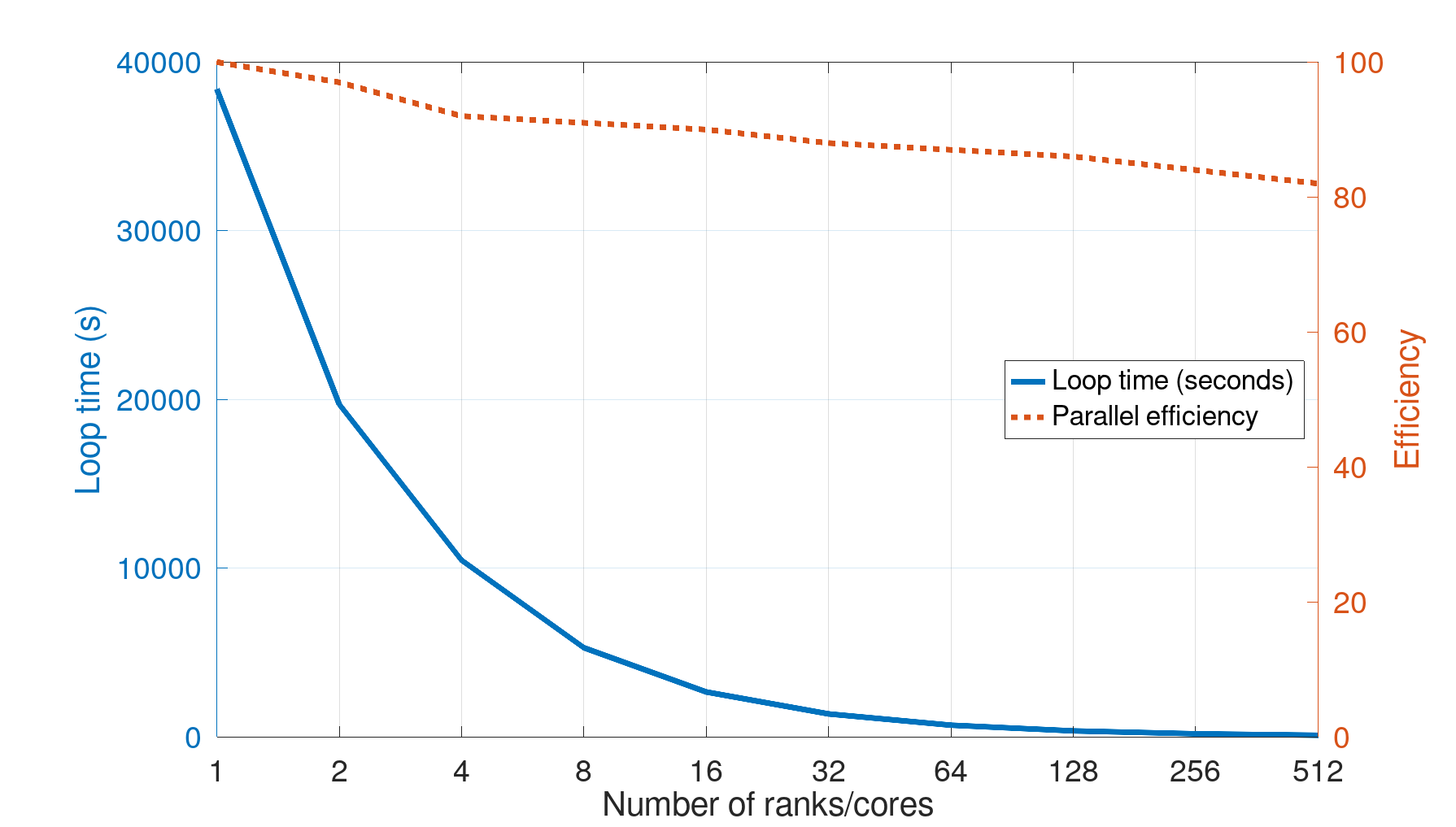}
     \caption{Strong scaling}
     \label{fig:lammps_strong_scaling}
 \end{subfigure}
 \caption{LAMMPS scaling tests}
 \label{fig:lammps_scaling}
\end{figure}
In this section, we evaluate the ExaNet interconnect running full MPI applications. The reported performance of an MPI application is the average performance of $15$ executions. The ExaNet-MPI runtime does not use the Allreduce accelerator described in Section \ref{sec:all_reduce}.
For all applications, we perform both weak- and strong-scaling tests. Apart from the actual performance achieved,
we also calculate weak ($E_{w}$) and strong scaling efficiency ($E_{s}$) through the following equation:
$E_{ \lbrace w,s\rbrace } = Sp_{N}^{\lbrace w,s\rbrace}/N$. In that equation, $N$ denotes the number of processes,
assuming that each process
runs on a separate core. $Sp_{N}^{\lbrace w,s\rbrace}$ is the speedup achieved when employing $N$ processes instead of a single 
one.
To be more precise, for the benchmarks where performance is evaluated through completion time, speedup for the
weak scaling tests is estimated through $Sp_{N}^{w} = N*\frac{t_{1}}{ t_{N}}$. In that equation, $t_{1}$ and $t_{N}$
denote the completion time for the case of a single and $N$ processes respectively.
For the strong scaling tests, the corresponding speedup is $Sp_{N}^{s} = \frac{t_{1}}{ t_{N}}$.

We first evaluate the platform by running the \textit{rhodopsin} protein benchmark of the Large-scale Atomic/Molecular Massively Parallel Simulator (LAMMPS) \cite{LAMMPS}, where we perform weak and strong scaling analysis.
In Figure \ref{fig:lammps_weak_scaling}, a weak scaling analysis is performed executing in parallel $1$ to $512$ processes.
Following the suggested configurations provided in \cite{lammpsuserguide},
$100$ timesteps are performed on a single core for a simulation of $32.000$ atoms. The simulation size when $512$
processes are executed in parallel is $16.384.000$ atoms. This figure presents the performance of LAMMPS depending on the number of cores used.  The corresponding parallel efficiency --which measures how well an application leverages additional cores-- is also shown. The left y-axis (\textit{loop time}) in the figure depicts the total simulation time. We run one MPI rank on each CPU core.
As this figure shows, the efficiency never drops below $84\%$ showing that, the application can scale well on many cores. However, there is a notable drop in the parallel efficiency when we have more than four MPI ranks. 
In particular, the parallel efficiency is $96\%$ and 
$89\%$ for
two and four ranks, respectively.
The main reason for this drop is that, the MPSoC includes four CPU cores. Thus, when all four cores are used, the contention to the main memory  
increases and the memory channel becomes the bottleneck. This is due to architectural limitations of the ZU9EG MPSoC, which supports a single memory channel to the main memory.

Figure \ref{fig:lammps_strong_scaling} presents the results from the strong scaling analysis.
As this figure shows, the LAMMPS \textit{rhodopsin} protein benchmark can still scale well as system resources increase.
Efficiency is above $80\%$ for all the system configurations. However, efficiency exhibits a notable drop 
as resources 
are added. This is a common case when running similar tests on similar size machines, as shown in \cite{lammpsuserguide}.
Through a custom profiler, we derive the share of overall running time spent in communication. For the strong scaling test,
average communication ratio is $0.028\%$ for the case of a single MPI rank and $12\%$ for the case of $512$ ranks.

\begin{figure}[b]
 \centering
 \begin{subfigure}{0.45\textwidth}
     \includegraphics[width=\textwidth]{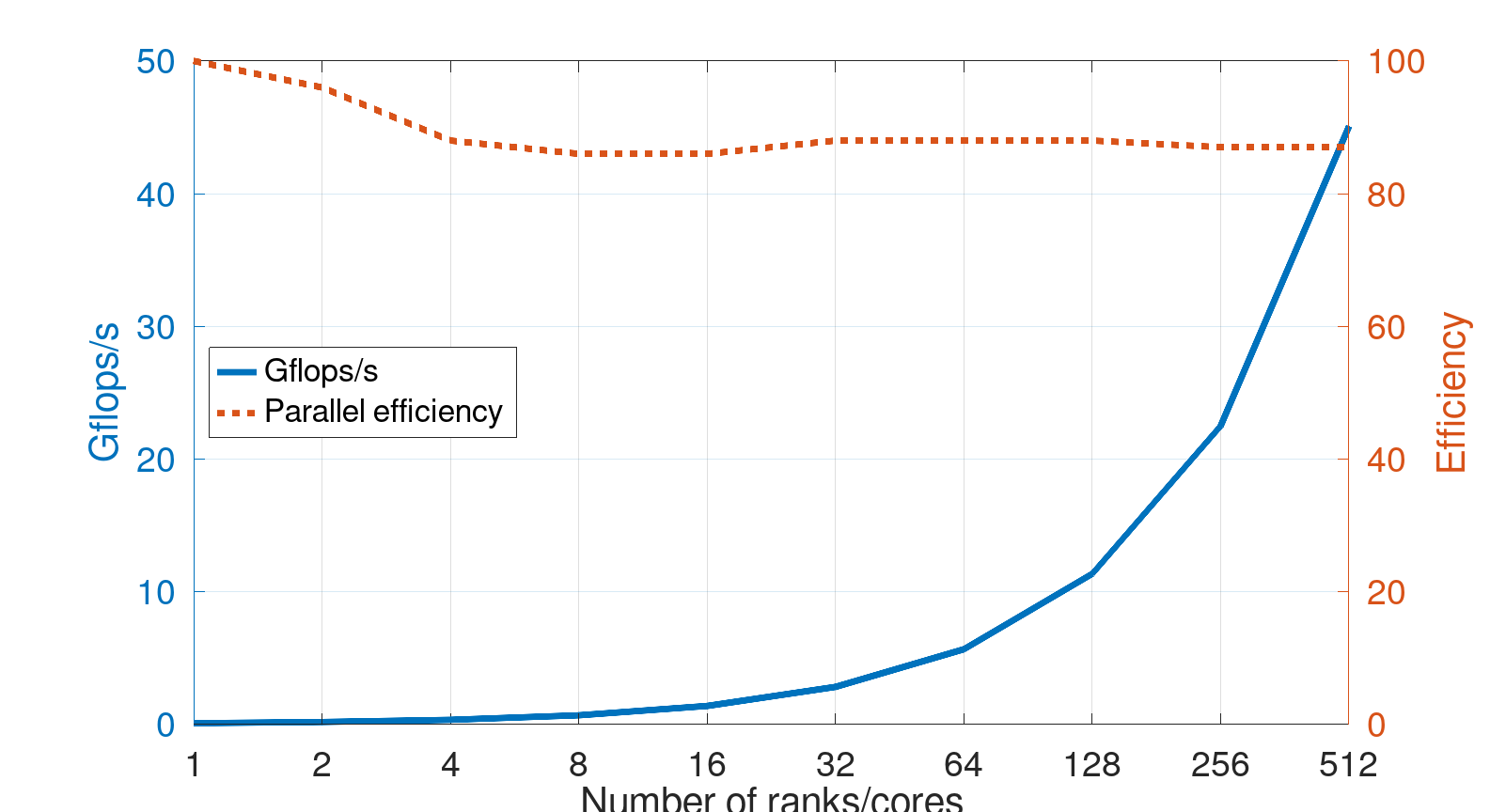}
     \caption{Weak scaling}
     \label{fig:hpcg_weak_scaling}
 \end{subfigure}
 \hfill
 \begin{subfigure}{0.45\textwidth}
     \includegraphics[width=\textwidth]{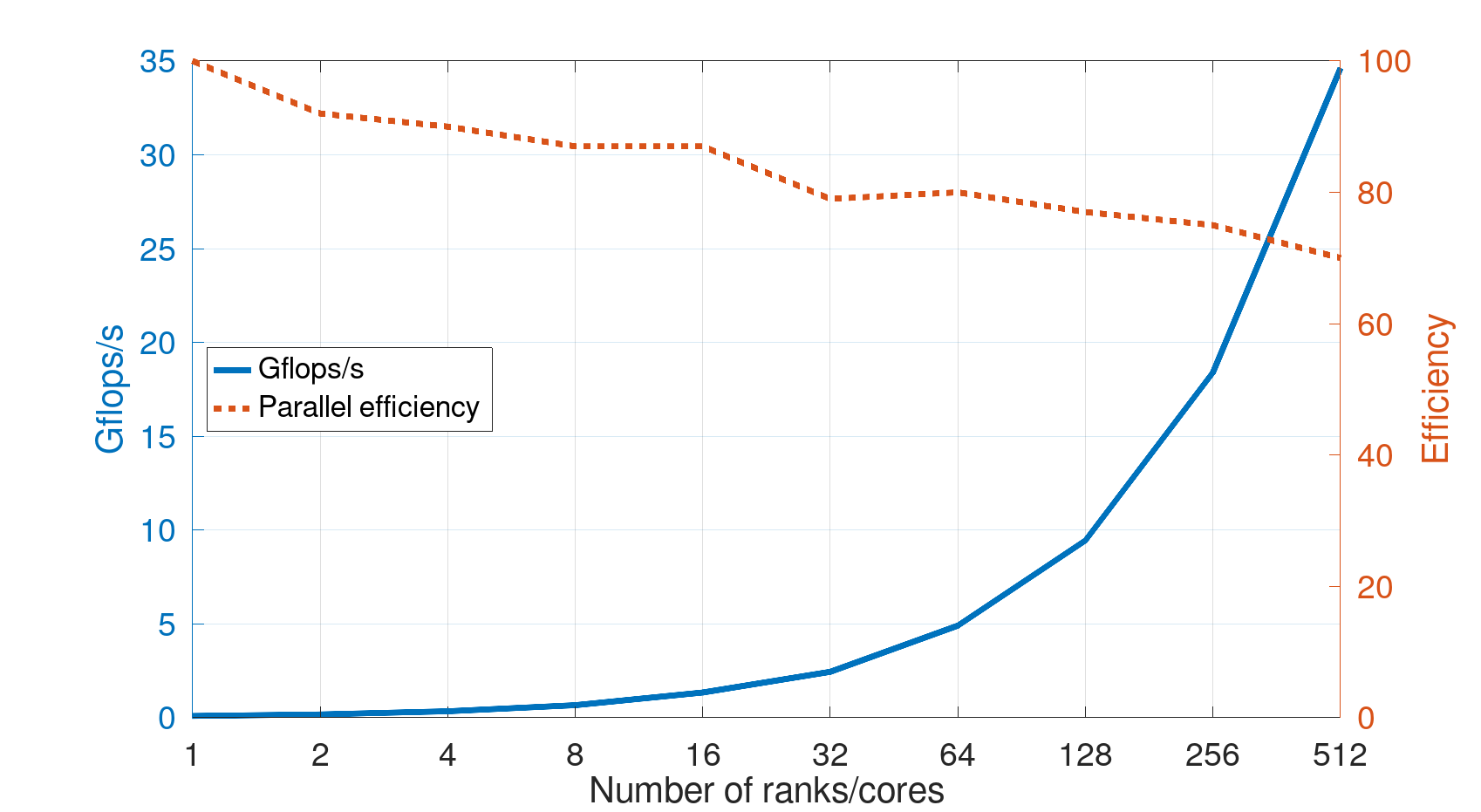}
     \caption{Strong scaling}
     \label{fig:hpcg_strong_scaling}
 \end{subfigure}
 \caption{HPCG scaling tests}
 \label{fig:hpcg_scaling}
\end{figure}
We next use the HPCG benchmark and perform both weak and strong scaling tests.
For the strong scaling tests, the local problem size is set to the
maximum problem size that fits in the memory of a single MPSoC for the case of a single run (nx=$256$, ny=$256$, nz=$128$).
For each different number of participating processes, we perform 15 different HPCG runs with duration for each run fixed
to $30$ minutes \cite{hpcgsite}. The results depicted in Figure~\ref{fig:hpcg_scaling} comprise average GFLOPS over $15$ different executions.
As Figure~\ref{fig:hpcg_strong_scaling} shows, efficiency in the strong scaling test becomes $92\%$, when using two instead of one core, and
is $70\%$ for the case of $512$ ranks.
Two indicative executions of HPCG, with two and $512$ MPI processes, reveal that the amount of time spent in communication is $0.7\%$ and $22.4\%$, respectively.
As far as the weak scaling test in Figure~\ref{fig:hpcg_weak_scaling} is concerned, local problem size is set to nx=$104$, ny=$104$, nz=$104$
for the case of a single process. For the runs with $512$ participating processes, the global problem size becomes nx=$832$, ny=$832$, nz=$832$.
For the case of a single MPI rank, memory occupancy reaches approximately $1$~GBy.
Efficiency is larger or equal than $87\%$ for all the weak scaling tests presented in Figure~\ref{fig:hpcg_weak_scaling}.

\begin{figure}[b]
 \centering
 \begin{subfigure}{0.45\textwidth}
     \includegraphics[width=\textwidth]{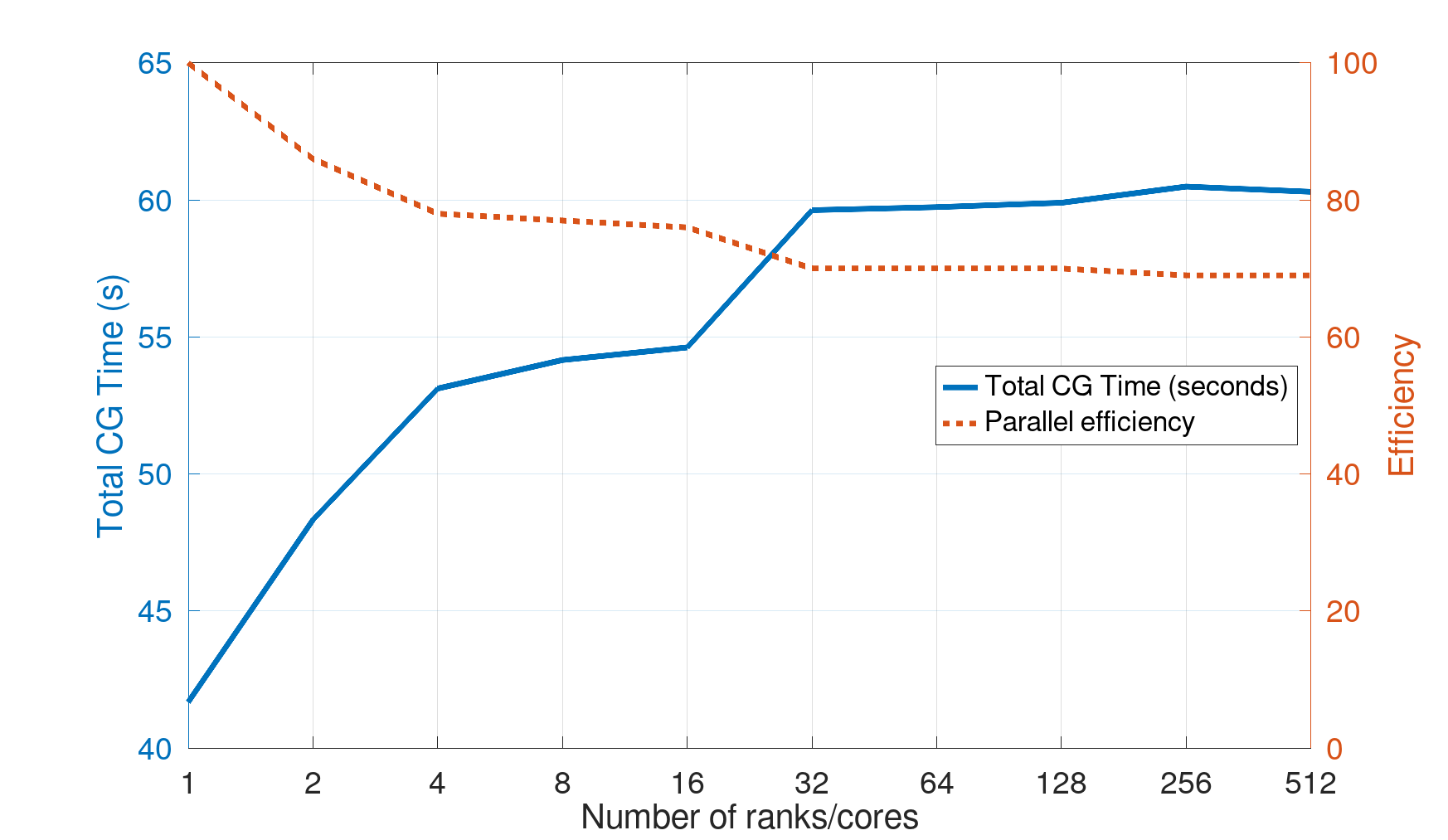}
     \caption{Weak scaling}
     \label{fig:minife_weak_scaling}
 \end{subfigure}
 \hfill
 \begin{subfigure}{0.45\textwidth}
     \includegraphics[width=\textwidth]{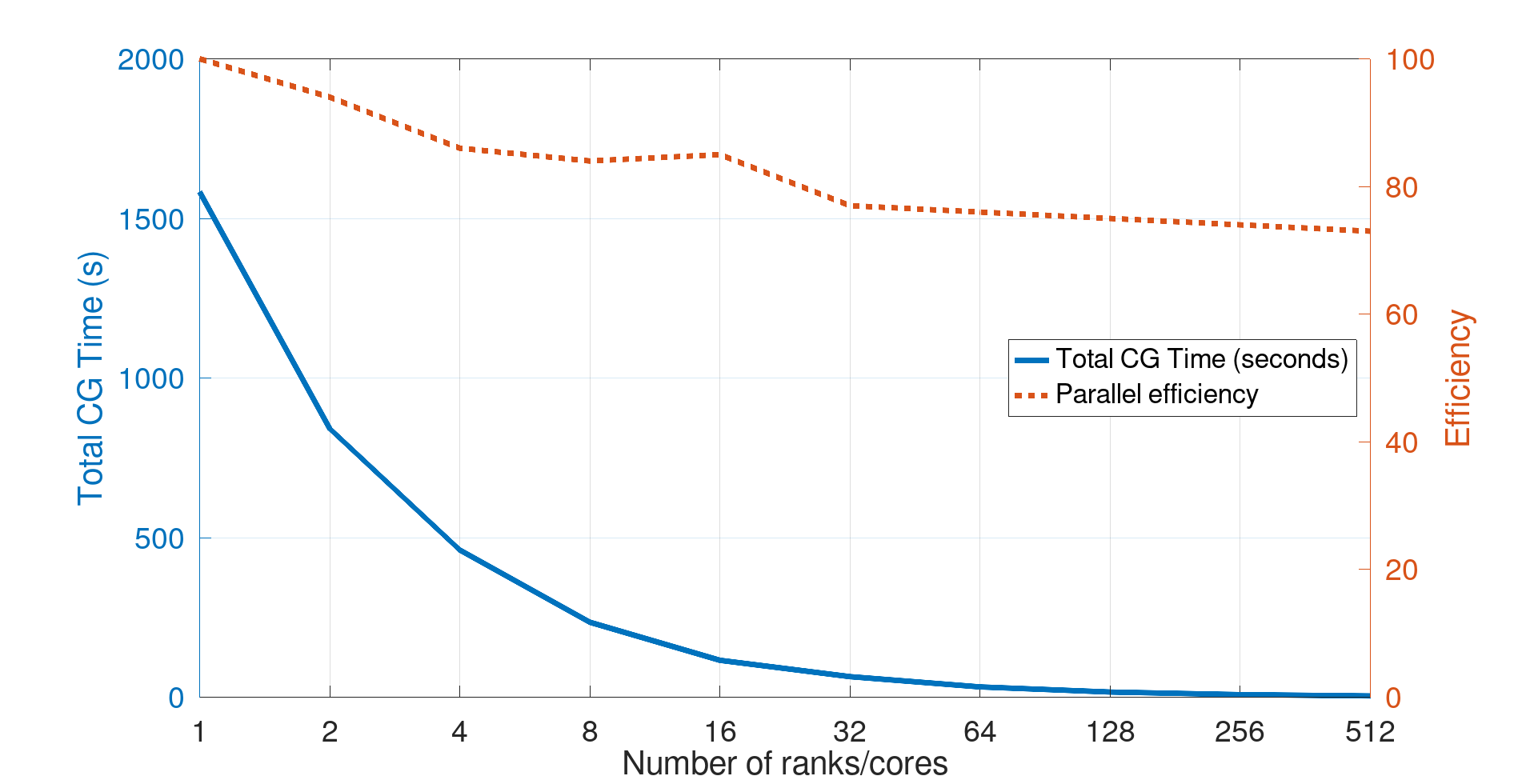}
     \caption{Strong scaling}
     \label{fig:minife_strong_scaling}
 \end{subfigure}
 \caption{miniFE scaling tests}
 \label{fig:minife_scaling}
\end{figure}
The final benchmark used to evaluate our platform is the (\textit{miniFE}) miniapp from the \textit{Mantevo} project  \cite{minifetechreport}.
For the strong scaling test, we fix the problem size to 
$(264, 264, 264)$
and the number of participating processes varies from a single one up to $512$ (i.e., full occupancy of the HPC prototype).
For the weak scaling, we modify the miniFE source code to perform $400$ iterations. The main reason is that for
a problem size of $200$ iterations and $256$ ranks, or more, final solution is not verified.
For the case of $512$ ranks, the miniFE global problem size becomes $(512,512,512)$.

Each one of Figures~\ref{fig:minife_weak_scaling}-\ref{fig:minife_strong_scaling}
depicts both average miniFE performance and the corresponding parallel
efficiency for the different setups applied. Performance is captured through the total Conjugate Gradient resolution
time (reported as \textit{Total CG Time}) and is the average of $15$ different runs.
For the strong scaling test, efficiency is as high as $73\%$ when the number of processes is increased to $512$.
The corresponding value is $69\%$ for the weak scaling test. It should also be noted that, compared to the other
benchmarks, miniFE spends a quite higher fraction of time spent in communication. In the weak scaling test and
the case of four processes for example, the amount of conjugate gradient resolution time spent in communication is
approximately $4.42\%$. As the number of participating processes increases, so does the communication time
reaching a ratio corresponding to $86\%$ for the case of $512$ processes. Even in this case though, efficiency is
almost $70\%$. 

As an overview, Table~\ref{tab:pareffsummary} presents minimum and maximum values of parallel efficiency for
the three MPI applications considered. These values correspond to configurations with 512 and two
processes respectively.
\begin{table}[]
\begin{tabular}{|c|c|c|c|c|}
\hline
          & \multicolumn{2}{c|}{Weak scaling} & \multicolumn{2}{c|}{Strong scaling} \\ \hline
Processes & 2              & 512             & 2               & 512              \\ \hline
Lammps    & 96\%             & 69\%              & 97\%              & 82\%       \\ \hline
HPCG      & 96\%             & 87\%              & 92\%              & 70\%       \\ \hline
MiniFE    & 86\%             & 69\%              & 94\%              & 72\%       \\ \hline
\end{tabular}
\caption{Parallel efficiency achieved by Exanet-MPI for Lammps, HPCG and MiniFE with two and 512
processes}
\label{tab:pareffsummary}
\end{table}

%% file: 7_hardware_accelerators.tex
\section{Hardware Acceleration}\label{sec:hardware_accelerators}

The high-speed AXI-based memory interfaces between the FPGA and the Processing System (PS) in the Xilinx ZU9EG MPSoC enable the deployment of hardware accelerators tightly coupled with the processor. To showcase the capabilities of the ExaNeSt prototype, High-Level Synthesis (HLS) was used to develop a Matrix Multiplication accelerator, as it is a very common kernel in HPC and AI applications -- for instance GEMM in BLAS Level-3 routines and also in Convolutional Neural Networks. In contrast to Section~\ref{sec:evaluation} that provided results for the complete ExaNeSt platform, this section focuses on a single QFDB, where all the FPGA resources are used by the accelerator.

Xilinx's Vivado HLS was used to develop a highly optimized matrix multiplication kernel tile using the directive-oriented style of HLS. This kernel tile is parameterizable with different sizes, in order to experiment with implementations that have different area \textit{vs.} performance trade-offs.
The kernel tile operates on single-precision floating point elements (FP32) and is tuned considering the finite reconfigurable resources (LUTs, Flip-Flops, DSPs) and the internal memory (BRAM) size limitations. Moreover, its design uses efficiently the provided AXI memory bandwidth.

The basic matrix multiplication algorithm is shown below:

\begin{lstlisting}
for i=0 to n do
  for j=0 to n do
    for k=0 to n do
      C[i][j] += A[i][k] x B[k][j]
\end{lstlisting}

\begin{figure}[b]
\centerline{\includegraphics[width=0.8\textwidth]{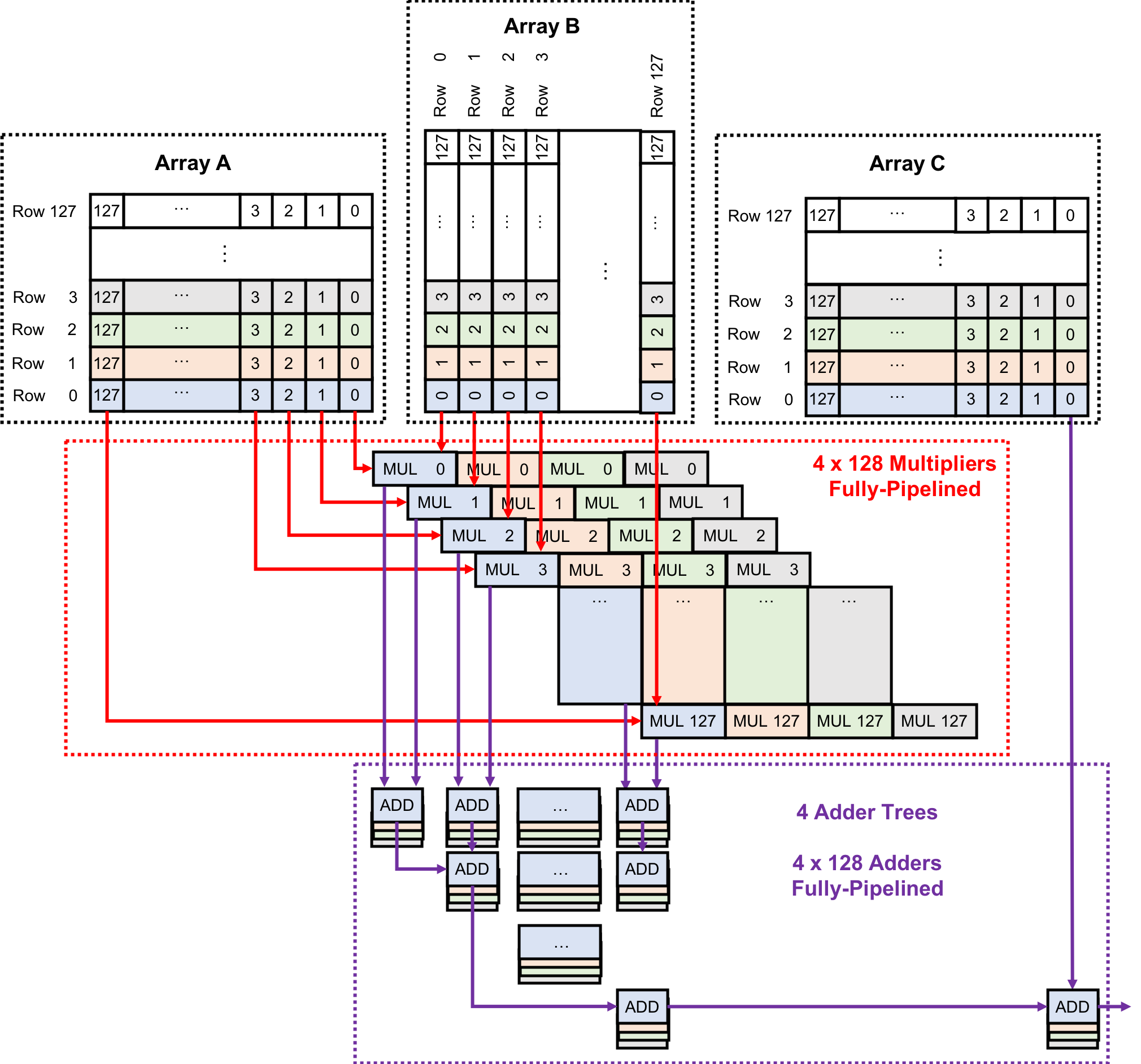}}
\caption{An overview of the matrix multiplication accelerator. Illustrating an $128 \times 128$ tile operating at 300~MHz and performing 512 multiplications and 512 additions per cycle.}
\label{fig:qfdb_sgemm_accel}
\end{figure}

Given the finite resources of the FPGA and the need to accelerate matrix multiplication of very large arrays that do not fit in internal memory (BRAM), a tiled approach was followed, and the optimized kernel was applied over the corresponding tiles of the original arrays. Based on the study of the associated area-speed trade-offs, a tile size of $128 \times 128$, as depicted in Figure~\ref{fig:qfdb_sgemm_accel}, operating at 300~MHz was selected. The kernel tile implementation unrolls the $k$ loop completely, performing 128 FP32 multiplications and 128 FP32 additions per cycle (fully-pipelined), which is similar to a vector unit operating on 4096 bits. Moreover, the $j$ loop is partially unrolled by a factor of 4, which is similar to having 4 vector units of 4096 bits. In total, the kernel tile performs 512 FP32 multiplications and 512 FP32 additions per cycle (fully-pipelined). Once the appropriate elements of the arrays are stored in on-chip accelerator memory (BRAMs), the tile execution lasts $\sim$4200 clock cycles. Three AXI HP ports (one per array) are utilized to load data into the local BRAMs, in order to continuously feed the kernel with data and match the required bandwidth. By performing a series of performance optimizations on the memory interfaces and load/unload techniques, each FPGA managed to achieve {\em 275 FP32 GFLOPS} and offload a matrix multiplication application from Linux. With all four FPGAs, a single QFDB can perform and sustain {\em more than 1 FP32 TFLOP/s}.

The matrix multiplication FP32 kernel tile of $128 \times 128$ operates at 300~MHz, and requires 153K LUTs, 300K Flip-Flops, 2057 DSP macros and 416 BRAMs. The resource utilization of each FPGA device implementing our kernel tile is 56\% of LUTs, 55\% of Flip-Flops, 46\% of BRAMs and 82\% of DSPs. Moreover, power consumption of the accelerator was measured using the power sensors residing on the QFDB.
Based on the readings of these sensors, encouraging power efficiency results were gathered.
These measurements show that the dynamic power consumption of the accelerator is 16.2~Watts, which yields 17 FP32 GFLOPS per Watt.

%% file: 8_conclusion.tex
\section{Conclusion} \label{sec:conclusion}

We presented the ExaNeSt FPGA-based prototype, 
which interconnects Xilinx Ultrascale+ MPSoC FPGAs 
using 10 to 16 Gb/s links in a 3-dimensional Torus topology. 
Each FPGA contains four ARM Cortex-A53 cores and reconfigurable logic, 
in which the firmware for the ExaNet interconnect is implemented. 
This interconnect provides user-level access 
to hundreds of RDMA channels and offers 
a fast, sub-microsecond engine for small transfers, 
and a throughput-optimized reliable-RDMA engine for MPI rendez-vous. 
The RDMA engine uses ARM's System-MMU for address translation, 
and segments transfers into blocks. 
We presented extensive evaluation results, 
using HPC codes running on up to 512 ARM Cortex-A53 cores/ranks.

To evaluate the ExaNeSt interconnect,
we run microbenchmarks and real HPC codes on a 512-core system.
The raw bandwidth inside the FPGA is 19.2 Gb/s,
while the MPI bandwidth inside one FPGA or between FPGAs in the same QFDB
is 13 Gb/s, dropping to 6.4Gb/s outside the QFDB.
Best-case latency between adjacent FPGAs is 470 ns,
split between hardware path and user-level access library.
For the weak ARM Cortex-A53 core,
the minimal MPI runtime adds almost 800 ns,
and the RDMA latency for the same message size is 5 µs.
MPI microbenchmarks reveal that broadcast latency scales as expected. Moreover,
application level results show that efficiency in all tests exceeds 69\%.
In spite of the inefficiencies introduced by the Cortex-A53 processors,
the minimal MPI port achieves good scaling efficiency
by leveraging the user-level primitives. 

We also designed a network hardware accelerator for MPI Allreduce,
and a matrix-multiplication accelerator that achieves 275 GFLOPS per MPSoC
and an energy efficiency of 17 GFLOPS per Watt for FP32 precision.
We implemented these in the FPGA logic, thus eliminating software overheads.

Overall, we consider the following to be the main contributions of the ExaNest prototype:
\begin{itemize}
    \item {\bf Multi-processor testbed 
    with configurable hardware support for communication} - 
    The first platform to integrate 256 FPGAs and offering 
    a multi-processor environment with configurable communication, as discussed in Section~\ref{sec:platform}.
    \item {\bf Dense node-level and system-level integration} - 
    At the node level, the ExaNest prototype integrates 
    4 FPGAs (ZU9EG MPSoCs), 4 DRAM modules (DDR4 SODIMMs), 
    and an M.2 Solid-State Disk (SSD) 
    on a 120mm x 130mm x 21mm compute node. 
    At the system level, the ExaNest prototype integrates 
    four such compute nodes 
    into a liquid-cooled thin (47mm thickness) blade.
    for optimal thermal management and high-density computing.
    \item {\bf Low-cost, low-latency Network Interface} 
    offering virtualization and fast completion notifications, 
    tightly coupled with the ARMv8 processing cores, as discussed in Section~\ref{sec:net_interface}.
    \item{\bf A custom MPI runtime} leveraging Network Interface's capabilities for user-level zero-copy transfers, and providing efficient communication for a variety of applications as discussed in Section~\ref{sec:evaluation}.
\end{itemize}

\section{Ongoing Work} \label{sec:ongoing_work}

We have been utilizing the ExaNeSt prototype as a platform to develop, test, and enhance High-Performance Computing (HPC) interconnect technologies. 
Within the {\em RED-SEA} EuroHPC project,  we are also developing caRVnet, the next generation of ExaNet.  caRVnet quadruples the data lane width (512 vs 128 bit) in the FPGA, and runs at 200MHz, thus achieving 100~Gb/s line rate.  
We are also implementing a testbed with a spine-leaf topology, consisting of an ExaNeSt mezzanine (4 QFDBs with 16 MPSoCs) and four central crossbars at the spine. Additionally, we are developing real-time monitoring and traffic generation tools that are capable of replaying arbitrary scenarios mixing victim and background flows. In other configurations, we are replaying background traffic to generate congestive episodes in parallel with MPI applications, and analyze the performance degradation under different congestion management schemes. We are further improving the network interface by incorporating a new hardware segmentation engine that reduces RDMA latency to below 0.5$\mu$s and significantly enhances the transfer rate for small messages. Finally, we are integrating the network interface with RISC-V processors using the CHI protocol and we are connecting it with BXI network switches~\cite{bxi}.

\section*{Acknowledgements}

The Prototype Rack described in this paper
was designed, built, and tested within the 
{\em ExaNeSt} (FET-HPC) project (2015-2019),
funded by the European Union’s
Horizon 2020 research and innovation programme
under grant agreement No 671553.
Its firmware and runtimes were improved
within the {\em EuroEXA} project (2017-2021),
similarly funded under grant agreement No 754337.
It is now being used
in large-scale congestion management experiments 
in the {\em RED-SEA} EuroHPC project (2021-2024), 
under grant agreement No 955776.

We wish to thank all the {\em ExaNeSt partners}
for their help throughout the project:
Iceotope Technologies Ltd. (UK),
Istituto Nazionale di Fisica Nucleare - INFN (IT),
the University of Manchester (UK),
EnginSoft Spa (IT),
Istituto Nazionale di Astrofisica - INAF (IT),
Exact Lab Srl (IT),
Universitat Politecnica de Valencia (ES),
MonetDB Solutions B.V. (NL),
Virtual Open Systems Sarl (FR),
Allinea Software Ltd. - now part of ARM (UK), and
Fraunhofer (DE).

We also wish to thank
the members of our CARV Laboratory at FORTH-ICS
who contributed to other related aspects of the project:
Antonis Psathakis, 
Babis Aronis, 
Panagiotis Peristerakis, 
Antonis Psistakis, 
Ioannis Vardas, 
Polyvios Pratikakis, 
Leandros Tzanakis, and
Kyriakos Paraskevas.

%% file: main.bbl
\begin{thebibliography}{10}

\bibitem{1592896}
N.~Adiga, G.~Almasi, G.~Almasi, Y.~Aridor, R.~Barik, D.~Beece, R.~Bellofatto,
  G.~Bhanot, R.~Bickford, M.~Blumrich, A.~Bright, J.~Brunheroto, C.~Cascaval,
  J.~Castanos, W.~Chan, L.~Ceze, P.~Coteus, S.~Chatterjee, D.~Chen, G.~Chiu,
  T.~Cipolla, P.~Crumley, K.~Desai, A.~Deutsch, T.~Domany, M.~Dombrowa,
  W.~Donath, M.~Eleftheriou, C.~Erway, J.~Esch, B.~Fitch, J.~Gagliano, A.~Gara,
  R.~Garg, R.~Germain, M.~Giampapa, B.~Gopalsamy, J.~Gunnels, M.~Gupta,
  F.~Gustavson, S.~Hall, R.~Haring, D.~Heidel, P.~Heidelberger, L.~Herger,
  D.~Hoenicke, R.~Jackson, T.~Jamal-Eddine, G.~Kopcsay, E.~Krevat, M.~Kurhekar,
  A.~Lanzetta, D.~Lieber, L.~Liu, M.~Lu, M.~Mendell, A.~Misra, Y.~Moatti,
  L.~Mok, J.~Moreira, B.~Nathanson, M.~Newton, M.~Ohmacht, A.~Oliner,
  V.~Pandit, R.~Pudota, R.~Rand, R.~Regan, B.~Rubin, A.~Ruehli, S.~Rus,
  R.~Sahoo, A.~Sanomiya, E.~Schenfeld, M.~Sharma, E.~Shmueli, S.~Singh,
  P.~Song, V.~Srinivasan, B.~Steinmacher-Burow, K.~Strauss, C.~Surovic,
  R.~Swetz, T.~Takken, R.~Tremaine, M.~Tsao, A.~Umamaheshwaran, P.~Verma,
  P.~Vranas, T.~Ward, M.~Wazlowski, W.~Barrett, C.~Engel, B.~Drehmel,
  B.~Hilgart, D.~Hill, F.~Kasemkhani, D.~Krolak, C.~Li, T.~Liebsch,
  J.~Marcella, A.~Muff, A.~Okomo, M.~Rouse, A.~Schram, M.~Tubbs, G.~Ulsh,
  C.~Wait, J.~Wittrup, M.~Bae, K.~Dockser, L.~Kissel, M.~Seager, J.~Vetter, and
  K.~Yates.
\newblock An overview of the bluegene/l supercomputer.
\newblock In {\em SC '02: Proceedings of the 2002 ACM/IEEE Conference on
  Supercomputing}, pages 60--60, 2002.

\bibitem{alveo-web}
I.~Advanced Micro~Devices.
\newblock Alveo adaptable accelerator cards for data center workloads, 2022.

\bibitem{10.1007/11602569_31}
S.~Agarwal, R.~Garg, and N.~K. Vishnoi.
\newblock The impact of noise on the scaling of collectives: A theoretical
  approach.
\newblock In D.~A. Bader, M.~Parashar, V.~Sridhar, and V.~K. Prasanna, editors,
  {\em High Performance Computing -- HiPC 2005}, pages 280--289, Berlin,
  Heidelberg, 2005. Springer Berlin Heidelberg.

\bibitem{ammendola2004apenet}
R.~Ammendola, M.~Guagnelli, G.~Mazza, F.~Palombi, R.~Petronzio, D.~Rossetti,
  A.~Salamon, and P.~Vicini.
\newblock Apenet: a high speed, low latency 3d interconnect network.
\newblock In {\em cluster}, page 481. Citeseer, 2004.

\bibitem{montblanc2020}
A.~Armejach, B.~Brank, J.~Cortina, F.~Dolique, T.~Hayes, N.~Ho, P.-A. Lagadec,
  R.~Lemaire, G.~López-Paradís, L.~Marliac, M.~Moretó, P.~Marcuello,
  D.~Pleiter, X.~Tan, and S.~Derradji.
\newblock Mont-blanc 2020: Towards scalable and power efficient european hpc
  processors.
\newblock In {\em DATE}, pages 136--141, 2021.

\bibitem{maxwell}
R.~Baxter, S.~Booth, M.~Bull, G.~Cawood, J.~Perry, M.~Parsons, A.~Simpson,
  A.~Trew, A.~McCormick, G.~Smart, R.~Smart, A.~Cantle, R.~Chamberlain, and
  G.~Genest.
\newblock Maxwell - a 64 fpga supercomputer.
\newblock In {\em Second NASA/ESA Conference on Adaptive Hardware and Systems
  (AHS 2007)}, pages 287--294, 2007.

\bibitem{redsea2022}
A.~Biagioni and et~al.
\newblock Red-sea: Network solution for exascale architectures.
\newblock In {\em The 25th Euromicro Conference on Digital System Design (DSD)
  and SEAA (Software Engineering and Advanced Applications) Conference}, volume
  Spain, Sept. 2022.

\bibitem{bittware-web}
BittWare.
\newblock Bittware fpga acceleration, 2019.

\bibitem{Blott16}
M.~Blott.
\newblock Reconfigurable future for hpc.
\newblock In {\em 2016 International Conference on High Performance Computing
  Simulation (HPCS)}, pages 130--131, July 2016.

\bibitem{noSQL}
B.~Brech, J.~Rubio, and M.~Hollinger.
\newblock Data engine for nosql - ibm power systems edition.
\newblock White Paper, 2015.

\bibitem{qfdb_paper}
F.~Chaix, A.~Ioannou, N.~Kossifidis, N.~Dimou, G.~Ieronymakis, M.~Marazakis,
  V.~Papaefstathiou, V.~Flouris, M.~Ligerakis, G.~Ailamakis, T.~Vavouris,
  A.~Damianakis, M.~Katevenis, and I.~Mavroidis.
\newblock Implementation and impact of an ultra-compact multi-fpga board for
  large system prototyping.
\newblock In {\em 2019 IEEE/ACM International Workshop on Heterogeneous
  High-performance Reconfigurable Computing (H2RC)}, pages 34--41, 2019.

\bibitem{chrysos2015discharging}
N.~Chrysos, L.~Chen, C.~Kachris, and M.~Katevenis.
\newblock Discharging the network from its flow control headaches: Packet drops
  and hol blocking.
\newblock {\em IEEE/ACM Transactions on Networking}, 24(1):15--28, 2015.

\bibitem{Cilardo18}
A.~Cilardo.
\newblock Htcomp: bringing reconfigurable hardware to future high-performance
  applications.
\newblock {\em {IJHPCN}}, 12(1):74--83, 2018.

\bibitem{Convey}
C.~C. Corp.
\newblock The convey hc-2 computer architectural overview (white paper).
\newblock
  https://www.micron.com/-/media/documents/products/white-paper/wp\_convey\_hc2\_architectual\_overview.pdf,
  2012.

\bibitem{minifetechreport}
P.~Crozier, H.~Thornquist, R.~Numrich, A.~Williams, H.~Edwards, E.~Keiter,
  M.~Rajan, J.~Willenbring, D.~Doerfler, and M.~Heroux.
\newblock Improving performance via mini-applications.
\newblock 01 2009.

\bibitem{6012897}
B.~Dally.
\newblock Power, programmability, and granularity: The challenges of exascale
  computing.
\newblock In {\em 2011 IEEE International Parallel Distributed Processing
  Symposium}, pages 878--878, 2011.

\bibitem{bxi}
S.~Derradji, T.~Palfer-Sollier, J.-P. Panziera, A.~Poudes, and F.~W. Atos.
\newblock The bxi interconnect architecture.
\newblock In {\em 2015 IEEE 23rd Annual Symposium on High-Performance
  Interconnects}, pages 18--25, 2015.

\bibitem{digilent-web}
{Digilent Inc. FPGA, Microcontrollers and Instrumentation}.
\newblock \url{http://www.digilent.com}, 2019.

\bibitem{doi:10.1177/1094342015593158}
J.~Dongarra, M.~A. Heroux, and P.~Luszczek.
\newblock High-performance conjugate-gradient benchmark: A new metric for
  ranking high-performance computing systems.
\newblock {\em The International Journal of High Performance Computing
  Applications}, 30(1):3--10, 2016.

\bibitem{Escobar16}
F.~A. Escobar, X.~Chang, and C.~Valderrama.
\newblock Suitability analysis of fpgas for heterogeneous platforms in hpc.
\newblock {\em IEEE Transactions on Parallel and Distributed Systems},
  27(2):600--612, Feb 2016.

\bibitem{catapultISCA}
A.~P. et~al.
\newblock A reconfigurable fabric for accelerating large-scale datacenter
  services.
\newblock In {\em {ACM/IEEE} 41st International Symposium on Computer
  Architecture, {ISCA} 2014, Minneapolis, USA, June 14-18}, pages 13--24, 2014.

\bibitem{catapultACM}
A.~P. et~al.
\newblock A reconfigurable fabric for accelerating large-scale datacenter
  services.
\newblock {\em Commun. {ACM}}, 59(11):114--122, 2016.

\bibitem{Ouyang14}
J.~O. et~al.
\newblock {SDA:} software-defined accelerator for large-scale {DNN} systems.
\newblock In {\em 2014 {IEEE} Hot Chips 26 Symposium (HCS), Cupertino, CA, USA,
  August 10-12, 2014}, pages 1--23, 2014.

\bibitem{cube}
M.~Y. et~al.
\newblock A performance evaluation of {CUBE:} one-dimensional 512 {FPGA}
  cluster.
\newblock In {\em Reconfigurable Computing: Architectures, Tools and
  Applications, 6th International Symposium, {ARC} 2010, Bangkok, Thailand,
  March 17-19, 2010. Proceedings}, pages 372--381, 2010.

\bibitem{exanesturl}
{European Exascale System Interconnect and Storage - ExaNeSt}.
\newblock \url{https://www.exanest.eu}.

\bibitem{meep}
A.~Fell, D.~J. Mazure, T.~C. Garcia, B.~Perez, X.~Teruel, P.~Wilson, and J.~D.
  Davis.
\newblock The marenostrum experimental exascale platform (meep).
\newblock {\em Supercomputing Frontiers and Innovations}, 8(1):62–81, Apr.
  2021.

\bibitem{7761639}
A.~D. George, M.~C. Herbordt, H.~Lam, A.~G. Lawande, J.~Sheng, and C.~Yang.
\newblock Novo-g\#: Large-scale reconfigurable computing with direct and
  programmable interconnects.
\newblock In {\em 2016 IEEE High Performance Extreme Computing Conference
  (HPEC)}, pages 1--7, 2016.

\bibitem{congestion18}
D.~Giannopoulos, N.~Chrysos, E.~Mageiropoulos, G.~Vardas, L.~Tzanakis, and
  M.~Katevenis.
\newblock Accurate congestion control for rdma transfers.
\newblock In {\em 2018 Twelfth IEEE/ACM International Symposium on
  Networks-on-Chip (NOCS)}, pages 1--8, 2018.

\bibitem{giannopoulos2018accurate}
D.~Giannopoulos, N.~Chrysos, E.~Mageiropoulos, G.~Vardas, L.~Tzanakis, and
  M.~Katevenis.
\newblock Accurate congestion control for rdma transfers.
\newblock In {\em 2018 Twelfth IEEE/ACM International Symposium on
  Networks-on-Chip (NOCS)}, pages 1--8. IEEE, 2018.

\bibitem{rivyera}
S.~GmbH.
\newblock Sciengines hardware, high performance reconfigurable computing, 2019.

\bibitem{4100410}
R.~L. Graham, G.~M. Shipman, B.~W. Barrett, R.~H. Castain, G.~Bosilca, and
  A.~Lumsdaine.
\newblock Open mpi: A high-performance, heterogeneous mpi.
\newblock In {\em 2006 IEEE International Conference on Cluster Computing},
  pages 1--9, 2006.

\bibitem{GROPP1996789}
W.~Gropp, E.~Lusk, N.~Doss, and A.~Skjellum.
\newblock A high-performance, portable implementation of the mpi message
  passing interface standard.
\newblock {\em Parallel Computing}, 22(6):789--828, 1996.

\bibitem{4536494}
T.~Hoefler, T.~Schneider, and A.~Lumsdaine.
\newblock Accurately measuring collective operations at massive scale.
\newblock In {\em 2008 IEEE International Symposium on Parallel and Distributed
  Processing}, pages 1--8, 2008.

\bibitem{hpcgsite}
{HPCG Benchmark}.
\newblock \url{https://hpcg-benchmark.org/index.html}.

\bibitem{7426807}
S.~Hunold and A.~Carpen-Amarie.
\newblock Reproducible mpi benchmarking is still not as easy as you think.
\newblock {\em IEEE Transactions on Parallel and Distributed Systems},
  27(12):3617--3630, 2016.

\bibitem{iceotope}
ICEOTOPE.
\newblock Iceotope: Precision immersion cooling from the cloud to the edge,
  2022.

\bibitem{trets20}
A.~D. Ioannou, K.~Georgopoulos, P.~Malakonakis, D.~N. Pnevmatikatos, V.~D.
  Papaefstathiou, I.~Papaefstathiou, and I.~Mavroidis.
\newblock Unilogic: A novel architecture for highly parallel reconfigurable
  systems.
\newblock volume~13, New York, NY, USA, Sept. 2020. Association for Computing
  Machinery.

\bibitem{Jones03impactsof}
T.~R. Jones, L.~B. Brenner, J.~M. Fier, T.~R. Jones, L.~B. Brenner, and J.~M.
  Fier.
\newblock Impacts of operating systems on the scalability of parallel
  applications.
\newblock Technical report, Lawrence Livermore National Laboratory, 2003.

\bibitem{GSAS2}
N.~D. Kallimanis, N.~Chrysos, and M.~Marazakis.
\newblock A flexible \& efficient shared memory abstraction with minimal hw
  assistance.
\newblock May 2018.

\bibitem{GSAS}
N.~D. Kallimanis, M.~Marazakis, and N.~Chrysos.
\newblock Gsas: A fast shared memory abstraction with minimal hardware support.
\newblock In {\em EuroExaScale workshop - HiPEAC 2019}, Valencia, Spain,
  January 2019.

\bibitem{5289226}
G.~Kalokerinos, V.~Papaefstathiou, G.~Nikiforos, S.~Kavadias, M.~Katevenis,
  D.~Pnevmatikatos, and X.~Yang.
\newblock Fpga implementation of a configurable cache/scratchpad memory with
  virtualized user-level rdma capability.
\newblock In {\em 2009 International Symposium on Systems, Architectures,
  Modeling, and Simulation}, pages 149--156, 2009.

\bibitem{7723536}
M.~Katevenis, N.~Chrysos, M.~Marazakis, I.~Mavroidis, F.~Chaix, N.~Kallimanis,
  J.~Navaridas, J.~Goodacre, P.~Vicini, A.~Biagioni, P.~S. Paolucci,
  A.~Lonardo, E.~Pastorelli, F.~L. Cicero, R.~Ammendola, P.~Hopton, P.~Coates,
  G.~Taffoni, S.~Cozzini, M.~Kersten, Y.~Zhang, J.~Sahuquillo, S.~Lechago,
  C.~Pinto, B.~Lietzow, D.~Everett, and G.~Perna.
\newblock The exanest project: Interconnects, storage, and packaging for
  exascale systems.
\newblock In {\em 2016 Euromicro Conference on Digital System Design (DSD)},
  pages 60--67, 2016.

\bibitem{katevenis2007interprocessor}
M.~G. Katevenis.
\newblock Interprocessor communication seen as load-store instruction
  generalization.
\newblock In {\em In The Future of Computing, essays in memory of Stamatis
  Vassiliadis, K. Bertels ea (Eds.), Delft, The Netherlands}. Citeseer, 2007.

\bibitem{10.1145/3149457.3149479}
R.~Kobayashi, Y.~Oobata, N.~Fujita, Y.~Yamaguchi, and T.~Boku.
\newblock Opencl-ready high speed fpga network for reconfigurable high
  performance computing.
\newblock In {\em Proceedings of the International Conference on High
  Performance Computing in Asia-Pacific Region}, HPC Asia 2018, page 192–201,
  New York, NY, USA, 2018. Association for Computing Machinery.

\bibitem{kornaros1998implementation}
G.~Kornaros, D.~Pnevmatikatos, P.~Vatsolaki, G.~Kalokerinos, C.~Xanthaki,
  D.~Mavroidis, D.~Serpanos, and M.~Katevenis.
\newblock Implementation of atlas i: a single-chip atm switch with
  backpressure.
\newblock In {\em Proceedings of the IEEE Hot Interconnects 6 Symposium}, pages
  85--96. Citeseer, 1998.

\bibitem{lammpsuserguide}
{Lammps user guide}.
\newblock \url{https://docs.lammps.org/Speed_bench.html}.

\bibitem{jcc2015}
P.~Lin, M.~Heroux, R.~Barrett, and A.~Williams.
\newblock Assessing a mini-application as a performance proxy for a finite
  element method engineering application.
\newblock {\em Concurrency and Computation: Practice and Experience}, 08 2015.

\bibitem{kexec}
{Kexec}.
\newblock \url{https://linux.die.net/man/8/kexec}, 2021.

\bibitem{hitechglobal-web}
H.~G. LLC.
\newblock Xilinx / altera fpga boards, design services, \& ip cores, 2019.

\bibitem{569696}
E.~Markatos and M.~Katevenis.
\newblock User-level dma without operating system kernel modification.
\newblock In {\em Proceedings Third International Symposium on High-Performance
  Computer Architecture}, pages 322--331, 1997.

\bibitem{maxelerMPC-web}
{Maxeler Technologies Inc. Maxeler Products}.
\newblock \url{https://www.maxeler.com/products/}, 2019.

\bibitem{amazon-ec2-f1}
{Amazon.com Inc. Amazon EC2 F1 Instances}.
\newblock \url{https://aws.amazon.com/ec2/instance-types/f1/}, 2019.

\bibitem{maxelerDFE-web}
{Maxeler Technologies Inc. Dataflow Computing}.
\newblock \url{https://www.maxeler.com/technology/dataflow-computing/}, 2019.

\bibitem{Oleksiak2019}
A.~Oleksiak, M.~Kierzynka, W.~Piatek, M.~vor~dem Berge, W.~Christmann,
  S.~Krupop, M.~Porrmann, J.~Hagemeyer, R.~Griessl, M.~Peykanu, L.~Tigges,
  S.~Rosinger, D.~Schlitt, C.~Pieper, U.~Janssen, H.~Rauchfuss, G.~Agosta,
  A.~Barenghi, C.~Brandolese, W.~Fornaciari, G.~Pelosi, J.~Pita~Costa,
  M.~Cecowski, R.~Plestenjak, J.~Cinkelj, L.~Cudennec, T.~Goubier, J.-M.
  Philippe, C.~Adeniyi-Jones, J.~Setoain, and L.~Ceva.
\newblock {\em M2DC---A Novel Heterogeneous Hyperscale Microserver Platform},
  pages 109--128.
\newblock Springer International Publishing, Cham, 2019.

\bibitem{osumicrobenchs}
{OSU Micro-Benchmarks}.
\newblock \url{http://mvapich.cse.ohio-state.edu/benchmarks/}.

\bibitem{paraskevas2018virtualized}
K.~Paraskevas, N.~Chrysos, V.~Papaefstathiou, P.~Xirouchakis, P.~Peristerakis,
  M.~Giannioudis, and M.~Katevenis.
\newblock Virtualized multi-channel rdmawith software-defined scheduling.
\newblock {\em Procedia Computer Science}, 136:82--90, 2018.

\bibitem{maxelerDFE}
O.~Pell and V.~Averbukh.
\newblock Maximum performance computing with dataflow engines.
\newblock {\em Computing in Science and Engineering}, 14(4):98--103, 2012.

\bibitem{maxelerDFE-SIGARCH}
O.~Pell and O.~Mencer.
\newblock Surviving the end of frequency scaling with reconfigurable dataflow
  computing.
\newblock {\em SIGARCH Comput. Archit. News}, 39(4):60--65, Dec. 2011.

\bibitem{1592958}
F.~Petrini, D.~Kerbyson, and S.~Pakin.
\newblock The case of the missing supercomputer performance: Achieving optimal
  performance on the 8,192 processors of asci q.
\newblock In {\em SC '03: Proceedings of the 2003 ACM/IEEE Conference on
  Supercomputing}, pages 55--55, 2003.

\bibitem{Plessl18}
C.~Plessl.
\newblock Bringing fpgas to hpc production systems and codes.
\newblock In {\em Fourth International Workshop on Heterogeneous
  High-performance Reconfigurable Computing}, workshop at Supercomputing, 2018.

\bibitem{ploumidisExacom}
M.~Ploumidis, N.~D. Kallimanis, M.~Asiminakis, N.~Chrysos, P.~Xirouchakis,
  M.~Gianoudis, L.~Tzanakis, N.~Dimou, A.~Psistakis, P.~Peristerakis,
  G.~Kalokairinos, V.~Papaefstathiou, and M.~Katevenis.
\newblock Software and hardware co-design for low-power hpc platforms.
\newblock page 88–100, Berlin, Heidelberg, 2019. Springer-Verlag.

\bibitem{psis_tpds}
A.~Psistakis, N.~Chrysos, F.~Chaix, M.~Asiminakis, M.~Gianioudis,
  P.~Xirouchakis, V.~Papaefstathiou, and M.~Katevenis.
\newblock Optimized page fault handling during rdma.
\newblock {\em IEEE Transactions on Parallel and Distributed Systems}, pages
  1--1, 2022.

\bibitem{psistakis2020part}
A.~Psistakis, N.~Chrysos, F.~Chaix, M.~Asiminakis, M.~Giannioudis,
  P.~Xirouchakis, V.~Papaefstathiou, and M.~Katevenis.
\newblock Part: Pinning avoidance in rdma technologies.
\newblock In {\em 2020 14th IEEE/ACM International Symposium on
  Networks-on-Chip (NOCS)}, pages 1--8. IEEE, 2020.

\bibitem{sequana-web}
A.~SE.
\newblock Bullsequana x supercomputers - atos, 2022.

\bibitem{summitNetwork}
C.~B. Stunkel, R.~L. Graham, G.~Shainer, M.~Kagan, S.~S. Sharkawi,
  B.~Rosenburg, and G.~A. Chochia.
\newblock The high-speed networks of the summit and sierra supercomputers.
\newblock {\em IBM Journal of Research and Development}, 64(3/4):3:1--3:10,
  2020.

\bibitem{9041710}
G.~Taffoni, L.~Tornatore, D.~Goz, A.~Ragagnin, S.~Bertocco, I.~Coretti,
  M.~Marazakis, F.~Chaix, M.~Plumidis, M.~Katevenis, R.~Panchieri, and
  G.~Perna.
\newblock Towards exascale: Measuring the energy footprint of astrophysics hpc
  simulations.
\newblock In {\em 2019 15th International Conference on eScience (eScience)},
  pages 403--412, 2019.

\bibitem{LAMMPS}
A.~P. Thompson, H.~M. Aktulga, R.~Berger, D.~S. Bolintineanu, W.~M. Brown,
  P.~S. Crozier, P.~J. in~'t Veld, A.~Kohlmeyer, S.~G. Moore, T.~D. Nguyen,
  R.~Shan, M.~J. Stevens, J.~Tranchida, C.~Trott, and S.~J. Plimpton.
\newblock {LAMMPS} - a flexible simulation tool for particle-based materials
  modeling at the atomic, meso, and continuum scales.
\newblock {\em Comp. Phys. Comm.}, 271:108171, 2022.

\bibitem{ZynqTRM}
Zynq ultrascale+ device technical reference manual, 2018.

\bibitem{amd-hacc}
A.~Xilinx.
\newblock Heterogeneous accelerated compute clusters, 2016.

\bibitem{xchipscope}
{Integrated Logic Analyzer v6.2}.
\newblock \url{https://docs.xilinx.com/v/u/en-US/pg172-ila}.

\end{thebibliography}
